\documentclass[pra,aps,twocolumn,showpacs]{revtex4}

\usepackage{palatino}

\usepackage[english]{babel}
\usepackage{multirow}

\usepackage{pifont}                           
\usepackage{amssymb}                           
\usepackage{amsmath}                           
\usepackage{graphicx}                          
\usepackage{bm}                                
\usepackage[pdftex]{color}

\newcommand{\NCphiZero}{\phi_0}
\newcommand{\NCphiSecond}{\phi_2}
\newcommand{\sGPphi}{\psi}
\newcommand{\sGPmu}{\mu}
\newcommand{\omegaZ}{\omega}
\newcommand{\ellZ}{\ell}
\newcommand{\Tc}{T_{c}}
\newcommand{\gOne}{g^{(1)}}
\newcommand{\gTwo}{g^{(2)}}
\newcommand{\BE}{\bar N}

\begin{document}

\title{A comparison between microscopic methods for finite temperature Bose gases}

\author{S.P. Cockburn$^1$}\email[E-mail: ]{s.p.cockburn@ncl.ac.uk}
\author{A. Negretti$^{2,3}$}
\author{N.P. Proukakis$^1$}
\author{C. Henkel$^4$}

\affiliation{
1. School of Mathematics and Statistics, University of Newcastle upon Tyne, Newcastle upon Tyne, 
NE1 7RU, United Kingdom\\
2. Lundbeck Foundation Theoretical Center for Quantum System
Research,\\
Department of Physics and Astronomy, Aarhus University,
8000 Aarhus C, Denmark\\
3. Institut f\"ur Quanteninformationsverarbeitung, Universit\"at Ulm, Albert-
Einstein-Allee 11, 89069 Ulm, Germany\\
4. Institut f\"ur Physik und Astronomie, Universit\"at Potsdam, Karl-Liebknecht-Str. 24-25, 14476 Potsdam, Germany}

\date{Tues 7 Dec 2010}


\begin{abstract}
We analyze the equilibrium properties
of a weakly interacting, trapped quasi-one-dimensional
Bose gas at finite temperatures and compare different
theoretical approaches.
We focus in particular on two stochastic theories: a
number-conserving Bogoliubov (ncB) approach and a
stochastic Gross-Pitaevskii equation (sGPe) that have been
extensively used in numerical simulations.
Equilibrium
properties like density profiles, correlation functions, and
the condensate statistics are compared to predictions
based upon a number of alternative
theories. We find that due to thermal phase fluctuations, 
and the corresponding condensate depletion,
the ncB approach loses its validity at relatively
low temperatures.
This can be 
attributed to the change in the Bogoliubov spectrum, as the
condensate gets thermally depleted, and to large
fluctuations beyond perturbation
theory. Although the two stochastic theories are built on
different thermodynamic ensembles (ncB: canonical, 
sGPe: grand-canonical), they yield the correct condensate
statistics in a large BEC (strong enough particle interactions).
For smaller systems, the sGPe results are prone to anomalously
large number fluctuations, well-known for the grand-canonical,
ideal Bose gas. 
Based on the comparison of the above theories to the modified
Popov approach, we propose a simple procedure for approximately extracting
the Penrose-Onsager condensate from first- and second-order
correlation functions that is computationally convenient.
This also clarifies the link between condensate and quasi-condensate 
in the Popov theory of low-dimensional systems.
\end{abstract}

\maketitle

\tableofcontents

\section{Introduction}


Following the first observation of Bose-Einstein condensation of
dilute gases, experimental and
theoretical efforts were mainly focused on the fundamental properties
of such degenerate quantum gases, including spatial and momentum
distributions, and collective excitations \cite{PethickBook,PitaevskiiBook}. 
Mean field theory was initially found
to be impressively successful in most cases. 
Indeed, at temperatures well below the phase transition, nearly all
atoms occupy one wave function that satisfies a nonlinear Schr\"odinger
equation, the celebrated Gross-Pitaevskii equation (GPe).
The nonlinear term describes the mean field potential experienced
by the atoms due to two-body interactions. In the language of quantum field theory, 
the GPe yields a zeroth-order approximation to the full matter-wave field
where both the non-condensed component of the gas and quantum
fluctuations are neglected.

The Bogoliubov theory provides an improved analysis of this system by
including small fluctuations around the condensate wave function. 
Its predictions include, e.g., the spectrum of collective excitations, the 
quantum depletion of the condensate, and correlation functions at both
zero and nonzero temperature. We focus
in this paper on a one-dimensional trapped gas as a model for a weakly interacting
quasi-one-dimensional system confined tightly in the radial direction. In such a 
system, the contribution of low-energy modes is significant.
From Bogoliubov theory, these modes
mainly affect the phase of the matter-wave field, the density fluctuations
being relatively weak. As a consequence, there is no Bose condensate
in the homogeneous limit. Still, a so-called quasi-condensate can be
identified
where long-range coherence manifests itself in the suppression of
density fluctuations,
while the phase is correlated only over distances smaller than 
the system size \cite{PopovBook,Petrov2000,Kagan2000,Andersen2002,AlKhawaja2002,Gies2004a}.
The situation is similar to a ``fragmented'' condensate where several 
low-energy modes appear with comparable weight \cite{Leggett2001}. 

The (quasi-)condensate
and atoms in excited states (``thermal cloud'') 
are often treated as two subsystems
that are coupled to each other by scattering processes that exchange
particles and energy. 
Approaches based on such a splitting between condensate and thermal
dynamics 
lead to a generalized GPe for the condensate dynamics
that differs from its $T = 0$ counterpart through the 
inclusion of the thermal cloud mean 
field (Hartree-Fock potential). In addition, a source term may describe the 
scattering of particles 
between the condensate and thermal cloud 
\cite{Proukakis1998a,Shi1998}.
The thermal cloud itself is described by a quantum Boltzmann equation
 \cite{Luiten1996,Jaksch1997,Holland1997} 
self-consistently coupled to the condensate
\cite{Zaremba1999,Walser1999,Bijlsma2000,Gardiner2000,Proukakis2001}.
In its kinetic formulation, the resulting
self-consistent, coupled Gross-Pitaevskii-Boltzmann approach,
which extends earlier work by Kirkpatrick and Dorfman \cite{Kirkpatrick1983},
and Eckern \cite{Eckern1984},
is often referred to as the ``ZNG'' scheme within the context of 
trapped atomic gases \cite{Zaremba1999,ZNGbook}.
This method reproduces the two-fluid hydrodynamics 
in the collisional, hydrodynamic regime \cite{Nikuni1999,Nikuni2001},
and has been tested successfully against experiment for collective 
modes \cite{Jackson2001,Jackson2002,Jackson2002b} and 
macroscopic excitations \cite{Jackson2007,Jackson2009}.
Since the ZNG approach is numerically formulated in a purely 3-dimensional 
context,
we shall not be considering it further within the present work, based on purely
one-dimensional simulations.

Despite their elegant formulation, kinetic theories based on mean field potentials
have the
drawback in lower dimensions of not fully  
capturing the large phase fluctuations in the quasi-condensate. 
In addition, so-called anomalous averages (or pair correlations) in the thermal 
cloud, typically omitted in such approaches,  
are expected to become particularly relevant at lower dimensions. It is not
entirely clear, however, how to obtain a gapless excitation spectrum for the 
system in the homogeneous limit, as required by the Hugenholtz-Pines 
theorem~\cite{Griffin1996, Proukakis1998b, Hutchinson2000, Andersen2002, 
AlKhawaja2002, Tommasini2005, Yukalov2006}. 
A modified mean field theory for low-dimensional quasi-condensates 
was developed by Stoof's group and one of the present co-authors
\cite{Andersen2002, AlKhawaja2002},
building on previous path-integral approaches pioneered by Popov
\cite{PopovBook,Kagan2000}.
In this ``modified
Popov theory'', the infrared divergences due to phase fluctuations  
are systematically removed, leading to a gapless, convergent and
computationally convenient scheme that applies in all dimensions, for
homogeneous and trapped systems. 
This approach has been used in particular
to construct the phase diagram of weakly-interacting
1D systems \cite{AlKhawaja2003}, and to investigate the interplay of 
density and phase fluctuations \cite{Proukakis2006c}.

The quasi-condensate dynamics at nonzero temperature is a challenging 
problem as the Bogoliubov approximation becomes invalid, at least at large times 
\cite{Sinatra2001,Sinatra2002}, and large thermal phase fluctuations have 
to be taken into account even at low temperatures where density
fluctuations are small \cite{Mora2003}.
Classical field methods have been developed to simulate numerically 
those modes of the system that feature a macroscopic occupation.
These methods rely upon the observation that for highly occupied modes,
the field operator can be replaced by a classical complex field which evolves
in time according to the GPe \cite{Svistunov1991,Davis2001,
Goral2002}. 
This description extends the $T=0$ GPe, by adding
stochastic elements that describe fluctuations of the
(quasi-)\-condensate modes;
these may be further coupled to the thermal cloud where the mean
occupation numbers are small. (In fact, too small to be treated in the 
classical field approximation, and more appropriately
described by quantum Boltzmann equations \cite{Stoof1999}.)
Within the class of classical field techniques, we mention
the projected GPe (pGPe) \cite{Davis2001,Blakie2005}, the truncated 
Wigner (tW) method \cite{Steel1998, Sinatra2000, Sinatra2001, Polkovnikov2003} 
the stochastic Gross-Pitaevskii equation (sGPe), when implemented 
in the classical limit
\cite{Stoof1999,Stoof2001,Duine2001,Gardiner2002,Gardiner2003,Proukakis2008,
Blakie2008}, 
and closely related classical field methods 
\cite{Goral2002, Berloff2002}.
Hybrid simulation techniques were also recently developed that attempt to
go beyond the classical limit \cite{Hoffmann2008,Deuar2009,Heller2009,Heller2010}.

These stochastic
approaches, the relation between them and other kinetic theories and their respective applications have been reviewed in Refs.\cite{Proukakis2008,Blakie2008,Brewczyk2007,Polkovnikov10}.
A key appeal is that they provide an approximation to
the full distribution function of the ultracold gas and give access to
physics beyond the mean field. They have been used, e.g., to probe the large
critical fluctuations near the phase transition
\cite{Davis2006,Weiler2008,Bezett2009b}, to
study dynamical effects of fluctuations on condensate growth 
\cite{Stoof2001,Duine2001,Proukakis2006a}
and macroscopic excitations \cite{Duine2004, Cockburn2010, Damski2010, Martin2010, Rooney2010}.
Another quantity of interest is the 
counting statistics of the condensate mode 
\cite{Kocharovsky2006, Svidzinsky2006, Bezett2009b, Idziaszek2009, Bienias2010}, 
which is analogous 
to the photon number distribution in quantum laser theory \cite{SargentScully}.

As the use of classical field simulations becomes more widespread, 
a quantitative study of their relative properties is essential.
The main purpose of this paper is to 
initiate such a quantitative study by
comparing two methods that can be implemented with reasonable effort,
each of which generates an ensemble of stochastic initial states to
mimic a finite-temperature Bose gas at equilibrium in a trap.
For simplicity, we focus on a
one-dimensional, weakly interacting system.

\begin{table*}[t]
  \begin{center}
    \begin{tabular}{ | c | l | c | c | l | c | c |}
       \hline 
  	\multicolumn{1}{|c|}{TEMPERATURE} &
  	\multicolumn{1}{|c|}{PHYSICAL PROPERTY} &
        \multicolumn{1}{|c|}{SGPE} & 
        \multicolumn{1}{|c|}{NCB} & 
        \multicolumn{1}{|c|}{`BENCHMARK' THEORY} & 
        \multicolumn{1}{|c|}{REF.} &
        \multicolumn{1}{|c|}{SEC.} 
        \\ 
  	\multicolumn{1}{|c|}{REGIME} &
  	\multicolumn{1}{|c|}{} &
        \multicolumn{1}{|c|}{} & 
        \multicolumn{1}{|c|}{} & 
        \multicolumn{1}{|c|}{FOR COMPARISON} & 
        \multicolumn{1}{|c|}{} &
        \multicolumn{1}{|c|}{} \\
        \hline
        \hline

\multirow{8}{*}{ $T < T_\phi / 2$ } 
        & Density profiles: Total & \ding{51} & \ding{51}
                             & modified Popov & \cite{Andersen2002,AlKhawaja2002}  
                             & \ref{subsec:total-density} \\ \cline{2-7}

         & Density profiles: condensate (Penrose-Onsager) & \ding{51} & \ding{51}
                             & modified Popov & \cite{Andersen2002,AlKhawaja2002} 
                             & \ref{subsec:PO-densities} \\ \cline{2-7}

         & \multirow{2}{*}{Spatial correlation function: $1^{\rm st}$ order}

         & \multirow{2}{*}{\ding{51}} & \multirow{2}{*}{\ding{51}} & modified Popov
                      & \cite{Andersen2002,AlKhawaja2002} 
                      & \multirow{2}{*}{\ref{s:g1}} \\ \cline{5-6}
                      
         & & & & Petrov {\it et al.} & \cite{Petrov2000} & \\ \cline{2-7}

         & Spatial correlation function: $2^{\rm nd}$ order & \ding{51} & \ding{51}  
                             & Kheruntsyan {\it et al.} & \cite{Kheruntsyan2005} 
                             & \ref{s:g2} \\ \cline{2-7}

         & Condensate statistics  & \ding{51} & \ding{51}
                             & Svidzinsky and Scully & \cite{Svidzinsky2006} 
                             & \ref{subsec:cond-stats} \\ \cline{2-7}

         & Pair anomalous average & \ding{51} & \ding{51}
                             & ($T=0$) Bogoliubov theory 
                             & \cite{PethickBook, PitaevskiiBook} 
                             & \ref{s:squeezing} \\ \hline

$ 0 < T < \Tc$ 
         & Quasi-condensate/condensate density profiles & \ding{51} & 
                             & modified Popov & \cite{Andersen2002,AlKhawaja2002}  
                             & \ref{subsec:quasi-vs-PO-condensate} \\ \hline

    \end{tabular}
  \end{center}
  \caption[]{Summary of `benchmark' theories used in analyzing the stochastic approaches.
             A tick in the `ncB' or `sGPe' columns indicates inclusion in the 
             comparison. 
             $T_\phi$ [Eq.(\ref{eq:def-T_phi-and-T_c})] is the critical temperature 
             above which the system's coherence length is smaller than its size
             due to phase fluctuations \cite{Petrov2000}.
             $T_{c}$ [Eq.(\ref{eq:def-T_phi-and-T_c})] is the critical temperature 
             for an ideal Bose
             gas in a one-dimensional harmonic trap \cite{Ketterle1996}.
             These temperatures are illustrated in
             Fig.\ref{fig:TN-plane} in relation to our parameters.
  }
  \label{tab:theory_sum}
\end{table*}

{\it Number Conserving Bogoliubov (tWncB):}
The first method is a Bogoliubov approach 
in which the total atom number $N$ is conserved
(formulated within the canonical ensemble)
\cite{Girardeau1959, Gardiner1997, Girardeau1998, Castin1998, Sinatra2000, Morgan2000}. 
This method
has been used as a starting point for dynamical
calculations within the truncated Wigner approximation 
(see Ref.\cite{Polkovnikov10} for a review). 
We henceforth denote this by tWncB.
The tWncB field contains both condensate and non-condensate modes,
calculated from the Bogoliubov-de Gennes equations. The modes
amplitudes are sampled to capture
both quantum and thermal fluctuations. 
We adopt here a formulation developed 
in a low-temperature expansion around the Gross-Pitaevskii 
mean field \cite{Castin1998, Sinatra2000}, assuming that 
$N_{\rm th}$, 
the number of atoms in non-condensate modes, is small 
compared to $N$.

{\it Stochastic Gross-Pitaevskii Equation (sGPe):}
The second method is the stochastic Gross-Pitaevskii equation (sGPe),
which prepares a grand-canonical ensemble dynamically by simulating 
a Langevin equation (see Refs.\cite{Stoof1999,Stoof2001,Duine2001} for
details of the scheme used here). 
We consider it here within the classical approximation 
where the mode occupations are large.
The stochastic field in the sGPe represents the low-lying
modes of the field which are coupled to a thermal cloud, 
treated as a heat bath.
The exchange of particles and energy through incoherent scattering
processes between these two sub-systems is represented
by a damping term and the Langevin force in the sGPe \cite{Stoof1999}.

We wish to show that there exist regimes where
the two methods are equivalent despite the physically very different pictures
behind them. To this end, we calculate and analyze relevant observables
like density profiles, spatial correlation functions, and condensate statistics.
Where feasible, we also compare to alternative finite temperature
theories, as detailed in Table \ref{tab:theory_sum}. 
This study is by no means complete (e.g., there is no comparison 
to `ZNG' or pGPe),
but we believe that it represents an important step towards benchmarking commonly used simulation methods for finite-temperature Bose gases. 
Other comparisons undertaken to 
date are summarized in Ref.\cite{Proukakis_ICPress}.

More specifically, we explain the origin of 
discrepancies between the two methods
considered here,
building on previous investigations of the 
validity conditions of the tWncB \cite{Sinatra2002}. 
We find in particular that the low-temperature (or small $N_{\rm th}$)
expansion behind the tWncB breaks down quite early, as the temperature
$T$ increases towards the characteristic temperature $T_\phi$
for phase coherence within a trap \cite{Petrov2000}:
\begin{equation}
	k_B T_\phi = N \frac{ (\hbar \omegaZ)^2 }{ \mu }
	< k_B \Tc =  \frac{ N }{ \ln 2 N } \hbar \omegaZ
	\label{eq:def-T_phi-and-T_c}
\end{equation}
($\omegaZ$ is the trap frequency and $\mu$ the chemical potential)
where $\Tc$ is the critical temperature for Bose-Einstein condensation 
in an ideal trapped Bose gas in 1D \cite{Ketterle1996}.
This failure, that also happens when the theory of 
Ref.\cite{Svidzinsky2006} is applied to an interacting,
trapped gas, is attributed to a distribution function for $N_{\rm th}$ that 
is broadened by thermal fluctuations beyond the limit set by the total number 
of particles $N$. 
In addition, these fluctuations are overestimated because of spurious
contributions from phase fluctuations.
We have observed that the breakdown of the tWncB approach is not completely
``cured'' by propagating the stochastic ensemble of wave fields under the
GPe.

At low temperatures and for smaller systems, we have found
large number fluctuations in the sGPe results, in particular in
the counting statistics of the condensate.
This is related to the anomalous number fluctuations
of the ideal gas in the grand-canonical ensemble 
\cite{Grossmann1996, Wilkens1997, Kocharovsky2006}. This feature 
does not occur for the canonical tWncB method, and it is removed
in larger condensates where particle interactions become important.
We emphasize that this agreement
illustrates how moments of the quantum field of very high order are 
correctly reproduced by the stochastic approaches.

In addition, we discuss the influence of the thermal
(non-condensate) density $n_{\rm th}( z )$ and the so-called anomalous
average (or pair correlation) $m( z )$ of the non-condensate 
field, by analyzing their back-action on the shape of the (Penrose-Onsager) 
condensate density. 
This anomalous average is related to both a renormalization of the
particle interactions due to the background field, and to the Landau and
Beliaev damping of condensate excitations (together with triple averages)
\cite{Bijlsma1997, Varenna_Burnett1998, Proukakis1998a, Shi1998, Rusch1999, Hutchinson2000, 
Morgan2000, AlKhawaja2002, Yukalov2006,
PitaevskiiBook, Zaremba1999, Pitaevskii1997, Giorgini1998, Giorgini2000, Fedichev1998}.
%

Following on from this, we discuss the connection between the condensate
mode, as obtained by applying the Penrose-Onsager criterion, and the quasi-condensate,
as predicted by the modified Popov theory of Andersen {\it et al.}
\cite{Andersen2002,AlKhawaja2002}. In stochastic theories, the
condensate mode is commonly extracted {\it a posteriori} from the total 
matter field ensemble, which can become a prohibitive computational task
in large systems. 
This is in stark contrast to kinetic theories based on symmetry breaking, where the
condensate is a separate quantity, obeying its own equation
of motion.
We extend the analysis of Ref.\cite{AlKhawaja2002erratum}
to construct an approximate formula for the condensate density involving
first- and 
second-order correlation functions which are straightforwardly obtained.
This illustrates the conceptual difference between the Penrose-Onsager condensate
and the quasi-condensate.

The paper is organized as follows:
We first describe in Sec.~\ref{sec:methods} 
the procedure of initial state generation
within each of the two selected stochastic approaches, 
highlighting the key 
conceptual differences between them. We also briefly review the
modified Popov theory that is used to benchmark 
a number of our results.
Sec.~\ref{sec:equilibrium_results} addresses
the equilibrium properties
such as density profiles and correlation functions,
comparing to other, pertinent theoretical results where appropriate.
In Sec.\ref{subsec:cond-stats}, the condensate statistics is discussed
and some features of the one-body density matrix in the 
quasi-condensate regime are illustrated. 
Sec.~\ref{sec:quasi-cond} uses the modified Popov approach to
address the physical meaning of the (Penrose-Onsager)
condensate mode, which may be extracted from the 
stochastic theories, and contrast this to the quasi-condensate concept.
Sec.~\ref{sec:tWncB-thermalization} shows 
that the ncB initial state under GPe evolution
does not lead to improved predictions for equilibrium properties.
Sec.~\ref{sec:conc} summarizes our results, with some additional 
material presented in two appendices for completeness.
%

\section{Simulation techniques under consideration}
\label{sec:methods}

Each of the stochastic simulation techniques we describe here are based on the mapping
of a quantum field theory of atoms to noisy c-number fields. 
In this section we discuss them in turn, paying particular attention to
two important practical elements: (i) the method of equilibrium 
initial state generation and (ii) the nature in which the GPe arises as an
energy and number conserving means to treat the system 
dynamics away from equilibrium.

\subsection{Truncated Wigner}
%
In the truncated Wigner (tW) method,
the temporal dynamics is governed by the familiar 
nonlinear Schr\"odinger or Gross-Pitaevskii equation (GPe):
\begin{equation}
  \begin{split}
    i\hbar\frac{\partial \psi }{\partial t} &=
    H_{\rm GP}[|\psi|^2] \psi - \mu \psi
    ,
    \\
    H_{\rm GP}[|\psi|^2] &= 
    -\frac{\hbar^2}{2m}\frac{\partial^2}{\partial z^2} + V(z) 
    + g |\psi(z)|^2
    ,
  \end{split}
\label{eqn:GPE_TW}
\end{equation}
where the nonlinear Hamiltonian $H_{\rm GP}$ contains the
trapping potential $V(z)$ and
the effective two-body interaction strength $g$,
and
$\mu$ is the chemical potential.
While Eq.\eqref{eqn:GPE_TW} looks identical to the usual $T=0$ GPe
equation, the meaning of $\psi$ is quite distinct:
(i) The complex field $\psi(z, t)$ represents the condensate, 
its elementary excitations, and the ``thermal cloud'' surrounding it.
(ii) The initial conditions are stochastic and include quantum and 
thermal fluctuations of the condensate and its excitations.
This is essential for incorporating spontaneous processes (scattering or decay)
that are not captured within mean field theory, see, e.g. Ref.\cite{Norrie2005,
Norrie_thesis}.
(iii) Averages of operator products are
first symmetrized before being mapped to classical fields, so that
the one-body density matrix becomes, for example,
\begin{equation}
\langle \hat\Psi^\dag( z ) \hat\Psi( z' ) \rangle \mapsto
\langle \psi^*(z) \psi( z' ) \rangle_W - n_q \delta_{z, z'}
	\label{eq:mapping-correlation-function}
\end{equation}
where the average $\langle \ldots \rangle_W$ is taken over the
initial conditions.
The second term is a $\delta$-function on a spatial grid and proportional
to the ``quantum density''
\begin{equation}
	n_q = \frac{1}{2 \Delta z} 
	\label{eq:def-quantum-density}
\end{equation}
where $\Delta z$ is the grid spacing. 
As a consequence of this ``Wigner symmetrization'', the density in the 
nonlinear term
in the Hamiltonian $H_{\rm GP}$ [Eq.(\ref{eqn:GPE_TW})] should appear
with a subtraction, $|\psi( z )|^2 - n_q$.
We have incorporated the corresponding
(small) energy shift $g n_q$ into the chemical potential $\mu$.
(See, e.g, Ref.\cite{Berg2009} how to
generalize the mapping~(\ref{eq:mapping-correlation-function}) 
to two-time correlations.)

\subsection{Number conserving Bogoliubov initial state}
\label{s:ncB-initial-state}

To obtain a thermal initial state for use 
in the tW simulations, we employ a stochastic 
sampling for the canonical density operator at thermal 
equilibrium, based on the (number-conserving) Bogoliubov (ncB)
approximation. In the usual Bogoliubov theory,
one shifts the Bose field operator by a c-number field (the order
parameter), which is equivalent to assuming that the system is in 
a superposition of states with different particle number (coherent
state). 
The number-conserving version of the Bogoliubov theory
\cite{Gardiner1997,Girardeau1998}. 
preserves the total number of atoms, $N$, and is
constructed to provide the correct counting statistics
for the condensate mode, in 
the limit of a small thermal component.
(The distribution function $P(N_{c})$ for
the number of atoms in the condensate is discussed in
Sec.\ref{subsec:cond-stats}.)
Here, we summarize a practical scheme 
to sample the canonical equilibrium
density operator for the quantum field at a fixed number of atoms $N$,
as explained in Refs.\cite{Sinatra2000,Sinatra2001}; a number of
technical details can be found in Appendix~\ref{a:tWncB}.

\subsubsection{Condensate mode}

The initial value for the classical field $\psi(z, 0)$ is split as
\begin{eqnarray}
	\psi(z, 0) &=& a_{c}  \phi_c(z) 
	+ \psi_{\perp}(z)
\,, 
	\label{eq:split-psi}
\end{eqnarray}
where the first term describes the condensate,
$a_{c}$ being the corresponding complex amplitude 
and $N_{c} = |a_c|^2$ the number of condensate atoms.
The condensate mode function $\phi_c(z)$ is normalized
to unity and is given in Eq.(\ref{eq:psi-c-in-ncB}). 
The splitting~(\ref{eq:split-psi}) is motivated from an expansion 
in the limits of large particle number $N$
and small interaction constant $g$ 
\cite{Gardiner1997,Castin1998,Morgan2000,GardinerS2007}. 
More precisely, 
the condensate mode and its excitations are calculated
self-consistently up to second order in $(N_{\rm th} / N )^{1/2}$,
where $N_{\rm th}$ is the number of non-condensed
particles, respectively. (This number also has a small quantum 
contribution.)

\subsubsection{Elementary excitations}

The non-condensate field $\psi_{\perp}( z )$ in Eq.(\ref{eq:split-psi})
arises in the next-order contribution of the ncB expansion. It is
expanded in the basis of the Bogoliubov modes:
\begin{eqnarray}
\psi_{\perp}( z ) = \sum_{k}%
\left[b_k\,u_{k}( z ) + b_k^*\,v^*_{k}( z )\right]
	.
	\label{Eqn:psiper}
\end{eqnarray}
The mode functions are the eigenvectors $(u_k, v_k)^T$
to the eigenvalue $E_k$ of the Bogoliubov-de Gennes
operator ${\cal L}_Q$ given in Eq.(\ref{eq:def-LQ}). 
The Bogoliubov spectrum $\{ E_k \}$ is positive and gives
the quasi-particle energies relative to the chemical potential. 
The Bogoliubov amplitudes
$b_k$, $b_k^*$ in Eq.(\ref{Eqn:psiper}) 
are sampled as independent complex Gaussian random numbers 
with zero mean and variance 
$\sigma_k^{2} = (1/2)\,\coth (\beta E_k/2)$
\cite{Sinatra2001,Sinatra2002}.
The mean population of a Bogoliubov mode is thus equal to
\begin{equation}
	\langle |b_k|^2 \rangle_{\rm W} = \sigma_k^2
	= \BE(E_k) + \frac{ 1 }{ 2 },
	\label{eq:ave-mode-amplitude-tW}
\end{equation}
the Bose-Einstein occupation 
number $\BE(E_k)$
plus an extra contribution $1/2$. This extra term appears due to the
symmetric ordering of the quantum operators:
\begin{equation}
	\langle |b_k|^2 \rangle_W \gets
	\frac12 \langle b_k^\dag b_k^{\phantom\dag} + 
	b_k^{\phantom\dag} b_k^{\dag} \rangle,
	\label{eq:symmetric-ordering-Bogoliubov-amplitudes}
\end{equation}
and represents quantum fluctuations.
Quantum fluctuations mimic spontaneous 
scattering into otherwise empty modes within a 
classical field approximation \cite{Blakie2008}. They also lead to 
the depletion of the condensate mode~\cite{Lee57}.
The condensate mode function $\phi_c( z )$, indeed, contains
a correction that depends on these amplitudes [the field
$\NCphiSecond( z )$ discussed after Eq.(\ref{eq:phi2-correction})].
Within the number-conserving scheme of Ref.\cite{Sinatra2002}, 
this correction reflects the change in the condensate number
(quantum and thermal depletion),
the interaction between condensate and non-condensate particles
via the Hartree-Fock potential 
$2 g \langle |\psi_\perp( z )|^2 \rangle_W$
and via the anomalous average 
$g \langle (\psi_\perp( z ))^2 \rangle_W$,
see Eqs.(\ref{eq:phi-2-equation}, \ref{eq:source-term-phi-2}).

\subsubsection{Condensate number}

\begin{figure}[htb]
\centerline{
  \includegraphics[width=85mm,clip]{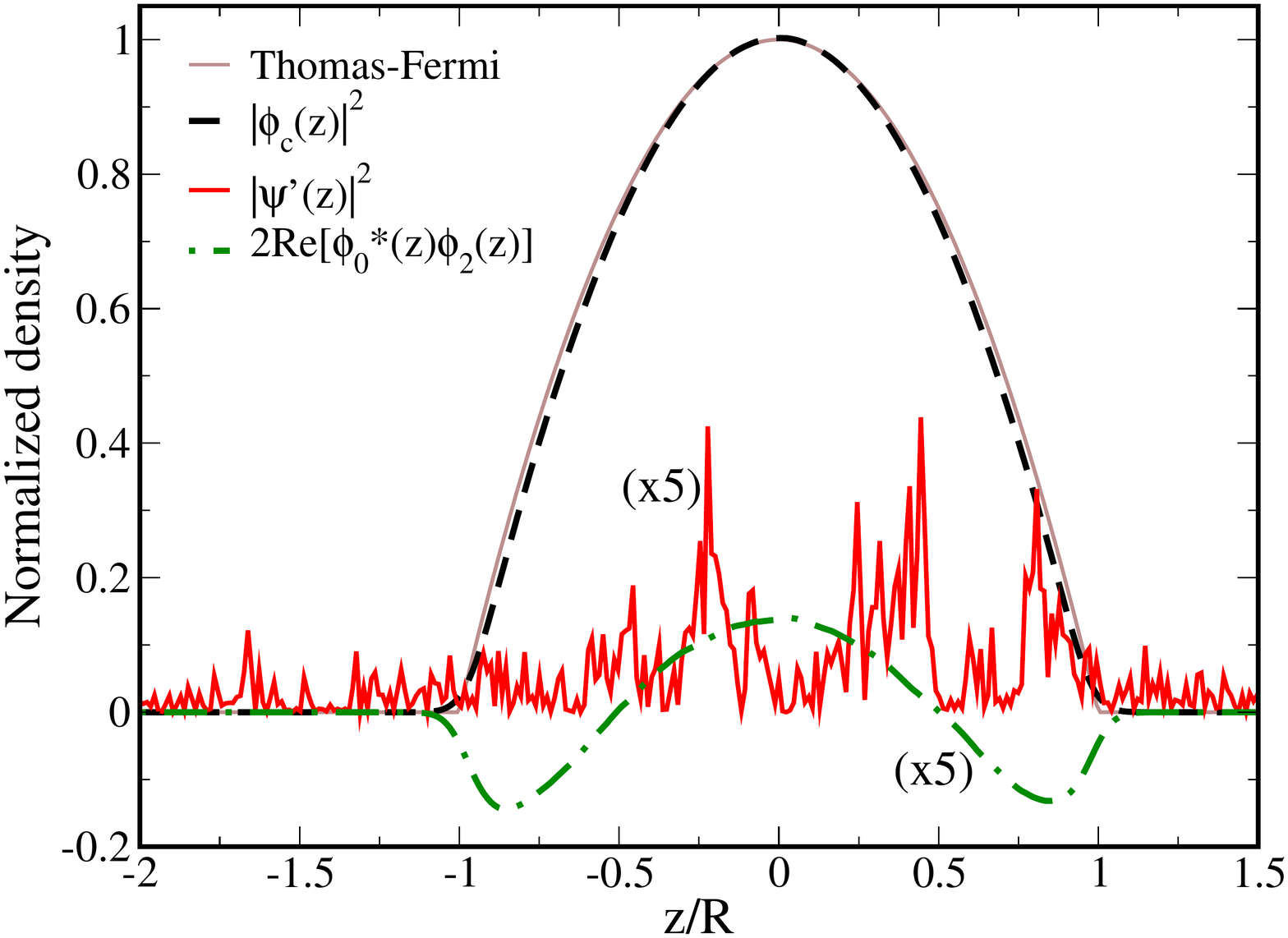}
}
\caption[]{%
(Color online)
Condensate density $N_{c}|\phi_c(x)|^2$ (dashed, black)
and a typical realization of the non-condensate
density $|\psi'(x)|^2 \equiv |\psi_\perp(x)|^2$ 
(noisy red curve; data is multiplied by 5 to be 
visible on this scale). 
The depletion correction to the condensate 
mode, $\NCphiSecond( z )$, is illustrated by plotting the `interference term'
$2\,{\rm Re}\,[\phi_c^*(x) \NCphiSecond( z ) ]$
(dot-dashed, green; also multiplied by 5). 
Note that the condensate
atom number $N_{c} \approx 19500$ depends on the realization. The 
zero-temperature
Thomas-Fermi shape for the same parameters (inverted parabola; thin, 
solid brown line) is also plotted. 
Here (and in most of the paper) we fix 
$\mu = 22.41 \, \hbar\omega$
(corresponding to $N = 20 000$ for the GPe at $T=0$),
choosing here $k_{B}T = 46\,\hbar\omegaZ$.
Densities are plotted in units of $g / \mu$.
}
\label{fig:ncB-one-sample}
\end{figure}

The stochastic ensemble of the non-condensate fields $\psi_\perp( z )$
now determines the sampling of the condensate number.
The number of condensed atoms $N_{c}$ is calculated 
from \cite{Sinatra2001,Sinatra2002}
\begin{eqnarray}
\label{Eqn:N0}
N_{c} 
&=& N 
- N_{\rm th}(\{ b_k \}) 
+ \mathcal{A}(\{ b_k \})
\,,
\\
N_{\rm th}( \{ b_k \}) &=&
\Delta z\,\sum_{z}|\psi_{\perp}(z)|^2 - 
{\cal M}/2 \;.
\end{eqnarray}
Here, $N_{\rm th}$ gives the number of non-condensed atoms,
while
${\cal M}$ is the number of terms in the 
expansion~(\ref{Eqn:psiper}) and depends on the computational
grid, and acts so as to
subtract the ``one half
atom per mode'' from the quantum fluctuations in $\psi_\perp$.
The quantity $\mathcal{A}(\{ b_k \})$,  given in 
Eq.(\ref{Eqn:A}), averages to zero and implements 
the ``canonical constraint''
at the level of variances:
it ensures that the fluctuations of the condensate number $N_{c}$ are 
anti-correlated to those of $N_{\rm th}$ (calculated in normal order), since 
the two number operators sum to the fixed total particle number $N$.
Once $N_{c}$ is calculated for each member of the ensemble,
the condensate amplitude $a_c$ in Eq.(\ref{eq:split-psi}) is taken
as $a_c = \sqrt{ N_{c} }$. 
The global phase is arbitrarily fixed here
but the ncB construction is actually
U(1)-invariant  
(see endnote~\cite{endnote137}),  
and the phase drops out in our observables of interest.
A typical example for a Wigner field is given in Fig.\ref{fig:ncB-one-sample}.

\subsubsection{Validity range}

We summarize a few issues that have to be considered for the
initial state preparation within the ncB expansion and the tW scheme.
There are two aspects here: First, the 
truncation made in order to obtain the evolution equation (\ref{eqn:GPE_TW})
assumes that third derivatives of the Wigner functional for the quantum
field are negligible~\cite{Steel1998,Sinatra2001,Blakie2008}.
This should be the case when the total number of particles
is much larger than the
number of relevant modes in the simulation,
i.e. $N \gg {\cal M}$.
Given this
restriction, however, the tW scheme gives approximate physical results 
even 
beyond
the time scale 
where the Bogoliubov theory, in its number-conserving form,
fails \cite{Sinatra2001}.

The second aspect is related to the low-temperature expansion of the 
number-conserving Bogoliubov approach that is
behind the initial state 
preparation \cite{Sinatra2000,Sinatra2001,Sinatra2002}.
The canonical distribution 
of the Bose gas is calculated by approximating the Hamiltonian of the
quantum field theory by the quadratic Bogoliubov Hamiltonian (as also done
in Refs.\cite{Kocharovsky00, Svidzinsky2006}). This
is valid when the number of non-condensate particles is small:
$N_{\rm th}\ll N$ which implies relatively low temperatures. 
The sampling of the condensate statistics, Eq.(\ref{Eqn:N0}), is also 
an approximation that has been discussed in Ref.\cite{Sinatra2001}:
in the quantum field theory, $N_{c}$ is restricted to be an integer (eigenvalue
of the number operator $\hat a_c^\dag \hat a_c^{\phantom\dag}$),
while the classical simulation returns continuous values for $N_{c}$. Both
schemes coincide when the counting statistics $P(N_{c} | \psi_{\perp})$
(conditioned on a given non-condensate field $\psi_{\perp}$) is broad enough,
and can be extended to a smooth function of $N_{c}$.
It has been shown that this condition can be met when $\psi_{\perp}$ is sampled
with sufficiently many modes (${\cal M}_{\rm th}$, say) having a significant 
thermal occupation (see \cite{Sinatra2002} for more details).
For a typical mode spacing $\hbar\omegaZ$, this gives a lower limit
on the temperature because we need
${\cal M}_{\rm th} \equiv k_B T / (\hbar\omegaZ) \gg 1$. 
Under these conditions, one can also justify the gaussian probability distribution 
for the non-condensate field $\psi_{\perp}$ that is used in the simulation
scheme \cite{Sinatra2002}.

In summary: since the Bogoliubov approximation 
is used for the representation of the $N$-body density operator,
the temperature should be not too high but at the same time also not too low 
because of the smoothness of the distribution function. 
This poses limits on the applicability of the state preparation protocol
within tWncB.
The above requirements
can be checked from the inequalities \cite{Sinatra2001}
\begin{equation}
	{\cal M}_{\rm th} \le \langle N_{\rm th} \rangle \ll \sigma^2( N_{c} )
	\label{eq:Sinatra2001-inequality}
\end{equation}
where $\sigma^2(N_{c})$ is the variance of the (unconditional) condensate 
statistics $P( N_{c} )$,
obtained from the sampling~(\ref{Eqn:N0}).
The first inequality precludes very low temperatures because
$N_{\rm th}$ would be too small. In addition, the condensate statistics 
must not be too broad,
\begin{equation}
	\sigma( N_{c} ) \ll \langle N_{c} \rangle = 
	N - \langle N_{\rm th} \rangle
	\label{eq:not-too-broad-counting-statistics}
\end{equation}
because otherwise the probability of returning negative values for 
$N_{c}$ would become significant. The method gives unphysical results
if a large fraction of negative values of $N_{c}$ is returned
and we find that this happens already at moderate
temperatures, when $\langle N_{\rm th} \rangle/N$ becomes of
the order of $\sim 0.2$.

%

\subsection{Stochastic Gross-Pitaevskii equation}

Within the stochastic Gross-Pitaevskii
equation, a finite temperature 
equilibrium state is obtained dynamically
by evolving a complex field $\psi( z, t )$ that is coupled to a 
thermal cloud which, when approximated as in thermal equilibrium,
acts as a heat bath (energy and particle reservoir)
\cite{Stoof1999,Stoof2001,Gardiner2003,Blakie2008,Cockburn2009}.

\subsubsection{System plus bath split}

We may physically motivate a division into two subsystems: 
the system is represented by the field $\psi( z, t )$ 
and describes the low-lying modes of the ultracold gas. These
are highly occupied,
therefore a classical field description is appropriate; 
the ``thermal cloud'' of atoms whose energies are
high above the typical energies of the condensate and its excitations,
obeys a separate quantum Boltzmann equation \cite{Stoof1999}.
Both subsystems are naturally
coupled to each other by exchanging energy and particles, hence the
description is given within the grand-canonical ensemble.
This leads to a nonlinear Langevin equation (see Eq.(\ref{eq:sGPe-0})),
often termed the stochastic GPe (sGPe).
The system dynamics now combines
deterministic aspects (encapsulated within the usual GPe)
and a stochastic coupling to the heat and particle reservoir of thermal atoms.

We note that there are two distinct formulations of such 
a nonlinear Langevin equation, 
which are motivated by the same physical ideas, but which arise from
very different formalisms (see Ref.\cite{Proukakis2008} for a review
and a discussion of subtle differences). 
The derivation of Stoof, 
which we shall adopt in this work, is based on the Keldysh 
non-equilibrium formalism \cite{Stoof1997,Stoof1999,Duine2001} and the resulting theory 
was first implemented numerically in Ref.\cite{Stoof2001}. 
An equation that differs in some details 
was formulated by Gardiner, Anglin and Fudge \cite{Gardiner2002} 
and cast into its current form by Gardiner and Davis \cite{Gardiner2003}, 
as a limiting case of the quantum kinetic theory put forward by Gardiner and Zoller 
\cite{Gardiner97a,Gardiner1998,Gardiner2000}. 
The stochastic equation which results, differs primarily in the use 
of a projector operator restricting dynamics to 
low energy modes, and is termed the stochastic projected
GPe (spGPe) \cite{Blakie2008}.

Technically, the particular sGPe discussed here 
is obtained
by expanding the system density matrix over coherent states using 
functional integration techniques, which leads naturally to the 
Keldysh non-equilibrium formalism. 
Ultimately, this is found to give a Fokker-Planck equation for
the Wigner distribution function of the entire atomic quantum field 
\cite{Stoof1997,Stoof1998}.
This procedure maps symmetrically ordered correlation functions of 
field operators onto stochastic field correlations, as is evident from the fact
that each mode $k$ of the classical field $\psi$ 
occurs in the stationary limit
with an occupation number $\BE( E_k ) + 1/2$  
\cite{Stoof1999,Polkovnikov10} (cf. also Eq.\eqref{eq:ave-mode-amplitude-tW}).
Within the modes which are predominantly classical, i.e. highly occupied,
$\BE( E_k )\gg1/2$, and this symmetrization is no longer important,
a common consideration in {\it all} classical field methods 
\cite{Svistunov1991,Davis2001,Goral2002,Blakie2008}.
This approximation will permit us to move from the
quantum relation Eq.~\eqref{eq:damping-term1} below
to its classical counterpart Eq.~\eqref{eq:damping-term2} which
makes the simulation scheme much simpler.

\subsubsection{Stochastic equation of motion}

Making a Hartree-Fock type {\it Ansatz} for the probability distribution
representing the entire trapped gas, leads to 
two separate probability distributions, 
representing separately the high- and low-lying system modes.
Integrating out the low-energy modes, one finds that the former may 
be treated by a quantum Boltzmann equation (qBe) \cite{Stoof1999}.
Integrating instead over the high-energy modes leads to
a nonlinear Langevin equation,
\begin{eqnarray} 
i \hbar \frac{ \partial \sGPphi}{ \partial t} 
&=  & 
\Big( 
H_{\rm GP}[ |\sGPphi|^{2} ] 
- \sGPmu
- i \hat R( z,t ) \Big) \sGPphi + \eta(z, t) \;,
\label{eq:sGPe-0}
\end{eqnarray}
where $\hat R( z,t )$ is a damping term, which should in general
be time dependent, and $\eta(z, t)$ a
stochastic ``force''. 
Note that it is for convenience in later discussions only
that we use the same symbol for the tWncB and sGPe
wavefunctions, despite the differences in physical content.
Assuming that the dynamics of the high-energy modes
may be neglected,
the physical picture underlying this equation 
is a splitting of the quantum field into low-lying modes (the ``system'',
described by $\sGPphi( z, t )$) and a ``thermal particle bath'',
which is considered to be at equilibrium and so, on average, 
Bose-Einstein distributed \cite{Duine2001} (since this is the
equilibrium solution to the qBE). 
The term $- i \hat R(z,t) \sGPphi(z, t)$ describes the particle exchange
due to collisions between system and bath atoms.
The real part of the operator $\hat R$ can be positive or negative, corresponding to loss or growth, as for the analogous operator appearing in the ZNG scheme \cite{ZNGbook}.
Since collisions occur randomly, the sGPe contains 
an associated `noise' term $\eta(z, t)$ in Eq.(\ref{eq:sGPe-0}).
The presence of both terms, dissipation and noise,
is essential to ensure that the fluctuation-dissipation theorem 
is satisfied: the system is thus guaranteed to reach
the correct equilibrium at a given temperature.

\subsubsection{Damping and noise}

The damping operator is given by the relation
\begin{equation}
	-i\hat R  = 
	 \frac{\hbar\Sigma^{K}}4  
	 \left( \BE(\hat \epsilon_{c} - \mu) 
	 + 1/2 \right)^{-1} 
	\label{eq:damping-term1}
\end{equation}
where $\hbar\Sigma^{K}$ is the so-called Keldysh self-energy 
(a complex quantity) and $\BE(\epsilon_{c} - \mu)$ is the 
Bose-Einstein distribution, representing
the occupation of low-lying modes as a function of mode energy
$\epsilon_c$. While this relation is exact at equilibrium, it 
cannot be easily implemented in this form
in numerical simulations for the following reasons: 
firstly, the
mode energy $\hat\epsilon_c$ actually corresponds to the nonlinear
differential operator 
appearing in the GPe~(\ref{eq:sGPe-0}); moreover, 
$\Sigma^{K}$ is determined by the thermal particle distribution, whose
accurate temporal representation 
would require solving a qBe. We therefore restrict the sGPe to its 
classical limit, as \emph{all} current numerical applications of this theory do. 
In the approach of Stoof, this means that the action of the damping operator 
takes the simple form
\begin{equation}
	-i\hat R  \sGPphi =
	 \frac{\hbar\Sigma^{K}}{ 4 k_B T } \left( 
	 H_{\rm GP}[|\sGPphi|^2] - \sGPmu \right) \sGPphi.
	\label{eq:damping-term2}
\end{equation}
where $\hbar\Sigma^{K}$ is still spatially
dependent in general \cite{Duine2001,Cockburn2010}, 
as the thermal particle energies are 
affected by the condensate mean-field.
This enables the equation to be cast in the form
\begin{eqnarray} 
i \hbar \frac{ \partial \sGPphi}{ \partial t} 
&=  & 
\left( 1 - i \gamma \right)\Big( 
H_{\rm GP}[ |\sGPphi|^{2} ] 
- \sGPmu
\Big) \sGPphi + \eta(z, t)\;,
\label{eq:sGPe-1}
\end{eqnarray}
where $\gamma = i \hbar\, \Sigma^{K} / 4 k_B T$.
The {\it form} of this equation is the same as the stochastic pGPe
(spGPe) implemented numerically by Davis and collaborators, 
except for the projector \cite{Blakie2008}
(see Ref.\cite{Proukakis2008} for a more detailed comparison).

In Eq.(\ref{eq:sGPe-1}), the term $\eta( z, t )$ is a complex, Gaussian, 
white-noise process with correlations
\begin{equation}
	\langle \eta^{*}(z,t) \eta( z', t') \rangle 
        = 2 \hbar \gamma k_{B} T 
	\delta (t-t') \delta ( z - z' )\;.
	\label{eq:sGPe-noise-spectrum-0}
\end{equation}
This relation can be read as a fluctuation-dissipation theorem for
the system, since the strength of fluctuations is proportional to
the damping parameter $\gamma$, on the one hand, and the
temperature $T$ of the heat bath, on the other.
This link is essential for the preparation of a thermalized
system. 
A similar stochastic scheme was formulated in 
Ref.\cite{Heller2009} for the ideal Bose gas where the noise was
filtered in order to preserve the total atom number (canonical
ensemble).

Since our primary interest in the present work is in generating an ensemble of 
equilibrium states, 
we make a further, numerically convenient, simplification
and treat  $\Sigma^{K}$ as a parameter independent of
time and, additionally, space.
This is also a standard approximation in spGPe simulations \cite{Blakie2008},
and we have tested that a spatially varying $\Sigma^{K}$ does not strongly 
affect the equilibrium state.
So, for our present purposes, we solve Eq.~\eqref{eq:sGPe-1} with  
the dimensionless quantity 
$\gamma = 0.01$.

\subsubsection{Validity range}

The main limitation behind the sGPe used here is the classical approximation
(highly occupied modes), as highlighted by 
the relation~\eqref{eq:damping-term2}.
In the case of a trapped system, this 
means that the applicability
of the theory varies spatially.
Indeed, the classical approximation is better suited to
the low energy, central region of the trap, in which there are many
particles due to the presence of a Bose-Einstein condensate.
In the outer trap regions, there are fewer atoms, and hence there
comes a point beyond which the classical approach is no 
longer well justified.
In general, this point is dependent upon the choice of grid spacing,
as a finer grid includes more high-energy modes.
For a given grid with spacing $\Delta z$, the accuracy of the classical 
approximation can be checked by comparing, for example,
the average density $\langle |\sGPphi(z)|^2 \rangle$ 
to the quantum density' $n_q = (2\Delta z)^{-1}$ (this value arises
from operator symmetrization
in Eq.(\ref{eq:mapping-correlation-function})
on the grid). We shall see that this
limits the applicability of the classical approximation to the sGPe
typically to the spatial range where the trapping potential is not 
too large, $V(z) - \sGPmu \le k_B T$. 
Within this study, we are interested primarily in the central region
$|z| < R$ where the condensate is present ($R$ is the Thomas-Fermi radius), 
as also highlighted in the
original sGPe numerical implementation \cite{Stoof2001}. 
We have verified that changes in the properties which
form the basis of our comparison are negligible over a range of grid spacings.
This is physically equivalent to the statement that our equilibrium results are 
unchanged for a range of cutoff energies, which mark the 
split between the low- (`classical') and high-lying (`thermal') modes
(see also Refs.\cite{Blakie2005,Blakie2008,Cockburn_PhD,Cockburn2009}).

\subsubsection{State preparation}
\label{s:sGPe-preparation}

\begin{figure}[t!]
  \centerline{
  \includegraphics[scale=0.256,clip]{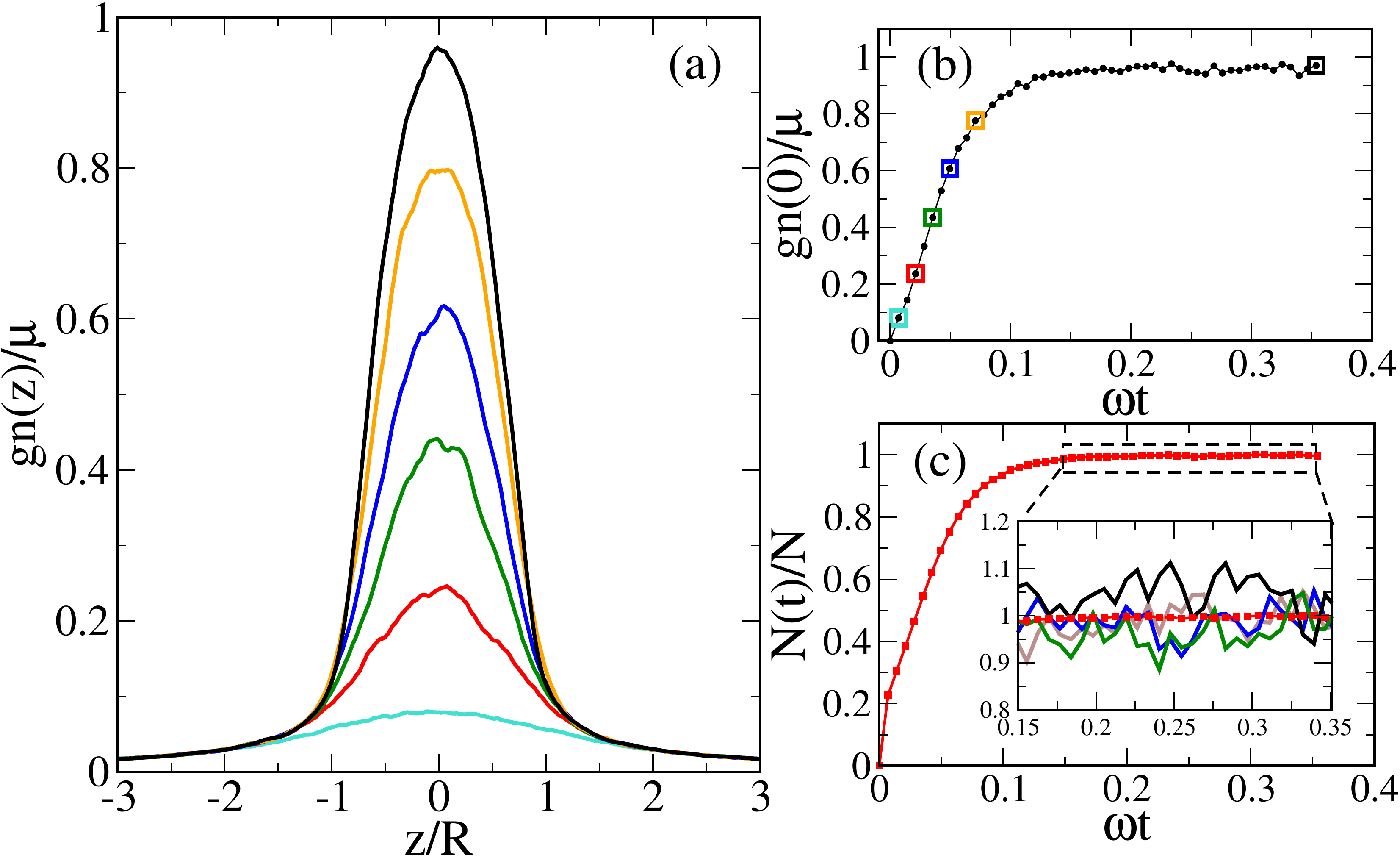}
}
  \caption[]{(Color online) Growth to equilibrium obtained by numerical
	   solution to the sGPe.
	   (a) Snapshots of the 
	   atomic density profile $\langle|\sGPphi(z,t)|^{2}\rangle$, 
	   with time increasing from bottom to top.
           (b) Growth in the average central density vs. time -- here the 
           times highlighted by colored squares 
	   label the corresponding colored density profiles in (a).
	   (c) Growth in the particle
	   number as the system approaches equilibrium.
           The inset shows trajectories from single numerical realizations,
           which illustrates the fluctuating particle number between these,
           compared to the result of the main plot in (c), 
           which was obtained by averaging over $1000$ such trajectories. 
           (Parameters: as stated in 
           subsection \ref{subsec:parameter-choice}, 
           but with $k_B T = 860\hbar\omegaZ$.)} 
  \label{fig:sGPe_growth}
\end{figure}

We now briefly explain how the sGPe~(\ref{eq:sGPe-1}) works in practice 
-- more details on this
can be found in recent reviews \cite{Blakie2008,Cockburn_PhD,Cockburn2009}.
As initial condition for the system, one
can start with $\psi(z, 0) \equiv 0$. 
The dissipative term $-i\gamma$ in 
Eq.(\ref{eq:sGPe-1}), leads to a change in the norm of $\psi(z, t)$, 
but this cannot increase a zero initial condition. 
It is the Langevin force $\eta(z, t)$ that `seeds' the field, 
as discussed in \cite{Stoof2001}.
The particle number $N(t) = \sum_i \Delta z |\sGPphi(z_i)|^2$ 
increases until
$\sGPphi( z, t )$ relaxes to the solution of the stationary GPe, 
at a given chemical potential
\begin{equation}
	H_{\rm GP}[|\psi|^{2}] \psi = \sGPmu \psi,
	\label{eq:statGPE-sGPe}
\end{equation}
as illustrated in Fig.\ref{fig:sGPe_growth}.
The state preparation in the sGPe is thus performed {\it dynamically}, 
as the system grows to equilibrium 
in contact with a heat bath at a specified temperature.
Once the dynamical equilibrium is reached,
the presence of the noise term $\eta( z, t )$ ensures that $N(t)$ 
fluctuates about its final value. 
The final, average atom number $\langle N \rangle$ depends on
the heat bath parameters (temperature $T$, 
chemical potential $\sGPmu$), on the trap parameters 
and the atomic species (through the interaction constant $g$). 
Although the subsequent dynamical evolution requires these noise terms to be maintained,
a simpler scheme, bearing close analogies to the truncated Wigner method,
can be based on dynamical propagation of the sGPe equilibrium state via the
GPe; such an approach
was first implemented in \cite{Proukakis2006a} to discuss quasi-condensate growth
on an atom chip.

It is clear that in the grand-canonical ensemble,
$N$ has a statistical distribution of non-zero width; this width is 
related in the case of the sGPe to the `history' of the particle transfer 
between system and heat bath as modelled by
the Langevin seed $\eta( z, t )$.
In the simulations, we observe indeed that $N$ differs quite
substantially from realisation to realisation, however the equilibrium value
is obtained with reasonable accuracy
after averaging over a few hundred of them.
The temporal variations in the ensemble-averaged particle number 
then become relatively suppressed. 
For smoother results and for improved accuracy, all results presented 
in this work are based on a sample of at least 
1000 individual realisations.

\subsection{Linking the theories}

At $T=0$, the Gross-Pitaevskii equation represents the dynamics
of a Bose-Einstein condensate by treating the system as made up of
a single, coherent mode. 
Within the tW approach, the GPe is instead used to propagate a
multi-mode system, the initial conditions being chosen at random
to sample the initial density operator of the system. The ensemble of
wave functions $\psi( x, t )$ represents \emph{all} modes of the
matter wave field, within the limits set numerically by the spatial grid,
which physically corresponds to the energy cut-off choice for the modes being probed.
It is for this reason that no explicit noise terms (Langevin forces) appear,
and that the total number of atoms (the norm of $\psi( x, t )$) is
conserved. The initial mode amplitudes combine quantum and thermal 
effects [see Eq.(\ref{eq:ave-mode-amplitude-tW})] consistent with the
Wigner mapping to symmetrically ordered operator products.
Polkovnikov has shown that 
the truncated Wigner approximation captures the next order 
correction beyond Gross-Pitaevskii 
in an expansion in $\hbar$ \cite{Polkovnikov2003}. The initial state that
we prepare here is based on a fixed number of atoms $N$ (canonical
ensemble), using the
number-conserving Bogoliubov approach, although alternative (grand-canonical)
schemes could be adopted as well \cite{Blakie2008}. 
Once we focus on low-lying modes, like the condensate mode, we recover
nevertheless a broad distribution for the non-condensed atoms $N_{\rm th}$
(the mirror image of the condensate statistics $P( N_c )$). In this 
perspective, we can even consider $\mu$ in Eq.(\ref{eqn:GPE_TW}) as a
``chemical potential'' for the Bogoliubov modes: the condensate plays
the role of a particle reservoir, as is 
perfectly reasonable if its population is large \cite{Castin2001}. 

In the sGPe approach, the wave function $\sGPphi(z, t )$ represents 
instead the low-lying modes of the system;
although higher-lying modes should in principle be described by
a quantum Boltzmann equation, here they are assumed to remain at equilibrium, 
thereby providing a heat bath to the low-energy 
sub-system under consideration. 
In the Bose-condensed phase, low lying system
modes are highly 
occupied and the classical approximation,
amounting to replacing the
Bose-Einstein by the Rayleigh-Jeans distribution,
is well justified. The dynamics of the low-lying modes is quite different, however, 
because particle exchange with the bath is allowed for; this method is
therefore a grand-canonical one, as can be seen by the 
growth plots of Fig.\ref{fig:sGPe_growth}.
The ``system--bath split''
can be applied to a trapped gas (a closed system) by choosing modes 
below a suitably small cutoff. 
For a fully self-consistent calculation, 
in which the thermal cloud dynamics are also accounted for, this cutoff
should be not lower than the global chemical potential \cite{Stoof1999}.
For a classical field method not taking the entire thermal cloud
dynamics into account, this should be chosen such that the 
highest modes simulated are macroscopically occupied \cite{Blakie2008}.
It has been proposed that a cutoff equal to $k_B T$ 
yields optimum results for the condensate statistics of an 
ideal gas \cite{Witkowska2009}.
For the purposes of our comparison, we have adopted a different 
choice here, and have taken for consistency the same spatial grid 
in the sGPe and the tWncB simulations,
which gives a cutoff of the order of
$E_{\rm max} \sim \hbar^2 / (m \Delta z^2)$.

Let us summarize the differences between the
initial state ensembles of the two methods:\\
(1)
The total atom number $N$ fluctuates in the sGPe
(grand-canonical), and is fixed in tWncB (canonical).\\
(2)
The system is thermalized either dynamically (sGPe) by
weakly coupling it to a heat bath, or by populating its excitation
modes with thermal statistics (tWncB). 
Low-lying thermal modes above the condensate 
equilibrate under the sGPe to the Rayleigh-Jeans
statistics (classical equipartition). This is actually the equilibrium
distribution for the finite temperature GPe, considered as a classical
field equation, as has been seen by studying thermalization in related 
classical field methods \cite{Svistunov1991,Davis2001,Goral2002}.
Within the tWncB scheme, these modes are populated according
to the usual Bose-Einstein statistics, with the addition of $1/2$
``quantum atom'' per mode. This coincides well with the Rayleigh-Jeans
statistics for modes with energies below $k_{B}T$ \cite{Blakie2007}, 
but gives a larger
contribution to high-energy modes, up to the numerical cutoff.
The tW dynamics
under the GPe redistributes these ``quantum atoms'' with the others,
leading, by the equipartition law, to an effectively higher temperature.
This restricts applications of the tW scheme to relatively short simulation 
times. 
In nearly integrable systems (like the
quasi-one-dimensional gas), thermalization can be quite slow, 
however \cite{Cazalilla2010, Mazets2010}.
\\
(3) 
The energy spectrum of the elementary excitations is calculated
in tWncB approximately, ignoring the thermal depletion of the condensate. 
Indeed, the Bogoliubov-de Gennes operator
[Eq.(\ref{Eqn:Lbg})] assumes that all particles are in the condensate
mode.

\subsection{Modified Popov theory}
\label{subsec:mod_Popov}

\begin{figure*}[htb!]
\centerline{
\includegraphics*[width=0.8\columnwidth]{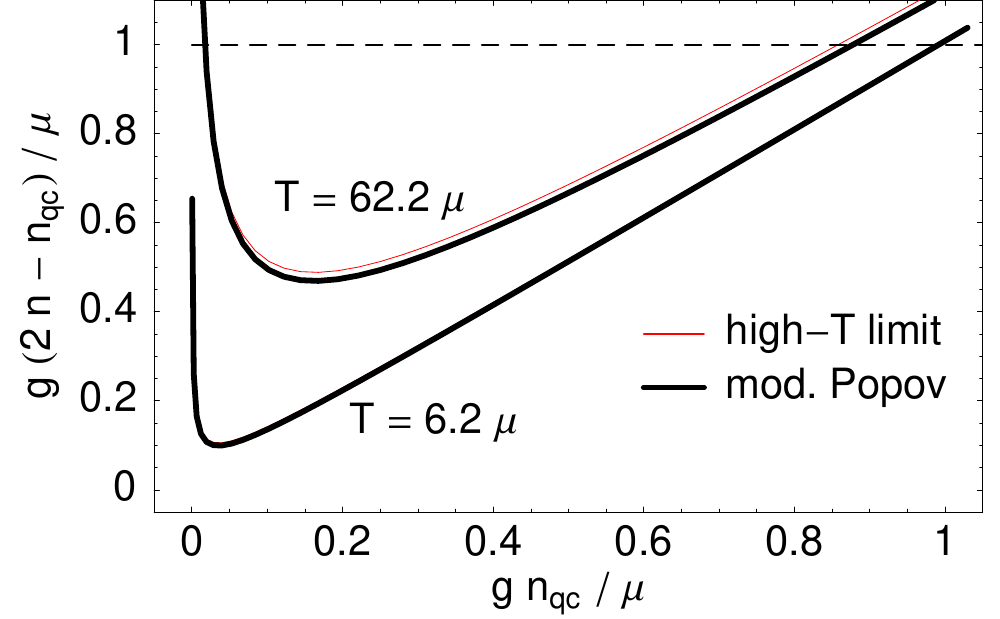}
\includegraphics*[width=0.8\columnwidth]{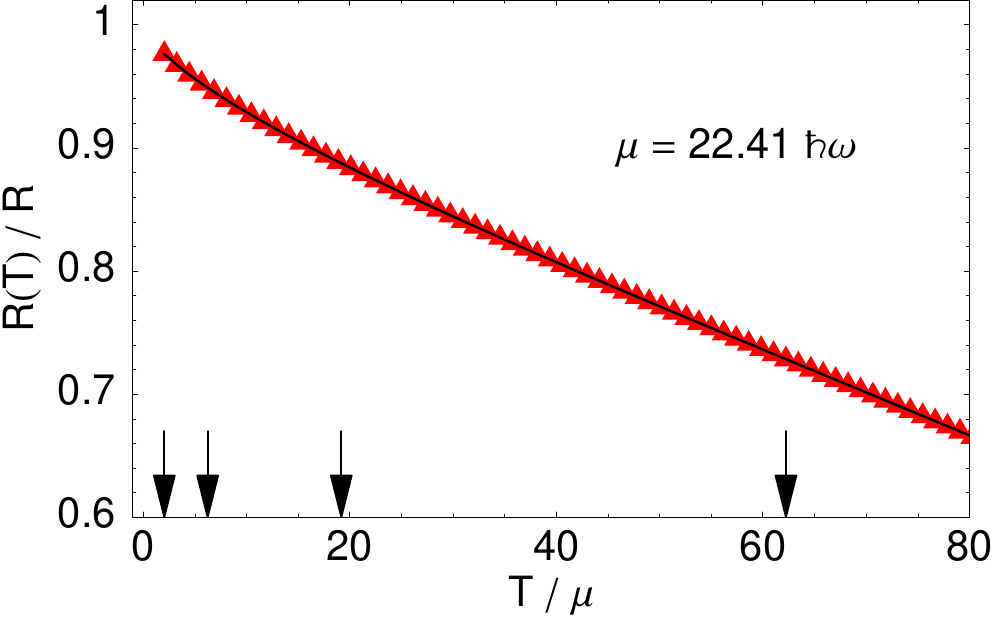}
}
\caption[]{(Color online) Illustration of the procedure for solving the modified Popov
theory. (left) Graphical determination of the self-consistent quasi-condensate
density: crossing point between the dashed and solid lines (lhs and rhs
of Eq.(\ref{eq:mP_mu}), $n_{\rm qc}$ is the quasi-condensate density
and $n$ the total density). Thin (red) line: classical (high-temperature)
approximation to Eq.(\ref{eq:mP_dens}). If we take 
$\mu$ as in the rest of the paper,
the two temperatures correspond to 
$N \approx 23\,800$, $T \approx 1.3\,T_\phi \approx 0.63\,\Tc$
and 
$N = 20\,000$, $T \approx 0.16\,T_\phi \approx 0.074\,\Tc$,
respectively (see Eq.(\ref{eq:def-T_phi-and-T_c})).
(right) Temperature-dependent Thomas-Fermi radius $R(T)$ determined
by numerically solving for the (local) chemical potential where
$\mu( z ) = 2 g n'_{\rm th}( z )$. The trap is harmonic, chemical 
potential at the center $\mu = 22.41\,\hbar\omegaZ$ as in the
rest of the paper, and $g = 0.01\,(\hbar^3\omegaZ/m)^{1/2}$. The arrows mark
the temperatures chosen for the simulations [Table~\ref{tab:temp_context}].
}
\label{fig:illu-mPopov}
\end{figure*}

\subsubsection{Motivation}

While the stochastic approaches discussed thus far are suitable
to describe both non-equilibrium and static properties of the Bose gas, 
we focus in this study on the detailed analysis of 
a partially condensed Bose gas at thermal equilibrium. 
Mean field theories \cite{Proukakis2008} are often applied to study the
thermodynamics in higher-dimensional Bose systems, and their solution is, 
in general, less involved than with stochastic theories.
In comparing the equilibrium properties of the sGPe and ncB
approaches, it will therefore prove useful to have an independent
method for comparison. 

In lower dimensions, mean field theories have to cope with infrared
divergences due to the enhanced role of fluctuations, as
predicted within the Mermin-Hohenberg-Wagner theorem \cite{Mermin1966,Hohenberg1967}.
The Popov approach \cite{PopovBook} where the fluctuations are
split into phase and density contributions, has proven useful to
treat phase fluctuations in low dimensions 
beyond second order around the mean field.
A consistent regularization scheme has been developed in
the modified Popov theory of Andersen {\it et al.} \cite{Andersen2002,AlKhawaja2002,StoofBook2009},
extending the work of Petrov {\it et al.} \cite{Petrov2000}
and Kagan {\it et al.} \cite{Kagan2000}. The resulting formulas
apply to any temperature and dimension, while simultaneously being
relatively straightforward to solve.

\subsubsection{Quasi-condensate density}

In low dimensions, a condensate does not arise in a homogeneous system, 
but still there is a temperature range $T_\phi \sim T < \Tc$
where a so-called quasi-condensate
can be identified whose density fluctuations are suppressed~\cite{PopovBook, Petrov2000, Kagan2000, Andersen2002, AlKhawaja2002}.
In the modified Popov theory, the quasi-condensate density $n_{\rm qc}$ 
may be obtained by solving 
self-consistently the following equations for the total density
[Eq.(4) from Ref.\cite{Andersen2002}]
\begin{equation}
  \begin{split}
    n = n_{\rm qc} + 
    \frac{1}{V}\sum_{{ p}}
    &\left[
    \bar{N}(E_{{ p}})\frac{\epsilon_{{ p}}}{E_{{ p}}}
    +
    \right.\\
    &\left. + 
    \frac{\epsilon_{{ p}} - E_{{ p}}}{ 2 E_{{ p}}}+
    \frac{ g n_{\rm qc}}{2\epsilon_{{ p}} + 2 \mu}
    \right],
  \end{split}
  \label{eq:mP_dens}
\end{equation}
and for the chemical potential
\begin{equation}
\mu = g(2n - n_{\rm qc})
.
	\label{eq:mP_mu}
\end{equation}
Here, $E_{{ p}} = [\epsilon_{ p}^{2} + 2gn_{\rm qc}\epsilon_{p}]^{1/2}$
is the Bogoliubov dispersion relation,  
$\epsilon_{ p} = p^2/2m$, $V$ is the system volume, and $g$ denotes the 
two-body T-matrix
(evaluated at $-2\mu$, which corresponds to the energy cost of exciting two atoms from the condensate).
Eq.\eqref{eq:mP_dens} is evaluated numerically, replacing the
sums over momenta by an integral.
(For a link with conventional mean field theories see Sec. \ref{cond_shape}). 

Eqs. \eqref{eq:mP_dens} and \eqref{eq:mP_mu} can also be applied in
a trap within a local density approximation, using a local chemical 
potential $\mu( z ) = \mu - V( z )$.
The quasi-condensate density is then found by solving
[Eq.(54) of Ref.\cite{AlKhawaja2002}]
\begin{equation}%
	\left(
	H_{\rm GP}[ n_{\rm qc} ] + 2 g n'_{\rm th}( z )
	\right)
	\sqrt{ n_{\rm qc}( z ) }
	= \mu \sqrt{ n_{\rm qc}( z ) }
	\label{eq:Popov-GPe}
\end{equation}
where $2 g n'_{\rm th}( z ) = 2 g ( n( z ) - n_{\rm qc}( z ) )$
is the Hartree-Fock potential due to the non-quasi-condensate particles.
%
%
%
%
The spatial point at which 
$2 g n'_{\rm th}( z ) = \mu(z)$
defines the temperature-dependent Thomas-Fermi radius, $R(T)$,
that gives an estimate for the thermal depletion of the (quasi-)condensate
(see Secs.~\ref{s:g1}, \ref{sec:quasi-cond}). 
We emphasize that this quantity is determined self-consistently
within the modified Popov theory.
For $|z| > R(T)$, we have $n_{\rm qc}( z ) = 0$, and adopting again
the local density approximation, the atomic density
corresponds to a thermal gas with a Hartree-Fock interaction,
\begin{equation}
\begin{split}
n(z) &= \int\!\frac{dp}{2\pi \hbar}
\BE( \varepsilon_{\rm HF}({p}, z) - \mu )
,
\\
\varepsilon_{\rm HF}( p, z ) & = \epsilon_p + V(z) + 2 g n( z )
.
\end{split}
\end{equation}
This procedure is illustrated in Fig.\ref{fig:illu-mPopov} where the left
panel shows both sides of Eq.(\ref{eq:mP_mu}) as a function of the
quasi-condensate density $n_{\rm qc}$. The crossing with the dashed
line determines the self-consistent $n_{\rm qc}( \mu, T)$. The (local) 
chemical potential can be lowered, which corresponds to moving
further from the trap centre, until a minimum $\mu_{\rm min}( T )$
below which the solution $n_{\rm qc} = 0$ must be taken \cite{AlKhawaja2003}. 
This defines the temperature-dependent Thomas-Fermi radius, 
\begin{equation}
	\mu \left( 1 - \frac{ R^2(T) }{ R^2 } \right) = \mu_{\rm min}( T )
	\label{eq:identify-RT}
\end{equation}
The right panel shows $R(T)$ in the temperature range of interest here:
the quasi-condensate is shrinking smoothly and is about $20\%$
smaller at $T \sim T_\phi$. (This number applies to the parameters
introduced in Sec.\ref{subsec:parameter-choice}.)

We recall that the central object within the modified Popov theory 
is the quasi-condensate, but this may be linked to the `true' condensate 
(if it exists, as defined by the Penrose-Onsager criterion), as is discussed
in Sec.~\ref{sec:quasi-cond}.
We will employ this modified Popov 
scheme in order to compare to various equilibrium
properties where appropriate. This has the advantages of simplicity and 
speed over the stochastic approaches, due to the relatively
straightforward manner in which the above equations may be solved.

%
\begin{figure}[htb!]
\centerline{
  \includegraphics[scale=0.8,clip]{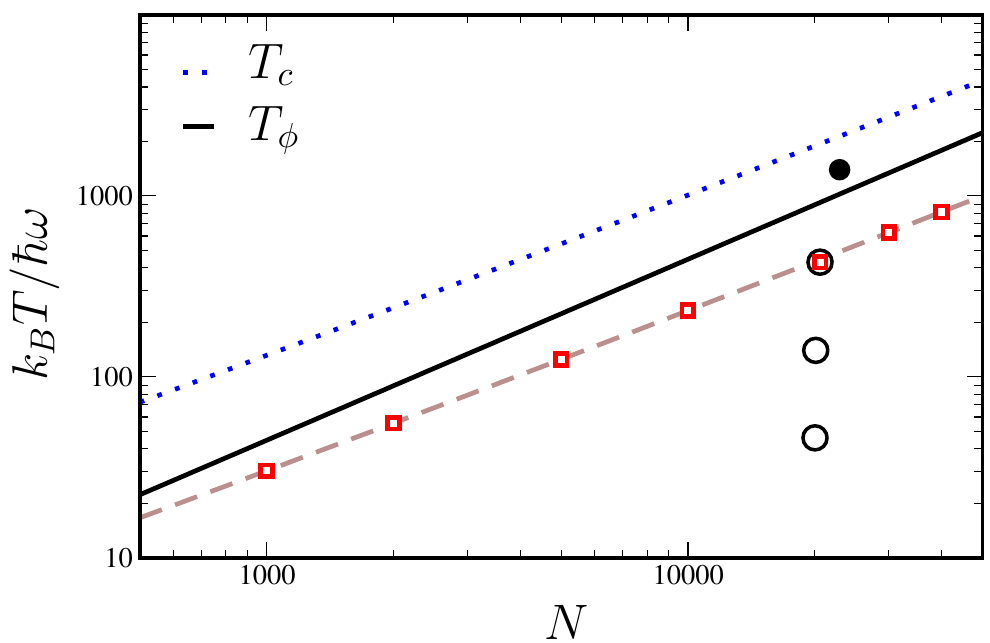}
}
\caption[]{(Color online) Characteristic temperatures and atom numbers
  of the sGPe and ncB simulations (hollow black circles).
  The filled black circle shows the higher temperature
  regime at which only sGPe simulations were undertaken. 
  The (1d) characteristic temperatures in a trap are
  $\Tc$ [Eq.(\ref{eq:def-T_phi-and-T_c}); 
  Bose-Einstein condensation in an ideal gas]
  and
  $T_{\phi} < \Tc$ [Eq.(\ref{eq:def-T_phi-and-T_c}); phase coherence],
  shown by the dotted blue and solid black lines respectively.
  The red squares mark the parameters chosen for the condensate statistics
  comparison of Sec.\ref{sec:cond_stats} at a fixed ratio $T/T_{c}=0.23$,
  as indicated by the dashed brown line. 
  }
  \label{fig:TN-plane}
\end{figure}

\begin{figure*}[htb]
\centerline{
\includegraphics[width=120mm,angle=0]{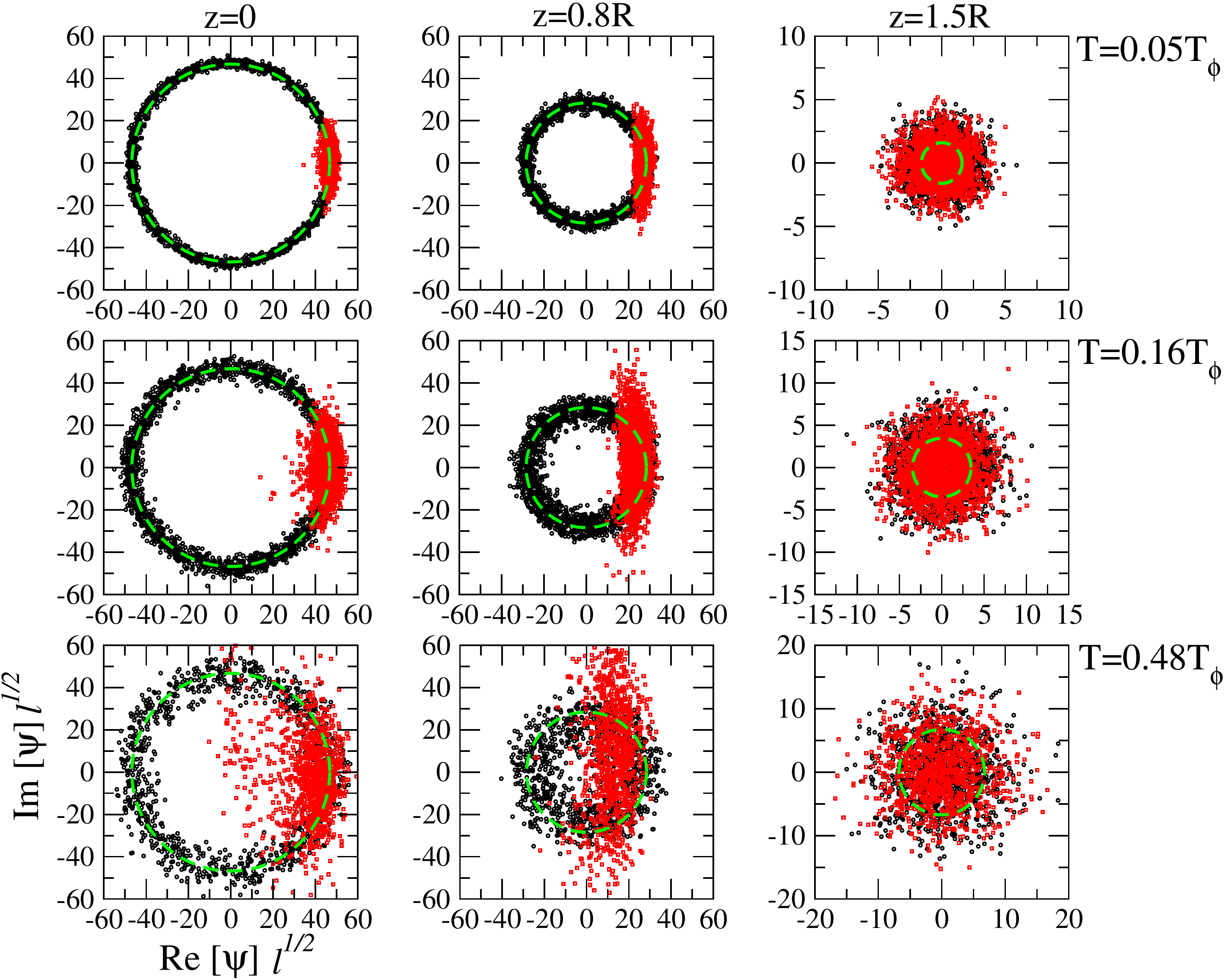}
}
\caption[]{(Color online) Data points representing the ensemble of 
stochastic field values $\psi( z )$ at three 
positions $z$ in the trap (from left to right, the average density decreases).
Temperature increases from top to bottom.
Black: sGPe simulation, red: ncB simulation. The dashed green 
circle indicates, for $|z| < R$,
the modulus of the Thomas-Fermi wavefunction, 
and for $z = 1.5\,R$ the square root of the Bose-Einstein 
density~(\ref{eq:Bose-Einstein-density}). In the trap center,
the Thomas-Fermi approximation yields for these parameters
$|\psi_{\rm TF}(0)| \approx 47 / \sqrt{\ellZ}$.
(Atom number $N \approx 20\,000$.)
}
\label{fig:point-cloud-1}
\end{figure*}
%
\subsection{Parameter choice for comparison}
\label{subsec:parameter-choice}

We wish to address the following issues:\\
(i) Initial state generation: how does the finite-temperature initial state 
compare within each method? \\
(ii) Ensemble choice: what role does 
the choice of thermodynamic ensemble play? \\
(iii) How does the quasi-condensate and condensate extracted
from the stochastic approaches compare to analogous quantities
within the modified Popov theory?

Our focus being on the relative merits of the stochastic methods 
as thermal field theories, we choose to work in 
a regime in which thermal effects dominate over quantum effects. 
We consider a quasi-one-dimensional confinement with a trap
frequency $\omegaZ$ (oscillator length $\ellZ$)
and take an effective
coupling constant $g = 0.01\,\hbar\omegaZ \, \ellZ$ which corresponds
to the weakly interacting regime.
We choose an (average) particle number, $N = 20\,000$, giving
a chemical potential 
$\mu = 22.41\,\hbar\omegaZ$ for the ground state of the 
GPe. 
Within the sGPe, the chemical potential is kept fixed as temperature is 
varied, leading to a small variation in the particle number;
this is however below $6\%$ for the three temperatures
at which we undertake the comparison. 
These are placed in context in 
Fig.\ref{fig:TN-plane}.
In particular, we probe 
at the lowest temperature a regime well suited to tWncB 
due to the requirement that $N_{\rm th} \ll N$,
an intermediate regime, and a higher temperature 
in which the `classical' sGPe is expected to be
most appropriate, due to the occurrence of more highly populated 
modes.
These are highlighted in Table \ref{tab:temp_context},
also showing some other simulation parameters.

\begin{table}[hbtp]
  \begin{center}
    \begin{tabular}{ | r || r | r | r | r | r | r |}
       \hline
	& $k_{B}T/\hbar\omegaZ$ 
        	& $T/T_{\phi}$ 
		        & $T/\Tc$   
				& $L / R$
				& ${\cal M}$
					& $\Delta z / \ellZ$ 
			\\ \hline
       \hline
       (a) \hspace*{\fill} ${\rm low}\,T$  	    
	& 46 
		& 0.052       
			& 0.025
				& 4.20
				& 127
					& 0.22
                        \\ \hline
       (b) \hspace*{\fill} ${\rm interm}\,T$   
	& 140     
		& 0.16       
			& 0.074
				& 12.0
				& 1023
					& 0.078
                         \\ \hline
      (c) \hspace*{\fill} ${\rm high}\,T$ 
	& 430        
		& 0.48       
			& 0.23
				& 17.0
				& 2047
					& 0.055
\\ \hline
\end{tabular}
\end{center}
    \caption[]{Simulation parameters: Atom number and chemical potential are
    fixed to $N \approx 20\,000$, $\mu = 22.41\,\hbar\omegaZ$, with
 the characteristic temperatures $\Tc$ and $T_\phi$ as in
 Eq.(\ref{eq:def-T_phi-and-T_c}). 
    The computational grid covers 
    $z = -L/2 \ldots +L/2$, with spacing $\Delta z$, number of
    points ${\cal M}$.  $R$ is the 
    Thomas-Fermi radius of Eq.(\ref{eq:def-TF-radius}) and 
    $\ell$ the size of the single-particle ground state.
}
  \label{tab:temp_context}
\end{table}

Length scales in the problem are scaled to the 
zero temperature Thomas-Fermi radius, 
\begin{equation}
	R = \sqrt{ \frac{ 2 \mu }{ \hbar \omegaZ }} \,\ellZ \approx 6.69\,\ellZ
	.
	\label{eq:def-TF-radius}
\end{equation}
Another relevant length scale is 
$z_{T} = \sqrt{2 (k_{B}T + \mu) / \hbar\omegaZ}
\, \ellZ$, 
which marks the boundary at which the thermal energy 
becomes comparable to the trap energy. We take a grid size
$L > 2 z_{T}$ [Table~\ref{tab:temp_context}] and a spacing
$\Delta z / \ellZ = \sqrt{2 \pi/({\cal M}+1)}$. The number of grid 
points, ${\cal M} + 1 = L / \Delta z$, is increasing with temperature
in order to resolve the 
thermal wavelength $\lambda_{T} =
\sqrt{ 2\pi\hbar / k_{B} T }$.
The grid spacing is chosen such that for an ideal gas,
the tWncB approach returns the correct
total particle number $\langle N \rangle$.
In addition, we check in the interacting case that doubling ${\cal M}$
does not change $\langle N \rangle$.
For time evolution via the (s)GPe, we use a Crank-Nicholson approach,
and a fixed time step $\omegaZ \Delta t = 10^{-4}$. Ensemble
averages are performed over at least $1000$ noise realizations.

A typical result of the stochastic methods is given in 
Fig.\ref{fig:point-cloud-1} where the data points give the realizations 
for the complex field $\psi( z )$ at selected values of position $z$ 
and temperature. The globally random phase of the sGPe data is quite
obvious, from the spread around the circle, 
while the ncB ensemble fixes the condensate mode to have
a real and positive amplitude. 
The overestimation of density fluctuations is also
quite visible in the ncB data at intermediate and high temperatures,
shown by the increased variation in the radius of the data points.
Outside the condensate region (Thomas-Fermi radius $R$), there are 
no significant differences between either method, where the
wavefunction represents an incoherent thermal gas.

\section{Equilibrium properties}
\label{sec:equilibrium_results}
 
We present here
an analysis of the initial states produced by the two stochastic
methods, that are expected to represent
thermal equilibrium in the trap. The accuracy of this equilibrium
state is important for modelling the finite-temperature dynamics when
perturbations take the system far away from equilibrium, e.g., changes
in the trapping potential.

\subsection{Density profile}
\label{subsec:total-density}

%
\begin{figure*}[hbt!]
  \centerline{
  \includegraphics[scale=0.4,clip]{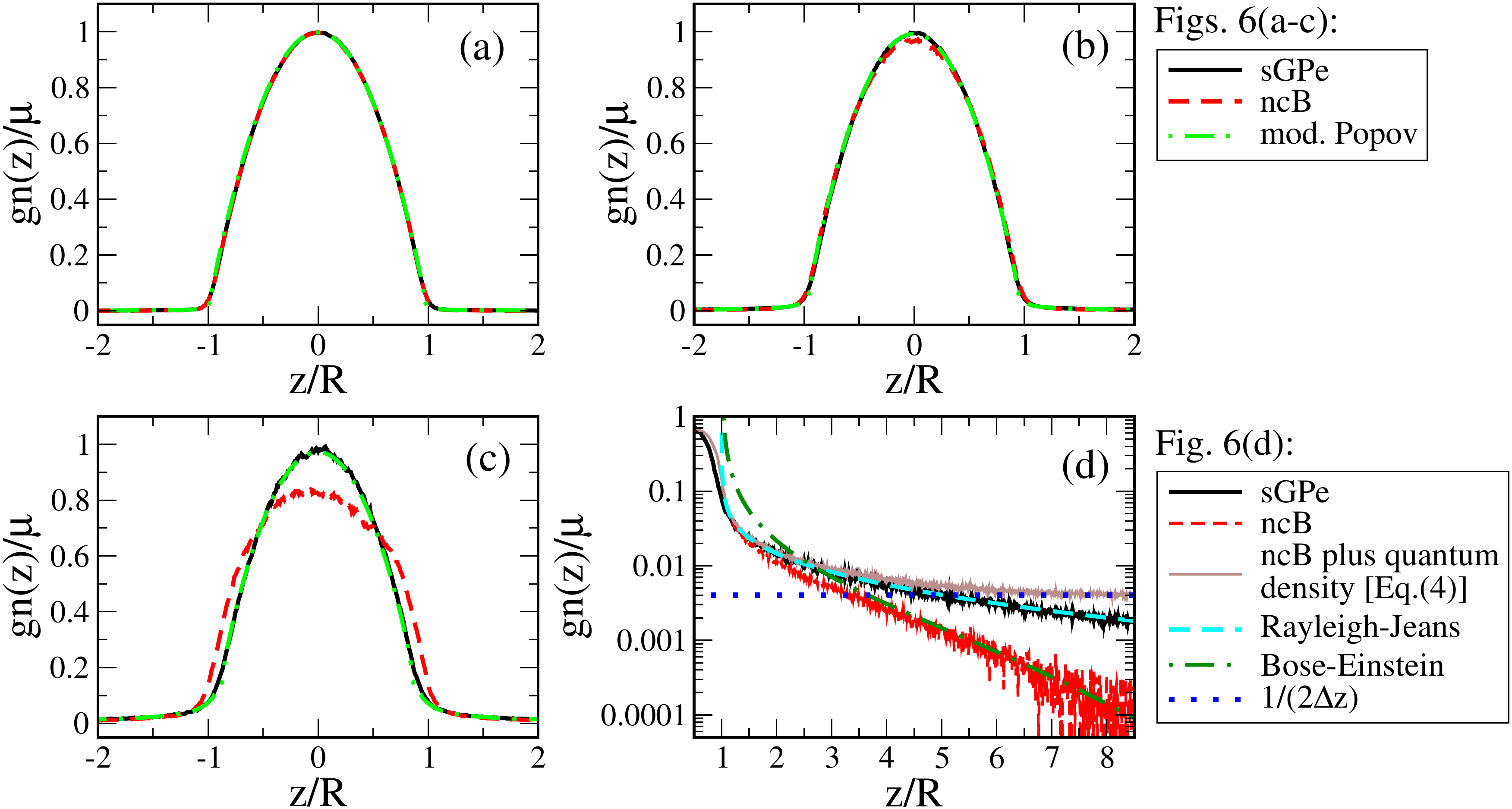}
}
  \caption[]{(Color online) Average density profiles $n( z )$, normalized as $g n( z ) / \mu$, 
 returned by
  the stochastic Gross-Pitaevskii equation (solid black) after an equilibration
  time $t_{\rm eq}$ and by the number conserving Bogoliubov expansion
  (dashed red). Plots (a)--(c) are for temperatures 
  $k_B T = 46,\, 140, \, 430\,\hbar\omegaZ$ listed in 
  Table~\ref{tab:temp_context}, $R$ is the Thomas-Fermi radius at $T = 0$ 
  [Eq.(\ref{eq:def-TF-radius})]. 
The dot-dashed green curves 
give the
prediction of the modified Popov theory, 
as developed by Andersen \& al \cite{Andersen2002}.
Plot~(d) analyzes, on a logarithmic scale, 
the density in the ``thermal wings'' of plot~(c) for
$R < z \lesssim z_{T}$ where $z_{T}$ (grey vertical line) is the point where
the trap energy becomes comparable to temperature,
$V( z_{T} ) = k_B T + \mu$.
The Rayleigh-Jeans density (dashed cyan) and the Bose-Einstein density
(dot-dashed green) are calculated for an ideal gas. The brown solid line
corresponds to the ncB density plus the quantum density level
which asymptotes to the latter,
$n_q$ [Eq.(\ref{eq:def-quantum-density})], indicated by the
blue dotted line.}
  \label{fig:total_densities}
\end{figure*}
We begin with the total density profile $n( z )$.
Similar to experimental data, this contains both the condensate in the 
trap center and thermally excited atoms that surround it. 
The equilibrium densities
in a harmonic trap are plotted in Fig.\ref{fig:total_densities}. 
Here and in the following figures, the sGPe densities (solid black) are
calculated as
$n( z ) = \langle |\psi(z, t_{\rm eq})|^2 \rangle$ where $t_{\rm eq}$ 
is the preparation time required for the system to reach a dynamical
equilibrium with the bath. In the ncB data (dashed red), we subtract the ``quantum
density'' $n_q = 1/( 2\Delta z )$ according to 
Eq.(\ref{eq:get-particle-density-in-Wigner}). 
This correction is small if the number of atoms per grid cell is large.
In addition, its impact on the mean field
is small since $g n_q \sim (1 \ldots 4)\times 10^{-3}\,\mu$.
%

For an independent benchmarking, the total density 
profiles are also compared to the total density of the
modified Popov scheme (green, dot-dashed).
At the low and intermediate temperature regimes, we find
good agreement
between all three theories. At the higher $T$, Fig.\ref{fig:total_densities}(c),
the ncB result deviates from both the sGPe and 
modified Popov density profiles (which are found to agree perfectly with each other \cite{AlKhawaja2002}).
Although the total density profiles also include a contribution from thermal atoms
within the condensate region, we defer their analysis in this region to 
Sec.\ref{subsec:PO-densities} below.

Instead, we focus first here on the representation of thermally excited atoms outside the 
condensate region. These atoms
populate the ``wings'' $R < |z| < z_{T}$, 
as illustrated in Fig.\ref{fig:total_densities}(d).
In this interval, the gas is still Bose degenerate with large occupation
numbers per mode.
Here, the data is well described
by a semi-classical ideal gas model. At each phase space point $(z, p)$
with effective single-particle energy $\varepsilon( p, z ) = 
p^2 / 2m + V( z )$, 
assuming the Rayleigh-Jeans law (equipartition)
for the occupation numbers, gives a density
\begin{eqnarray}
	n_{\rm RJ}(z) &=&
	\int\limits_{-p_{\rm max}}^{+p_{\rm max}}\!\frac{ {\rm d}p }{ 2\pi\hbar }
	\frac{ k_{B} T }{ \varepsilon( p, z )  - \mu }
	\label{eq:Rayleigh-Jeans-density}
\\
&=&
	\frac{  \sqrt{ 2 }  }{ \pi \,\ell} 
	\frac{T}{\sqrt{(V(z) - \mu) \hbar\omegaZ} }
	\arctan\left[\sqrt{ \frac{ E_{\rm max} }{
	V( z ) - \mu  } }\right]
\nonumber
\end{eqnarray}
Here, $E_{\rm max} = p_{\rm max}^2 / 2m$ is a cutoff energy that
depends on the maximum kinetic energy on the grid. As shown in 
Fig.\ref{fig:total_densities}(d) (dashed cyan),  
a good match to the sGPe data is obtained for 
$E_{\rm max} = 2\pi ( \ellZ /\Delta z)^{2} \hbar\omegaZ$ 
{}[ or $p_{\rm max} = 2\sqrt{\pi} \, \hbar / \Delta z$].
The divergence at $z \to \pm R$ is an artefact due to an infrared divergence
of the semiclassical approximation. 

Repeating this analysis with the Bose-Einstein distribution gives a density
\begin{eqnarray}
	n_{\rm BE}( z ) &=& 
	\int\!\frac{ {\rm d}p }{ 2\pi\hbar }
	\BE( \varepsilon( p, z ) - \mu )
	\label{eq:Bose-Einstein-density}
\\
&=&
	\frac{  1 }{ \ell }
	\sqrt{ \frac{ k_{B} T  }{ 2 \pi \, \hbar \omegaZ }  }\,
	g_{1/2}( x )
\nonumber
\end{eqnarray}
where $x = \exp[ ( \mu - V(z) ) / k_{B} T ]$ and
the Bose function has the asymptotics 
$g_{1/2}( x ) \approx x$ for $x \ll 1$. Good agreement with the
ncB data is obtained in the limit of infinite cutoff, see
Fig.\ref{fig:total_densities}(d) dot-dashed green). Indeed, it is
easy to see that the contribution of momenta above $p_{\rm max}$
is exponentially small provided $E_{\rm max} \gg k_{B} T$, as is the
case here. 

At positions beyond $z_{T}$, the gas enters a non-degenerate
regime where the occupation numbers are small for a large range
of momenta. Indeed, the fugacity satisfies
$\exp[ ( \mu - V(z) ) / k_{B} T ] \ll {\rm e}^{-1}$ for $|z| \gg z_{T}$,
and one can make the Boltzmann approximation,
$n( z ) \propto \exp[ - V( z ) / k_{B} T ]$. This corresponds to 
the ``thermal cloud'' familiar from the bimodal density distributions
of a partially condensed Bose gas in a trap~\cite{PitaevskiiBook}. 
The ncB data
provides a smooth crossover into this region provided the subtraction
of the quantum density is performed.
The brown solid line in
Fig.\ref{fig:total_densities}(d) shows the non-subtracted density
that reduces to a flat background of quantum density,
$n( z ) = n_q$, in the ``Boltzmann tail''. In this region, the actual number
of atoms per mode is much smaller than unity, and classical field methods
are no longer strictly justified.

By comparing the Rayleigh-Jeans and the Bose-Einstein densities
[Fig.\ref{fig:total_densities}], it is
clear that the sGPe overestimates the number of atoms in the thermal 
wings.
The numbers that one gets by integrating $n_{\rm RJ}( z )$ and
$n_{\rm BE}( z )$ in this region,
are quite small, however, when compared to the total number
of atoms. This is summarized in the last two columns of
Table~\ref{tab:excess-atoms}.
\begin{table}[hbt]
	\begin{center}
	\begin{tabular}{ | r || r | r | r | r | r | r | r |}
        \hline 
  	\multicolumn{1}{|c|}{} &
  	\multicolumn{1}{|c|}{$k_{B} T / \hbar \omegaZ$} &
        \multicolumn{2}{|c|}{$\langle N \rangle $ $[-z_T, z_T]$} & 
        \multicolumn{1}{|c|}{$\langle N \rangle\, {\rm wing}$} &
        \multicolumn{1}{|c|}{$\langle N \rangle\, {\rm tail}$} 
        \\ \cline{3-4}
  	\multicolumn{1}{|c|}{} &
  	\multicolumn{1}{|c|}{} &
        \multicolumn{1}{|c|}{sGPe} & 
        \multicolumn{1}{|c|}{ncB} & 
        \multicolumn{1}{|c|}{${\rm RJ} - {\rm BE}$} & 
        \multicolumn{1}{|c|}{${\rm BE}$} \\
        \hline
        \hline
	(a) & $46$ & 	
	$20\,007$ & $19\,994$ &
	$30.9$ & $7.5$
\\
		\hline
	(b) & $140$ & 
	$20\,132$ & $19\,981$ &
	$114.9$ & $24.6$
\\
		\hline
	(c) & $430$ & 
	$20\,795$ & $20\,498$ &
	$338.6$ & $78.5$
\\
		\hline
	\end{tabular}\end{center}
\caption[]{
Average atom numbers in the ``classical region'' outside the condensate
and beyond.
First and second column: average number within $|z| < z_T$ where
$V( z ) \le \mu + k_{B} T$, as obtained from the numerical 
simulations (including the Wigner subtraction for the tWncB
data). 
Columns three and four are based on the ideal gas densities
obtained in the classical approximation
(Rayleigh-Jeans law) and using Bose-Einstein occupation numbers.
The column ``wing'' gives the excess atoms present 
in the non-condensate, but still highly populated
region $R \le |z| \le z_T$ outside the condensate
(Thomas-Fermi radius $R$). The column ``tail''
gives the atoms that are located in the ``tails''
$z_T \le |z| < \infty$ of an ideal gas with the Bose-Einstein distribution.
For the Bose-Einstein density, an infinite momentum
cutoff is taken, as in Eq.(\ref{eq:Bose-Einstein-density}).
For the Rayleigh-Jeans density~(\ref{eq:Bose-Einstein-density}), 
we take a kinetic energy cutoff
$E_{\rm max} = 2\pi ( \ellZ /\Delta z)^{2} \hbar\omegaZ$ 
with grid spacing $\Delta z$ as given in 
Table~\ref{tab:temp_context}.
}
\label{tab:excess-atoms}
\end{table}
In the following, we calculate the total atom number by integrating the
density of each method between $z = - z_{T} \ldots z_{T}$, where the 
classical approximation is valid.

Consider now the significant difference between sGPe and ncB in 
Fig.\ref{fig:total_densities}(c), that occurs within the condensate region
at the highest temperature. We attribute this to the large thermal
amplitudes of the Bogoliubov modes that are no longer small compared
to the condensate mode (see Fig.\ref{fig:point-cloud-1}), and therefore 
the calculation of the non-condensate density has to be done more carefully.
In the modified Popov theory of Ref.\cite{Andersen2002}, fluctuations are
split into contributions to the density and phase, and phase fluctuations
are systematically discarded when calculating the average density.
We find that the sGPe data agrees well with this total density 
and are therefore confident that it captures the correct result.

\begin{figure*}[bth] 
  \centering
  \includegraphics[scale=0.375,clip]{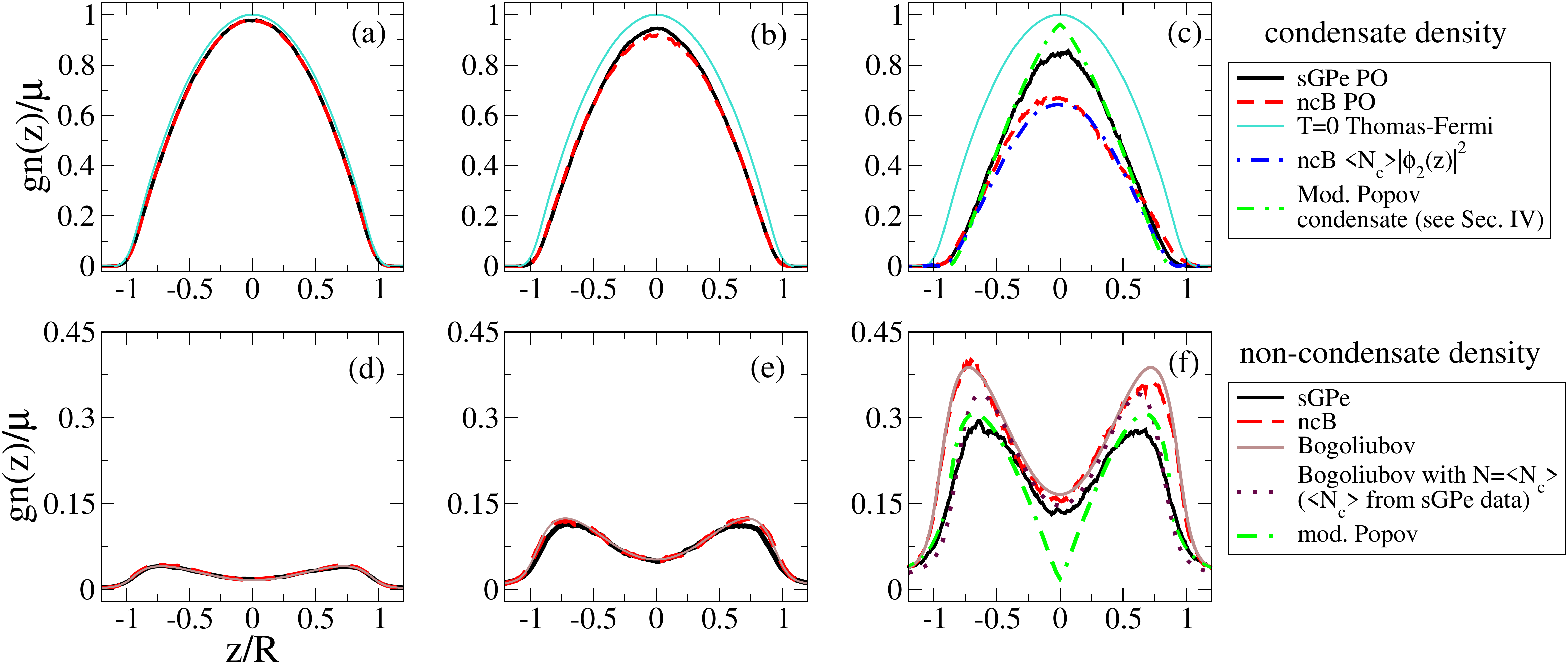}
  \caption{(Color online) 
Condensate (top row) and thermal cloud (bottom row) densities for temperatures
(a),(d) $k_{B} T = 46\,\hbar\omegaZ$;
           (b),(e) $k_{B} T = 140\,\hbar\omegaZ$; and
           (c),(f) $k_{B} T = 430\,\hbar\omegaZ$. 
Condensate density $n_c( z ) 
           = \langle N_c \rangle | \phi_c( z ) |^2$ 
           obtained from Penrose-Onsager analysis of 
           the one-body density matrix
           [sGPe: solid, black; tWncB: dashed, red] . 
           Also shown for reference is the $T=0$ stationary solution to the
           GP equation with eigenvalue $\mu$ (thin solid, turquoise), which coincides with the zeroth
           order condensate mode within ncB. In (c) the 
           condensate density $\langle N_c \rangle | \phi_2( z ) |^2$
           following the 2nd order correction within the ncB expansion is also shown (dot-dashed, blue),
           as well as the modified Popov {\it condensate} 
           density (dot-double dashed, green) [see Sec.\ref{sec:quasi-cond} 
           Eq.\eqref{eq:qc_to_PO_trap} for details].
           Bottom row: thermal density $n_{\rm th}( z )= n(z) - n_c( z ) $ [sGPe: solid black; 
           tWncB: dashed red].
           As in Fig.\ref{fig:total_densities}, the Wigner correction was made 
           for the tWncB case. Shown in (d--f) is the $T=0$ Bogoliubov prediction, Eq.\eqref{Eqn:Lbg}, with $N$ 
           (solid brown) and, in (f) only, also $N=N_{PO}$ of the sGPe 
           (dotted maroon), alongside the modified Popov 
           result (dot-double dashed, green).
           }
  \label{fig:PO_densities}
\end{figure*}

\subsection{Condensate and thermal excitations}
\label{subsec:PO-densities}

Having considered the total atomic density profiles, a further 
important temperature-dependent quantity is the condensate fraction.
The classical field methods under consideration here provide a 
unified description of the lowest modes of a trapped Bose gas,
and so further analysis is necessary to isolate the
condensate fraction, as in experiment. We choose to compare the
coherent and incoherent phases of the gas, and so focus on
a partitioning based on those atoms within the Penrose-Onsager ground state,
and those in orthogonal states \cite{Blakie2008}.

\subsubsection{Density profiles and depletion}

The phase coherent fraction of the gas is identified by making use of 
the Penrose-Onsager (PO) criterion for Bose-Einstein condensation \cite{Penrose1956}.
The stochastic simulations provide us, again through the operator-classical field 
correspondence, with the so-called one-body-density matrix $\rho( z, z' ) = 
\langle \psi^*( z ) \psi( z' ) \rangle$.
This matrix is hermitian and positive; evaluating its trace by integrating 
spatially, we get $\langle N \rangle$ in the sGPe method, and
$N + {\cal M}/2$ within the ncB approach. The PO criterion states
that Bose-Einstein condensation has occurred when the largest eigenvalue
of $\rho( z, z' )$, denoted here by $\langle N_c \rangle$, is comparable to
$N$, the other eigenvalues being much smaller \cite{Leggett2001}.
The eigenvector $\phi_c$ corresponding to $\langle N_c \rangle$ gives us the 
(PO) `condensate mode', 
whose spatial width characterizes the long-range phase coherence of the degenerate Bose gas. 
This mode provides a numerical way to implement the splitting
in Eq.(\ref{eq:split-psi}) between condensate and excitations,
since we get the condensate amplitude from
\begin{equation}
	a_c = \Delta z \sum_z \phi_c^*( z ) \psi( z )
	,
	\label{eq:def-ac-by-projection}
\end{equation}
and $\psi_\perp( z ) := \psi( z ) - a_c \phi_c( z )$ is by construction
orthogonal to $\phi_c$.

Figure~\ref{fig:PO_densities} plots the condensate
density $n_c( z ) = \langle N_c \rangle |\phi_c( z )|^2$ and the `thermal 
density'
$n_{\rm th}( z ) =  n( z ) - n_c( z )$, with corresponding 
thermal fractions 
 given in Table~\ref{tab:thermal_depletion}.
At the lowest
temperature (left images), 
the condensate dominates [Fig.\ref{fig:PO_densities}(a)] 
and the thermal density is globally weak [Fig.\ref{fig:PO_densities}(d)]. 
The two broad peaks at $z \sim \pm R$ arise because
the repulsive interaction with the condensate pushes the thermal component
out of the trap center. 
In the intermediate temperature data (middle images),  
the thermal fraction is larger,
yet the two methods give good agreement, 
with only a marginal difference in the 
peak value of the condensate density
which now becomes noticeably smaller than the $T=0$
solution for the same $\mu$ (thin blue line).
At the relatively higher temperature $k_{B}T = 430\,\hbar\omegaZ$
(right images), the two methods
disagree: their condensates are similar in axial extent, 
but the tWncB mode (dashed red) has a lower peak than the sGPe(solid black), 
so contains fewer atoms, 
and as a result the thermal fraction is higher.
For all temperatures, the approximate condensate mode constructed
in the ncB theory [$\phi_c( z )$ as given in Eq.(\ref{eq:psi-c-in-ncB});
dot-dashed, blue]
agrees well with the PO condensate extracted from the
density matrix of the ncB simulations (dashed red): 
we show data in Fig.\ref{fig:PO_densities}(c) only
as these quantities become indistinguishable at 
lower temperatures.

Although the modified Popov theory inherently solves for the 
quasi-condensate density, $n_{\rm qc}(z)$
(rather than the phase coherent, Penrose-Onsager 
condensate mode plotted here),
we also show in Fig. \ref{fig:PO_densities}(c) 
a prediction for the phase coherent condensate
which may be extracted from the modified Popov approach
(dot-dashed, green curves), and whose calculation does not require
the full one-body density matrix, as we discuss in more detail in
Sec.\ref{sec:quasi-cond}.
At this moderate temperature, this prediction 
agrees well with the sGPe, except for a small 
region near the centre; at lower temperatures, we have found an
even better agreement.

The difference between the ncB and sGPe results is 
likely due to the overestimation of the thermal density
in the condensate region within ncB. Consider the complex values of $\psi( z )$
in the bottom images of Fig.\ref{fig:point-cloud-1} for $z \lesssim R$:
the density is determined by the modulus of $\psi( z )$, and the tWncB
data clearly have more points with larger modulus.
Another reason may be the way the Bogoliubov energy 
spectrum is calculated: indeed, it is based on a condensate wave function,
denoted $\NCphiZero$ in Eq.(\ref{eq:psi-c-in-ncB}),
which contains all the particles of the system,
whereas an improved approach would account for the depletion of the 
condensate in calculating the Bogoliubov spectrum. To illustrate this, we 
consider as a first step an improved spectrum based upon a
a condensate with the same number of atoms as contained within 
the sGPe PO condensate. The thermal density which results is 
shown as the dotted maroon line in Fig. \ref{fig:PO_densities}(f), 
and is already in better agreement with both the sGPe
and modified Popov results, despite the 
quite rudimentary nature of the modification to 
the Bogoliubov spectrum.

\begin{table}[btp]
  \begin{center}
    \begin{tabular}{ | r || r | r | r |}
       \hline
       & $ k_{B} T / \hbar\omegaZ$ & {\rm sGPe} & {\rm tWncB}
\\ \hline\hline
	(a)              & 46      & 0.0474 & 0.0468
\\ \hline
	(b) 	          & 140     & 0.141    & 0.143
\\ \hline
	(c) 		& 430      &  0.365    & 0.450
\\ \hline
    \end{tabular}
  \end{center}
    \caption[]{Thermal fraction 
    $\langle N_{\rm th} \rangle / \langle N \rangle$ 
    versus temperature for the sGPe 
    and tWncB initial
	     states for the three chosen temperatures,
where $\langle N_{\rm th} \rangle$ is the integral over $n_{\rm th}( z )$ in the region
	     $|z| < z_{T}$, and
             the correction due to symmetric operator ordering is applied to 
             the tWncB data as
             for Fig.\ref{fig:total_densities}.}
\label{tab:thermal_depletion}
\end{table}

The thermal fraction, $\langle N_{\rm th} \rangle / \langle N \rangle
= 1 - \langle N_c \rangle / \langle N \rangle$,
varies slightly depending on the simulation method,
as shown in Table \ref{tab:thermal_depletion}. 
Since the total atom number $N$ varies across the statistical
ensembles, we calculate the condensate fraction from
$\langle N_{\rm c} \rangle / \langle N \rangle$. 
We observe again larger values for this quantity in the tWncB method at higher
temperatures (row (c)). We suggest that the depletion in this range
is not small enough to warrant the ncB expansion around a ``large''
condensate. Indeed, if we identify from Eq.(14), Ref.\cite{Castin1998},
$(N_{\rm th} / N_c)^{1/2}$ as a small expansion parameter,
within the ncB calculation this reaches the
value 
$\approx 0.90$ 
that is clearly not small.

\subsubsection{Condensate shape}
\label{cond_shape}
%
\begin{figure}[hbt]
\centerline{
\includegraphics*[scale=0.4]{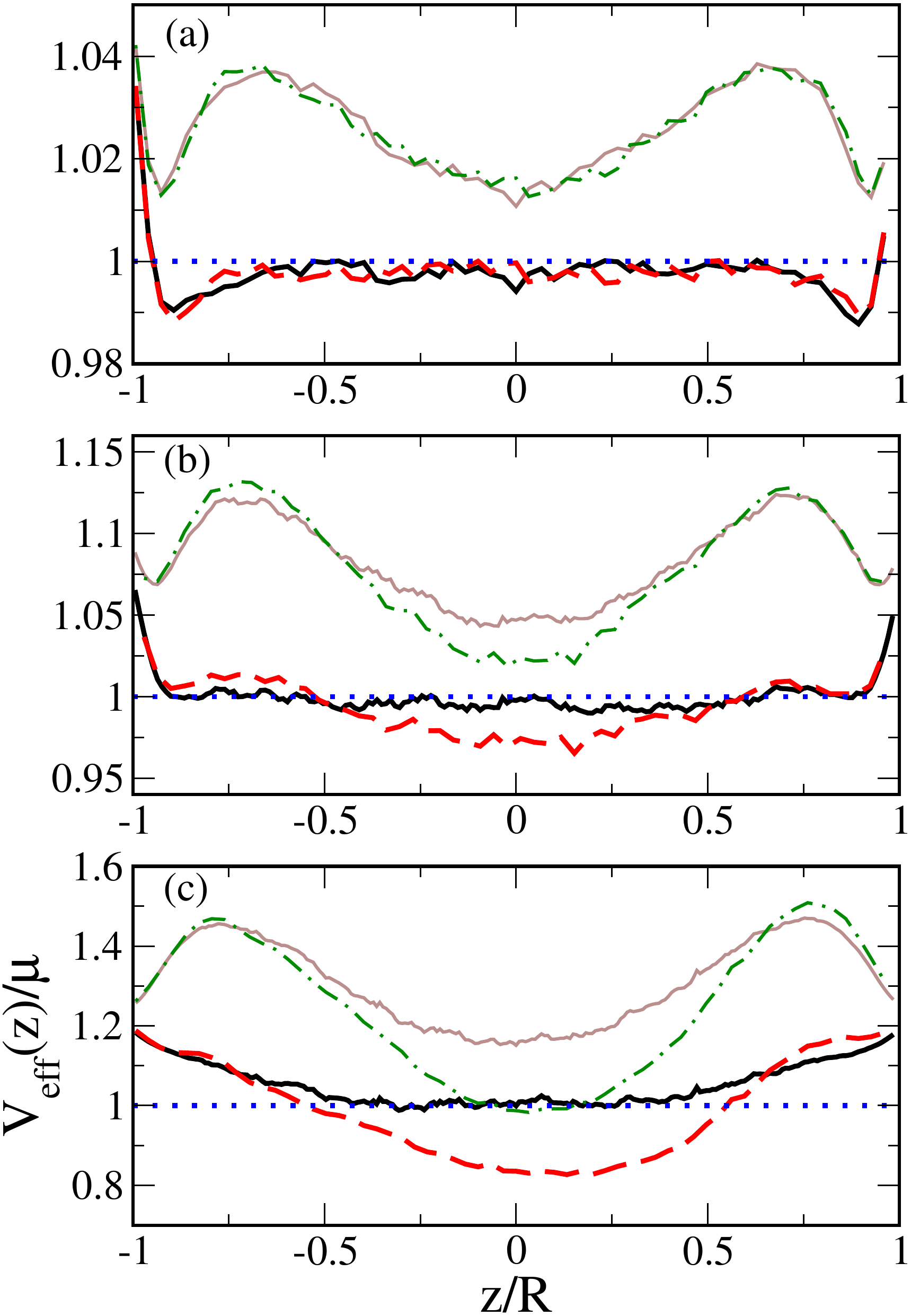}
}
\caption[]{(Color online) Analysis of the back-action of the non-condensate atoms on
the shape of the condensate mode $\phi_c$.
We plot the effective potential of Eq.(\ref{eq:HFB-potential})
with (lower curves, HFB theory) and without 
the last term (upper curves, HF theory), 
normalized to $\mu$. The condensate mode function $\phi_c( z )$ is
taken from a Penrose-Onsager analysis of the one-body density matrix.
Solid black/brown lines: sGPe data, (dot-)dashed red/green lines: tWncB data.
The HFB theory that includes
the back action from the anomalous average 
is closer to the actual chemical potential.
}
\label{fig:check-HFB}
\end{figure}
The back action of the non-condensate particles can be made visible by
a careful analysis of the shape of the condensate wave function $\phi_c( z )$,
the results of which are summarized in Fig.\ref{fig:check-HFB}.
The simplest generalization of the GPe that applies to nonzero temperature
is a Hartree-Fock (HF) potential due to non-condensate particles
(analogous to Eq.(\ref{eq:Popov-GPe}))
\begin{equation}
	\mbox{HF:} \quad
	\left(
	H_{\rm GP}[ N_c |\phi_c|^2 ] + 
	2 g n_{\rm th}( z )
	\right) 
	\phi_c = \mu \phi_c( z )
	\label{eq:GPe-with-HF-potential}
\end{equation}
where the thermal density is
\begin{equation}
	n_{\rm th}( z ) = \langle \hat\psi_{\perp}^\dag( z ) 
	\hat\psi_{\perp}( z ) \rangle
	\label{eq:def-thermal-density}
\end{equation}
One could however also 
take into 
account the anomalous average $m( z )$ due to non-condensate modes.
If the condensate field $a_c \phi_c$ is chosen real,
the anomalous average is simply given by
\begin{equation}
	m( z ) = \langle \hat\psi_{\perp}( z ) \hat\psi_{\perp}( z ) \rangle
	.
	\label{eq:def-anomalous-average}
\end{equation}
(A definition not based on U(1) symmetry breaking can be found
in Eq.(\ref{eq:def-squeezing}) below.)
Within the Bogoliubov approximation, one has
\begin{eqnarray}
	n_{\rm th}( z ) &=&
	\sum_{k} \{
	\BE( E_k ) |u_k( z )|^2 
	+ (\BE( E_k ) + 1 ) |v_k( z )|^2
	\}
	, \qquad
	\label{eq:nth-with-Bogoliubov}
\\
	m( z ) &=& \sum_{k} (2 \BE( E_k ) + 1 ) u_k( z ) v_k^*( z )
	.
	\label{eq:m-with-Bogoliubov}
\end{eqnarray}
(In a homogeneous gas of dimensions $D \ge 2$, the sums in
Eqs.(\ref{eq:nth-with-Bogoliubov}, \ref{eq:m-with-Bogoliubov}) are 
ultraviolet 
divergent~\cite{Bijlsma1997,Proukakis1998a,Shi1998} 
and
are regularized routinely 
by a renormalized coupling constant.
This is not needed in the one-dimensional case considered here,
due to the decay of $v_k(x)$ for $E_k \gg \mu$.)
This leads to the following `Hartree-Fock-Bogoliubov (HFB)
extension of Eq.(\ref{eq:GPe-with-HF-potential}):
\begin{eqnarray}
	\mbox{HFB:}\quad
	\left(
	H_{\rm GP}[ N_c |\phi_c|^2 ] + 
	2 g n_{\rm th}( z )
	\right) 	\phi_c &&
\label{eq:GPe-with-HFB-potential}
\\
	{} +
	g m( z ) \phi_c^*
	 &=& \mu \phi_c( z )
\nonumber
\end{eqnarray}
where we allowed momentarily for a complex-valued condensate
wave function to illustrate the $U(1)$ invariance of the theory
\footnote{%
The global U(1) symmetry holds with the transformation rules
$\phi_c \mapsto {\rm e}^{ {\rm i}\theta } \phi_c$,
$u_k \mapsto {\rm e}^{ {\rm i}\theta } u_k$,
$v_k \mapsto {\rm e}^{ -{\rm i}\theta } v_k$,
$n_{\rm th} \mapsto n_{\rm th}$,
$m \mapsto {\rm e}^{ 2{\rm i}\theta } m$.\label{endnote137}}.

The data shown in Fig.\ref{fig:check-HFB}, taken from
both stochastic methods, suggests that Eq.(\ref{eq:GPe-with-HFB-potential}) 
is more appropriate for the (PO) condensate mode, at least in the central
region of the trap.
This effect, along with the role of higher anomalous averages,
has been studied in detail in the context of
the microcanonical pGPe theory in \cite{Wright_preprint}.
There are significant 
differences at higher temperatures in the tWncB data, similar to those
appearing in the average density. 
%
\begin{figure*}[htb]
\centering
\includegraphics[scale=0.28,clip]{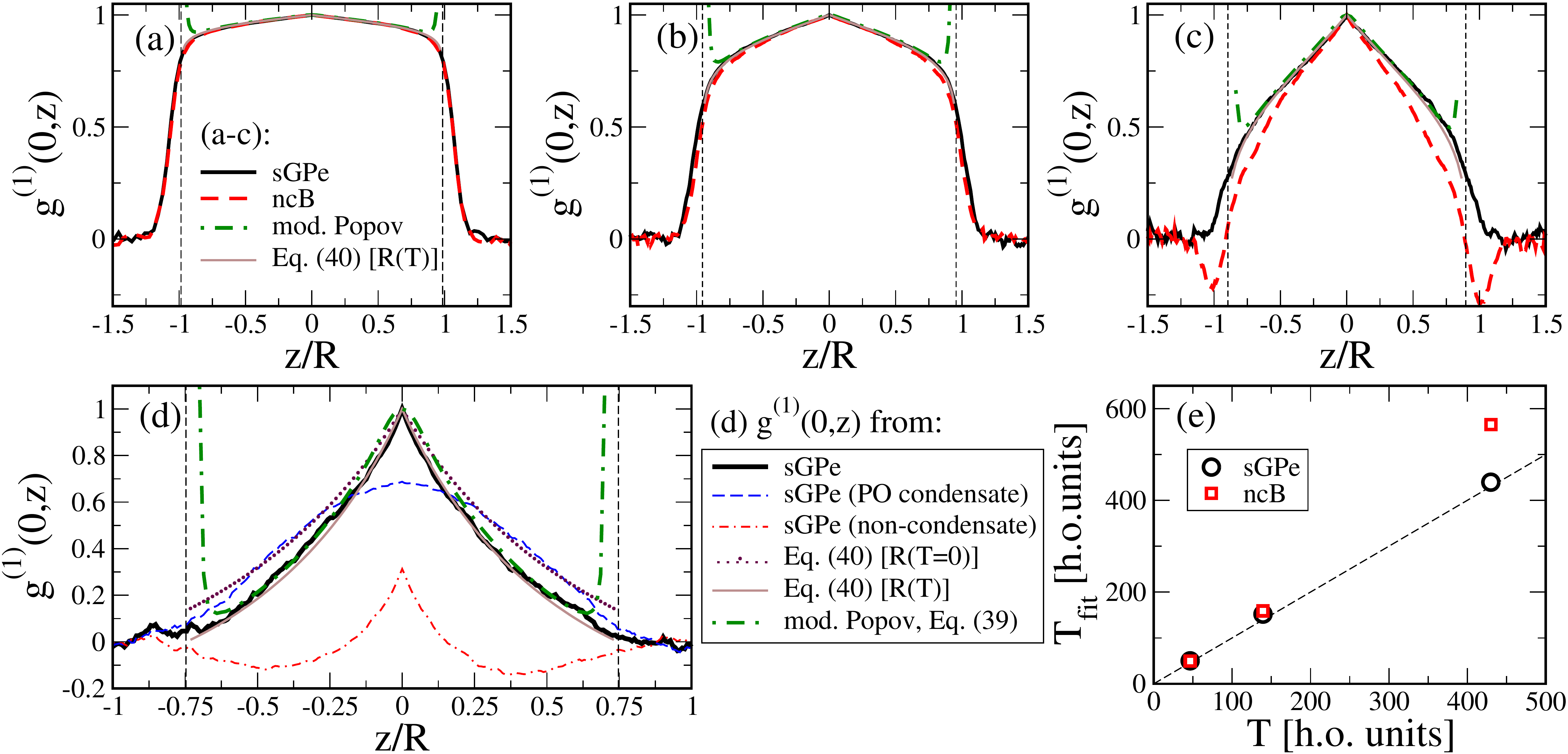}
  \caption{(Color online) 
First order correlation function
$\gOne( z ) = \gOne( 0, z )$, as defined in Eq.(\ref{eq:def-g1}),
from the sGPe data (solid black) and the ncB data (dashed red).
Temperatures in (a--c) as given in Table~\ref{tab:temp_context},
in (d), we have $T \approx 1.3\,T_{\phi}$ (only sGPe data shown).
We plot the real part of $\gOne( z )$, the imaginary part being 
negligibly small. Dot-dashed green lines:
Eq.(\ref{eq:mod-Popov-g1}), based on modified Popov theory. 
Brown, dotted lines: Eq.(\ref{eq:log_g1}), 
In (d), we plot Eq.(\ref{eq:log_g1}) using $R(T) < R$ 
(solid thin brown) and $R(T) = R$ (dotted orange).
The contributions from the sGPe Penrose-Onsager condensate
mode (dashed blue) and from non-condensate modes (dot-dashed
red) are shown separately, from Eq.(\ref{eq:break-g1-in-c-and-nc}).
In (a--d) the vertical dashed thin lines indicate $R(T)$.
(e) Estimated temperature based on fitting $\gOne( z )$ 
near $z = 0$ to Eq.\eqref{eq:approx-model-for-g1-correlation}
(dashed line) for sGPe (black circles) and ncB (red squares).
%
}
\label{fig:g1}
\label{fig:PO_dens_g1cond}
\end{figure*}
We recall that the HFB theory has been put into question because it
leads, for a homogeneous system, to a gapped excitation spectrum,
in contradiction to the Hugenholtz-Pines theorem \cite{Griffin1996, 
Proukakis1998b,Hutchinson2000,
Tommasini2005, Yukalov2006,Proukakis2008}.

One should also compare to the
modified Popov theory (Sec.\ref{subsec:mod_Popov}).
Indeed, we can interpret
the HFB potential in Eq.(\ref{eq:GPe-with-HFB-potential})
for the condensate mode, as a modification of the thermal density.
Going back to a real-valued condensate $\phi_c( z )$, the 
effective potential in Eq.(\ref{eq:GPe-with-HFB-potential}) takes the
form
\begin{equation}
	V_{\rm HFB} = V + g N_c \phi_c^2 + 2 g n_{\rm th} + g m
	\label{eq:HFB-potential}
\end{equation}
while the last term is missing in Eq.(\ref{eq:GPe-with-HF-potential}).
Now in Bogoliubov theory, we have (adopting real mode functions
normalized according to Eq.\eqref{eq:uv-orthogonal})
\begin{eqnarray}       
&& n_{\rm th}( z ) + m( z ) =   \sum_{k} \{
         \BE_k (u_k( z ))^2 +
        (\BE_k + 1 ) (v_k( z ))^2
\nonumber\\[-0.5em]    
&& \qquad {} +
         (2 \BE_k + 1 ) u_k( z ) v_k( z )       
\}
\nonumber\\    
&&=     \sum_{k} \{
         \BE_k (u_k( z ) + v_k( z ))^2
        + (u_k( z ) + v_k( z )) v_k( z )
        \}
        \hspace*{2em}  
\label{eq:retrieve-modified-mean-field}
\end{eqnarray}
with the shorthand $\BE_k = \BE( E_k )$. 
Equation~(\ref{eq:retrieve-modified-mean-field}) is closely related to
the thermal density $n_{\rm th}'$ of 
the modified Popov theory [Eq.(\ref{eq:mP_dens})]
which can be expressed in terms of Bogoliubov amplitudes 
(normalized to $u_p^{2}-v_{p}^{2}=1$) as
\begin{equation}
  \begin{split}
	n'_{\rm th}= n-n_{\rm qc} &=
        \frac{1}{V} \sum_{p} \{ 
	\BE_p (u_p + v_p)^2
	\\
	&
	+ (u_p + v_p) v_p
        + \frac{gn_{\rm qc}}{2\epsilon_{p} + 2\mu}
	\} 
  \end{split}
\end{equation}
This clearly contains an additional term;
this term is, however, small
in the temperature regime considered 
for our numerical simulations,  
leading to a good analogy
when the above re-interpretation of mean field potentials is made. This is in fact 
somewhat analogous to the work of Ref.\cite{Lee2003}.

%
The analogy cannot be pushed further, since the modified Popov
theory does not directly deal with the condensate mode (in the PO sense),
but with the quasi-condensate. See Sec.\ref{sec:quasi-cond}
for a link between the two quantities.


\subsection{Correlation functions}
\label{s:coherence-functions}

The focus in this section is upon spatial coherence in phase and
density. These show a rich physics in weakly interacting
one-dimensional Bose systems due to the separate 
characteristic temperatures for the suppression of
phase and density fluctuations.

\subsubsection{First-order coherence: phase fluctuations.}
\label{s:g1}

To study the phase coherence, we begin with the first-order coherence 
function (a normalized one-body density matrix)
\begin{equation}
	\gOne( z', z ) = \frac{ \langle \psi^*( z' ) \psi( z ) \rangle }{
	[ n( z ) n( z' ) ]^{1/2} }
	\label{eq:def-g1}
\end{equation}
where $n(z)$ is the average density and 
the normalization gives $\gOne( z, z ) = 1$.
For simplicity, we fix one position in the trap centre, $z' = 0$,
and write $\gOne( z ) = \gOne( 0 , z )$.  As illustrated in Fig.\ref{fig:g1},
$\gOne( z )$ scales roughly linearly in the centre and drops quickly
to zero towards the border of the 
condensate mode ($z \sim R$).
The slope in the centre agrees well between the two methods
(sGPe vs. ncB: solid black vs. dashed red),
but for $z \sim R/2$, differences appear at the highest 
temperature. Quite striking are the negative values for
$\gOne( z )$ which then occur, within the ncB formalism,
in the region of the Thomas-Fermi radius 
(anti-correlation between central region and condensate edge).

At the low temperatures probed, the behaviour of 
$\gOne( z )$ 
compares well with the theory of phase
fluctuations in weakly interacting Bose gases:
One starts from the \emph{Ansatz}
\begin{equation}
	\gOne( z ) = 
	\exp[ - \langle (\hat{\theta}(z) - \hat{\theta}(0))^2 \rangle/2 ]
	\label{eq:model-for-g1-correlation}
\end{equation}
and works out the thermal statistics for the 
phase operator $\hat{\theta}(z)$.

From this point, the required exponent may be calculated within 
the modified Popov theory \cite{AlKhawaja2002, StoofBook2009, Proukakis2006b}.
Making the classical approximation $\BE(E_j) \approx k_B T/E_j$, we can
write this exponent as
\begin{equation}
  \begin{split}
    \langle [ \hat\theta(z) - & \hat\theta(0) ]^{2} \rangle
    \\
    &=\frac{ 4 T }{ 3 T_{\phi} }
    \sum_{j > 0} 
    \bigg[
      \frac{2j+1}{j(j+1)}\left( P_{j}(z/R(T)) - P_{j}(0) \right)^{2}
      \\
      & - 
      \frac{ (2j+1)(\hbar\omega)^{2} }{ 8\mu^2 }\left(
		\frac{ P_{j}(z/R(T)) }{ 1 - z^2/R(T)^2 } - P_{j}(0)
      \right)^{2}
      \bigg].
  \end{split}
  \label{eq:mod-Popov-g1}
\end{equation}
where $P_{j}(z)$ are Legendre polynomials of order $j$.
Here $k_{B}T_\phi  
\approx 40 \,\mu$ is the characteristic temperature for phase fluctuations
[Eq.(\ref{eq:def-T_phi-and-T_c})].

The first term of Eq.(\ref{eq:mod-Popov-g1})
was derived by Petrov {\it et al.}~\cite{Petrov2000}
who, focussing on the temperature range $\hbar \omegaZ \ll k_B T$,
did not explicitly include in their expression
the temperature dependence of the Thomas-Fermi radius $R(T)$.
For the parameters considered, the second term within the 
sum of Eq.(\ref{eq:mod-Popov-g1}) typically gives a small contribution,
but leads to a rounding off of the central peak in $\gOne(0,z)$.
This additional term leads to a divergence in $g^{(1)}(0,z)$, 
however, as $|z|\rightarrow R(T)$,
due to the assumption that the condensate density is parabolic.
The results of Eq.\eqref{eq:mod-Popov-g1} are 
shown by the dot-dashed, green curves in Fig.~\ref{fig:g1}. In particular, 
at the highest temperature (panel (d)), the shrinking of the quasi-condensate
($R(T) < R$) is clearly seen, due to the increased thermal component of 
the density.

A closed form for $\gOne$(0,z) 
may be obtained by neglecting the second term 
of Eq.(\ref{eq:mod-Popov-g1}), 
since the mode summation for the first 
term can be performed analytically,
as pointed out previously in \cite{Petrov2000,Luxat2003}.
This gives
\begin{equation}
  \begin{split}
    \langle [ \hat{\theta}({z}) - \hat{\theta}({0}) ]^{2} \rangle
     = \frac{ 4T }{ 3T_{\phi} } \left|\log\left(\frac{1 + z/R(T)}{1 - z/R(T)}\right)\right|,
  \end{split}
  \label{eq:log_g1}
\end{equation}
where we have kept the temperature-dependent Thomas-Fermi radius,
thereby generalizing the expressions of Refs.\cite{Petrov2000,Luxat2003}. 
We note that Ghosh considered a quantized hydrodynamic
approach to the correlations of a quasi-1d Bose gas \cite{Ghosh2006}, 
which also extends the work of Petrov {\it et al.} in Ref.~\cite{Petrov2000}.
Eq.(\ref{eq:log_g1}) is shown with $R(T)$ taken from the modified 
Popov simulations
by the solid, thin brown lines in 
Figs.~\ref{fig:g1}(a--d) and agrees well with the numerical data. 
At the highest temperature [Fig.~\ref{fig:g1}(d)],
we see a clear deviation between an approach like the modified Popov theory
that takes condensate depletion into account (taking $R(T) < R$,
solid, thin brown), and
the original expression of Ref.\cite{Petrov2000} where the condensate
size is fixed at a constant Thomas-Fermi radius (dotted orange).


In the central trap region $|z| \ll R$, we can approximate Eq.\eqref{eq:log_g1} 
as 
\begin{equation}
        \langle (\theta(z) - \theta(0))^2 \rangle 
	\approx
	\frac{2 z }{ L_\phi( T ) },
	\quad
	L_\phi( T ) = \frac{ 3 T_\phi }{ 4 T } R(T),
	\label{eq:def-spatial-phase-diffusion}
\end{equation}
where $L_\phi(T)$ is the phase correlation length. This expression
illustrates that as $T \ge T_\phi$, the system is (first-order) coherent 
over a scale significantly shorter than the Thomas-Fermi 
radius \cite{Petrov2000}.
%
Assuming $R(T) \approx R$
and using the Thomas-Fermi formula for the condensate profile,
we can re-write the product $T_\phi R$ in 
Eq.(\ref{eq:def-spatial-phase-diffusion}) in a model-independent way
\begin{equation}
	|z| \ll R: \quad
	\gOne( z ) \approx 1 - \frac{g}{2 \hbar \omegaZ\, \ellZ }
	\,
	\frac{ z \, k_{B} T }{ \ellZ\, \mu }
	.
	\label{eq:approx-model-for-g1-correlation}
\end{equation}
Based upon Eq.\eqref{eq:approx-model-for-g1-correlation}, 
the slope of $\gOne(z)$ in the region $0 < z \alt R/2$ 
yields an independent measure of the temperature. We compare the
temperature extracted in this way (denoted $T_{\rm fit}$) 
to the input temperature of the simulations.
Both simulations give coherence functions $\gOne( z )$ that are 
consistent with Eq.(\ref{eq:approx-model-for-g1-correlation}), 
except for the ncB data at $T \approx 0.25\, T_\phi$
where the phase coherence is decaying faster. 
The ensemble prepared by ncB in the latter case appears not
only to be at a higher temperature (as in Fig.\ref{fig:PO_densities}(c)), 
it is actually not even stationary, as we illustrate in 
Sec.\ref{sec:tWncB-thermalization}.
\begin{figure*}[htb!]
  \centering
  \includegraphics[scale=0.28,clip]{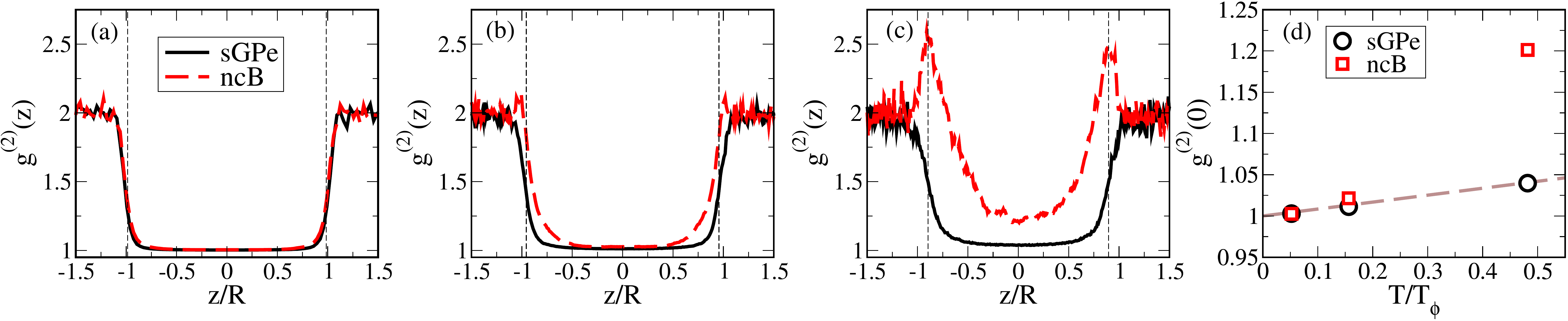}
  \caption{(Color online) Density correlation function 
  $\gTwo( z )$ [Eq.(\ref{eq:def-g2})]
           from the sGPe simulations (solid black) and
           the ncB simulations (dashed red). For the ncB case, the corrections
           of Eq.(\ref{eq:g2-with-subtractions}) are applied, so that in the
           quantum field theory, $\gTwo( z )$ has the meaning of a second-order
           coherence function.
           The vertical dashed thin lines indicate $R(T)$ at the 
           temperatures in (a--c), which are as in Table~\ref{tab:temp_context}.
(d) Comparison of results for $\gTwo(0)$ (black circles: sGPe, red squares: tWncB) 
with Lieb-Liniger theory,
Eq.(\ref{eq:g2-Kheruntsyan-our-units}), 
               taken from Ref.\cite{Kheruntsyan2005} (dashed brown).}
\label{fig:g2}
    \label{fig:g2-Kheruntsyan}  
\end{figure*}
Let us come back to the link between the phase coherence function
$\gOne$ and the one-body density matrix.
As explained in Sec.\ref{subsec:PO-densities}, the latter can be expanded
in orthogonal modes, with the (PO) condensate mode $\phi_c( z )$ giving the
dominant contribution. In the notation of the ncB approach 
{}[Eq.(\ref{eq:split-psi})], one can decompose the stochastic 
wavefunction in condensate and excitation parts, 
$a_c\phi_c( z ) + \psi_{\perp}( z )$,
where $\psi_{\perp}( z )$ represents all modes orthogonal 
to the PO mode. The
first order correlation function may then be broken down into
\begin{equation}
  \begin{split}
  \langle\psi^{*}(0)\psi(z)\rangle 
  = & \langle N_c \rangle
  \phi_c^{*}(0)\phi_c(z)
  +
  \langle\psi_{\perp}^{*}(0)\psi_{\perp}(z)\rangle
  .
  \end{split}
	\label{eq:break-g1-in-c-and-nc}
\end{equation} 
This gives two contributions to $\gOne( z )$ that are
 illustrated in Fig.\ref{fig:PO_dens_g1cond}(d). 
The condensate mode (dashed blue) provides the largest 
contribution and is positively correlated.
The contribution due to modes orthogonal to the PO mode 
(dot-dashed red) becomes negative towards the condensate
border because these modes have additional nodes (only
even mode functions contribute). This reduces $\gOne( z )$ below
the condensate contribution. The ``spike'' near the centre
is also due to higher modes and contains the approximately
exponential decay due to phase fluctuations 
{}[Eqs.(\ref{eq:model-for-g1-correlation}, \ref{eq:def-spatial-phase-diffusion})]
that becomes narrower as $T \sim T_\phi$. 


\subsubsection{Second-order coherence: density fluctuations}
\label{s:g2}
We now consider correlations of order four, namely 
fluctuations of the atomic density. These are captured by the 
coherence function
\begin{equation}
	\gTwo( z ) = \frac{ \langle |\psi(z)|^4 \rangle }{ [n( z )]^2 }.
	\label{eq:def-g2}
\end{equation}
In terms of field operators, we actually consider the
probability of detecting two atoms at $z$,
$\gTwo( z ) \propto
\langle\hat{\mathrm{\Psi}}^{\dagger}\hat{\mathrm{\Psi}}^{\dagger}
\hat{\mathrm{\Psi}}\hat{\mathrm{\Psi}}\rangle$.
This operator ordering is mapped to the following combination of tW data:
\begin{equation}
\langle\hat{\mathrm{\Psi}}^{\dagger}\hat{\mathrm{\Psi}}^{\dagger}
\hat{\mathrm{\Psi}}\hat{\mathrm{\Psi}}\rangle
=
\langle |\psi(z)|^4 \rangle_W - 4\langle |\psi(z)|^2 \rangle_W n_q
+ 2 n_q^2
	\label{eq:g2-with-subtractions}
\end{equation}
where $n_q$ is the quantum density level on the 
grid [Eq.(\ref{eq:def-quantum-density})].
The local density $n(z)$ used for normalization is also Wigner-corrected
and given by Eq. \eqref{eq:get-particle-density-in-Wigner}. The sGPe
data are taken as in Eq.~(\ref{eq:def-g2}).

It is well known that for a single-mode coherent field, 
$\gTwo( z ) = 1$, while for a chaotic (multi-mode) field with Gaussian
statistics, $\gTwo( z ) = 2$ \cite{MandelWolf,Dodd1997}.
Anti-bunching, $\gTwo( z ) < 1$,
is a non-classical effect that we do not expect within the classical
stochastic theories used here; it occurs indeed at lower 
temperatures, see Refs.\cite{Kheruntsyan2005,Eckart2008}.
Any value in between the limits
$1$ and $2$ is thus a measure of how many modes effectively
contribute to the density. 
The data shown in Fig.\ref{fig:g2} follows the expected 
behaviour \cite{Dodd1997, Blakie2005, Proukakis2006b, Bisset2009a}: 
a flat plateau in the center and a jump from $1$ to $2$ at the border 
of the condensate. (The oscillations outside the center are statistical
errors that are enhanced by the normalization in Eq.(\ref{eq:def-g2}).)
The jump at the Thomas-Fermi radius
becomes more gradual as the temperature rises, and the 
single-mode region shrinks \cite{Proukakis2006b}. 
At the highest temperature, the ncB theory gives anomalously large
values $\gTwo( z ) > 2$ near the condensate border.

Since the lowest excitation modes of the condensate carry mainly phase
fluctuations~\cite{PitaevskiiBook}, we expect a significant deviation
from $\gTwo( z ) = 1$ to set in at a higher temperature compared to 
$\gOne( z )$. This can be made more precise by comparing to 
Ref.\cite{Kheruntsyan2005} where Kheruntsyan \emph{et al.} 
use exact solutions of the Lieb-Liniger model, within the 
local-density approximation, to calculate the density correlation in 
a trapped gas. Their result for the trap center,
within the weakly interacting regime  
$\mu \ll 2 k_{B}T \ll k_{B} T_{d} = N\, \hbar\omegaZ$ of interest here,
is (Eq. (5.10) of \cite{Kheruntsyan2005}):
\begin{equation}
	\gTwo( 0 ) 
	\approx 1 + \frac{ 4\sqrt{ 2 } \, T }{ 3 T_{d} }.
	\label{eq:g2-Kheruntsyan-our-units}
\end{equation}
The linear increase in temperature is in good agreement
with the results of our classical field simulations, 
see Figure \ref{fig:g2-Kheruntsyan}(d), and \cite{Proukakis2006b}.
The high-temperature ncB data, however,
is much too large compared to both
the sGPe and Lieb-Liniger theory.
This suggests either a higher temperature, consistent with the findings of 
previous tests, or a significant overestimation of density fluctuations.
We mention that values $\gTwo( z ) \to 3$ \cite{Gardiner2001a}
would arise when the field
$\psi( z )$ has a fixed phase and behaves like a real-valued random
number. This is related to a large contribution from the 'squeezing 
correlation' $m( z ) \sim \langle \psi( z ) \psi( z ) \rangle$ 
that we discuss now.

\begin{figure*}[htb!]
  \centering
  \includegraphics[scale=0.345,clip]{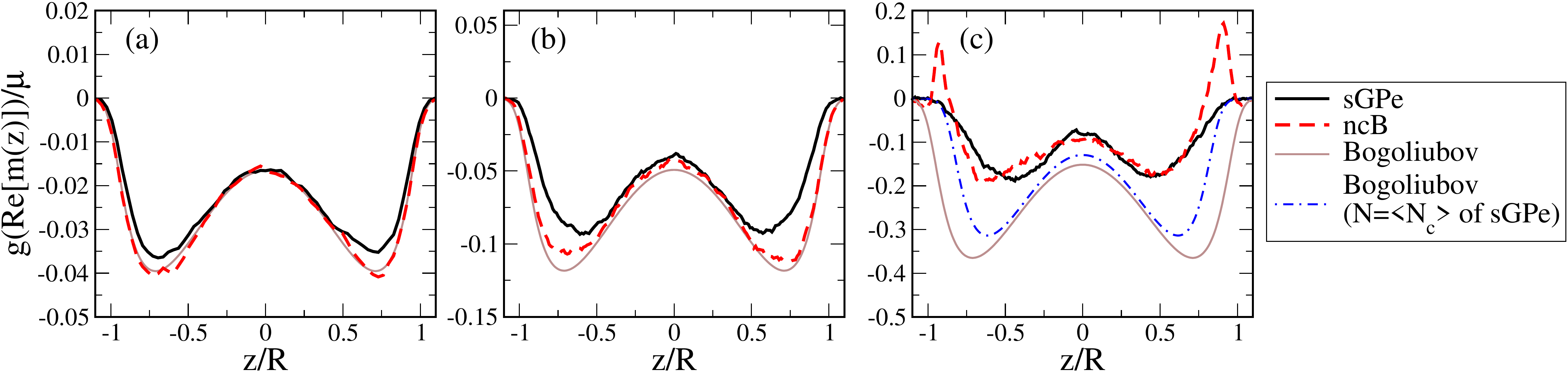}
  \caption{(Color online) Real part of the squeezing correlation (anomalous average)
  $m( z )$, as defined in Eq.(\ref{eq:def-squeezing}), for the temperatures
  of Table~\ref{tab:temp_context}. Solid black: sGPe result, dashed red: ncB result.
  The Bogoliubov result (solid brown, Eq.(\ref{eq:m-with-Bogoliubov}))
  is calculated with the mode functions of the $T = 0$ BdG operator~(\ref{eq:def-LQ}),
  and also using the Bogoliubov spectrum calculated for a condensate number equal
  to the sGPe PO condensate number (blue, dot-dashed).
  }
\label{fig:squeezing-correlation}
\end{figure*}

\subsubsection{Squeezing and anomalous average}
\label{s:squeezing}

We finally consider the anomalous average of the non-condensate
field defined as (cf. \cite{Wright2010})
\begin{equation}
	m( z ) = \bigg\langle\frac{ 
	 [ (a_c \phi_c( z ))^* \psi_\perp( z ) ]^2 
	}{  |a_c\phi_c( z )|^2 }\bigg\rangle 
	\label{eq:def-squeezing}
\end{equation}
where $a_c \phi_c( z )$ is the component of the matter wave field
along the condensate mode [Eq.(\ref{eq:def-ac-by-projection})]
and $\psi_\perp( z )$ is the perpendicular component.
Note that $m( z )$ as defined in Eq.(\ref{eq:def-squeezing})
is invariant under global phase transformations
of both the total field $\psi( z )$ and the condensate mode function
$\phi_c( z )$. It vanishes if the condensate amplitude $a_c$ and 
$\psi_\perp( z )$ have no fixed phase relation: we thus probe the
phase locking between the condensate and non-condensate fields.
We use for our data analysis the Penrose-Onsager condensate mode
introduced in Sec.\ref{subsec:PO-densities}.
This interpretation of the anomalous average can be re-phrased in the 
squeezing language of quantum optics~\cite{MeystreSargent}: 
the interference term between condensate and non-condensate fields
is split in two quadrature fields (both have the dimension of a density)
\begin{equation}
	(a_c \phi_c( z ) )^* \psi_\perp( z ) =
	X_n( z ) + {\rm i} X_\theta( z )
	.
	\label{eq:introduce-quadratures}
\end{equation}
The real part $X_n$ indeed gives the (local) density fluctuation on top of 
the condensate density $|a_c \phi_c( z )|^2$, while the imaginary part
$X_\theta$ describes phase fluctuations (if these are small). 
On average, these quadratures are zero, and the difference of their
variances is
\begin{equation}
	\langle X_n^2( z ) \rangle - \langle X_\theta^2( z ) \rangle = 
	{\rm Re} \, \langle
	[ (a_c \phi_c( z ))^* \psi_\perp( z )]^2 \rangle
	\label{eq:interpret-m-as-squeezing}
\end{equation}
which is just the real part of $m( z )$ defined in Eq.(\ref{eq:def-squeezing}).
The sum of these variances equals the normalization factor in Eq.(\ref{eq:def-squeezing}) times the non-condensate density $n_{\rm th}( z ) =
\langle |\psi_{\perp}( z )|^2 \rangle$.

Within the Bogoliubov approximation, we calculate $m( z )$ by choosing 
a phase reference where the condensate field $\phi_c( z )$ is
real-valued. By expanding $\psi_\perp( z )$ over Bogoliubov modes
with operator amplitudes
$\beta_k = a_c^* b_k / |a_c|$ instead of $b_k$, global phase
invariance holds (see Refs.\cite{Castin1998, Kocharovsky2006,endnote137}).
In the Bogoliubov limit, condensate number fluctuations can be ignored,
$|a_c|^2 = \langle N_c \rangle$, 
and we recover Eq.(\ref{eq:m-with-Bogoliubov})
\begin{eqnarray}
	m( z ) &=& 
	\sum_k \langle
	( u_k( z ) \beta_k^{\phantom\dag} 
		+ v_k^*( z ) \beta_k^{\dag} )  
	( u_k( z ) \beta_k^{\phantom\dag} 
		+ v_k^*( z ) \beta_k^{\dag} )
		\rangle
\nonumber
	\\
	&=& 
	\sum_k 
	( 2 \BE(E_k) + 1 ) \,
	u_k( z ) v_k^*( z )
	\label{eq:result-Bogo-squeezing}
\end{eqnarray}
where $\BE(E_k) 
= \langle \beta_k^\dag\beta_k^{\phantom\dag} \rangle$.
This quantity is thus sensitive to the `anomalous' or `hole' part
$v_k( z )$ of the Bogoliubov modes. 
In particular, we note that the anomalous average
shows a quite precise linear scaling in $T$ in the temperature range of 
interest. This illustrates the relative dominance of highly populated modes
that are well described within the classical approximation.

The data in Fig.\ref{fig:squeezing-correlation} show a reasonable
qualitative agreement between the stochastic simulations 
and Bogoliubov theory:
$m( z )$ has a large negative real part. Beyond this, there are
clearly differences on a quantitative level, 
particularly as temperature is increased.
Note that the anomalous average
is of comparable magnitude to the non-condensate
density $n_{\rm th}( z )$, which points towards a strong enhancement of the
phase fluctuation quadrature $X_\theta( z )$ relative to density fluctuations.
The agreement between sGPe (solid black) and ncB (dashed red) remains reasonable
at all temperatures considered in our comparison, however both theories deviate from the
($T=0$) Bogoliubov prediction (brown, solid line). 
Also shown in
Fig. \ref{fig:squeezing-correlation} is the result of
Eq.\eqref{eq:result-Bogo-squeezing},
calculated with Bogoliubov mode functions  
for a condensate with a reduced number of atoms: we have replaced
in the Bogoliubov--de Gennes operator $N |\NCphiZero|^2$ by
$\langle N_c \rangle |\phi_c|^2$, where $\langle N_c \rangle$
is obtained from the sGPe. Note that this does not
significantly improve the agreement, unlike the case of the thermal density 
of Fig.\ref{fig:PO_densities}(f).
\begin{figure*}[htb!]
\centerline{
\includegraphics[width=120mm]{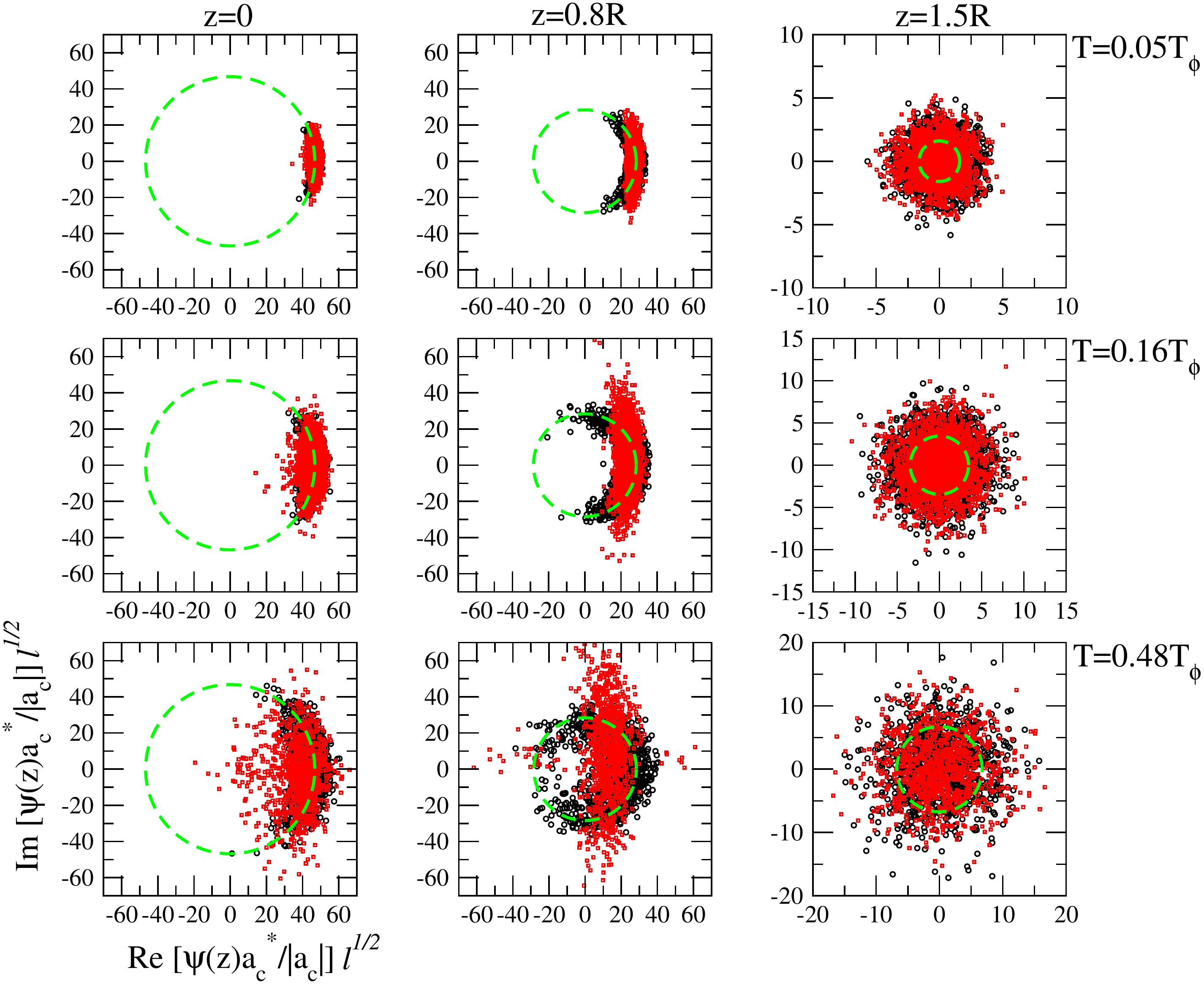}
}
\caption[]{(Color online) Representation of the ensemble of stochastic
fields $\psi( z )$, locked to the phase of the condensate mode 
(we multiply with $a_c^* / |a_c|$ where $a_c$ is the 
amplitude~(\ref{eq:def-ac-by-projection}) of the (Penrose-Onsager)
condensate mode). Black: sGPe simulation, red: ncB simulation.
Trap position and temperatures as in 
Fig.\ref{fig:point-cloud-1}. Note the elliptical shape of the point cloud,
illustrating the relative enhancement of phase fluctuations (imaginary
part). The sGPe data deform into a ``crescent'' shaped cloud at 
intermediate and high temperatures, while the ncB distribution remains
aligned to orthogonal quadratures. 
The dashed green circles represent the Thomas-Fermi ($|z| < R$) and 
Bose-Einstein ($z = 1.5\,R$) predictions for the square root of the
density at each trap position and temperature, 
as in Fig.\ref{fig:point-cloud-1}.
}
\label{fig:point-cloud-2}
\end{figure*}
The enhancement of phase fluctuations
is nicely illustrated in Fig.\ref{fig:point-cloud-2} where the realizations
of the complex field $\psi( z ) a_c^* / |a_c|$ are plotted. 
The additional phase factor, relative to the data of 
Fig.\ref{fig:point-cloud-1}, 
removes the random phase of the condensate mode,
and reveals the squeezed distribution of the complex field, 
with enhanced fluctuations in the quadrature
orthogonal to the condensate mode. At higher temperatures, the sGPe
data show that these fluctuations are channeled into a 
``crescent''-shaped region, maintaining the relative suppression of
density fluctuations. The ncB expansion does not take this into account,
and the fluctuations keep their alignment to orthogonal quadratures
so that
density fluctuations (in the radial direction) become too large. This
clearly happens when the phase difference across different points in the
system becomes comparable (in standard deviation) to $\pi/2$ 
so that linearization procedures break down
(``quasi-condensate regime'').

\subsection{Condensate statistics and fragmentation}
\label{subsec:cond-stats}

We analyze in this section the one-body density matrix 
$\langle \psi^*( z ) \psi( z' )$ in more detail. Its eigenvector with
the largest eigenvalue corresponds to the condensate mode 
$\phi_c( x ) $ in the
sense of Penrose and Onsager, as explained in Sec.\ref{subsec:PO-densities}.
The distribution function of the corresponding complex amplitude
$a_c$ [Eq.(\ref{eq:def-ac-by-projection})]
provides us with the probability of finding $N_c = |a_c|^2$ atoms
in the condensate, the so-called ``counting statistics''. We emphasize
that this quantity depends on moments (correlation functions) of
arbitrarily high order of the stochastic field. We also discuss the relative
importance of non-condensate modes whose occupation grows as the 
temperature increases, illustrating a phenomenon similar to 
fragmentation~\cite{Leggett2001} for $T \sim T_\phi$ and above.

\subsubsection{Counting statistics}
\label{sec:cond_stats}

The statistics $P( N_c )$ of the number of condensate atoms 
has been well studied in the context of laser theory \cite{SargentScully}, 
and Bose-Einstein condensation \cite{Grossmann1996, Wilkens1997, 
Kocharovsky00, Svidzinsky2006, Shelankov08, Idziaszek2009, Bisset2009b, Bienias2010}.
It is worth mentioning that number distributions for
an ideal Bose gas provide an example where the canonical
and grand-canonical ensembles of thermodynamics are not 
equivalent: only the average atom numbers coincide, while all higher 
moments of the (total) atom number are anomalously large
in the grand-canonical ensemble
\cite{Wilkens1997, Kocharovsky2006}. This anomaly is removed in
an interacting gas due to the energetic cost of adding particles to 
the condensate. 
\begin{figure*}[htb]
  \centering
  \includegraphics[scale=0.29,clip]{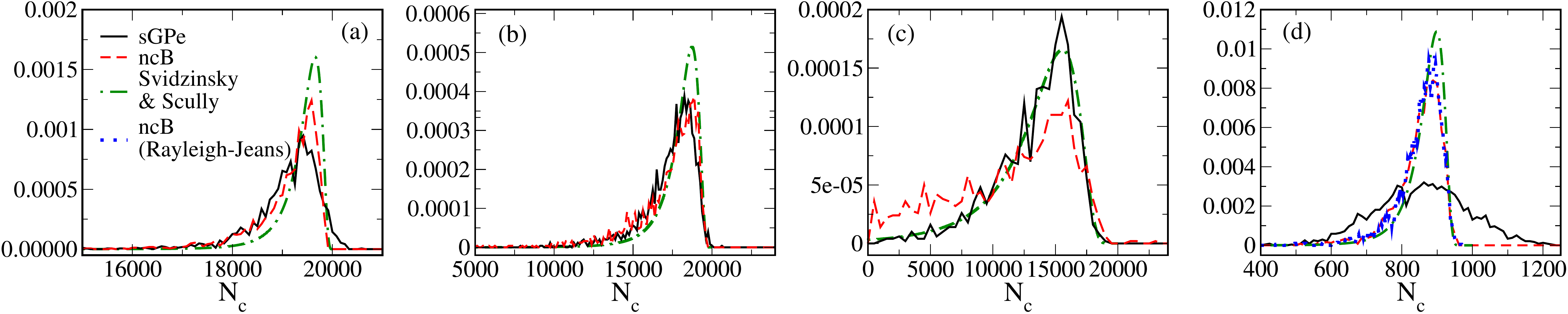}
  \caption[]{
  (Color online) Counting statistics $P(N_c)$ for the condensate number,
  with $N_c = |a_c|^2$, and the condensate mode in
    Eq.(\ref{eq:def-ac-by-projection})
  obtained from the Penrose-Onsager criterion.
  Solid black: sGPe, dashed red: ncB, dot-dashed green: theory of 
  Ref.\cite{Svidzinsky2006}.
  Temperatures in (a--c) increase from left to right, 
  (as in Table~\ref{tab:temp_context}), 
  ($N \approx 20\,000$, $\mu = 22.41\,\hbar\omegaZ$).  
  Panel (d) 
  has $N \approx 1\,000$ ($\mu = 3.11\,\hbar\omegaZ$)
  and a temperature $T \approx 0.23\,\Tc$, the same ratio as in
  (c). Dotted blue line: ncB with Rayleigh-Jeans instead of Bose-Einstein
  occupation numbers, i.e., in Eq.(\ref{eq:ave-mode-amplitude-tW}),
  $\sigma^2_k \mapsto k_{B} T / E_k$.
}
\label{fig:counting-statistics}
\label{fig:cond_stats_summary}
\end{figure*}
The counting statistics $P( N_c )$ is found from the stochastic
data by drafting a histogram of the values for $N_c = |a_c|^2$ 
across the ensemble of realizations, with
$a_c$ calculated from Eq.(\ref{eq:def-ac-by-projection}).
Obviously, the $N_c$ need not be integers
here, due to the replacement of operators with classical fields.
Figure~\ref{fig:counting-statistics} compares these data with the
theory of Scully and co-workers (S\&{}Co), in particular 
Ref.\cite{Svidzinsky2006}.
This is developed within the canonical ensemble, and treats the
non-condensate modes either within Bogoliubov theory (low temperatures), 
or extrapolated to higher temperature with the help of a rate equation
approach. We outline the main steps in Appendix~\ref{a:Scully-theory}.

\begin{figure}[b!]
  \centering
  \includegraphics[scale=0.326,clip]{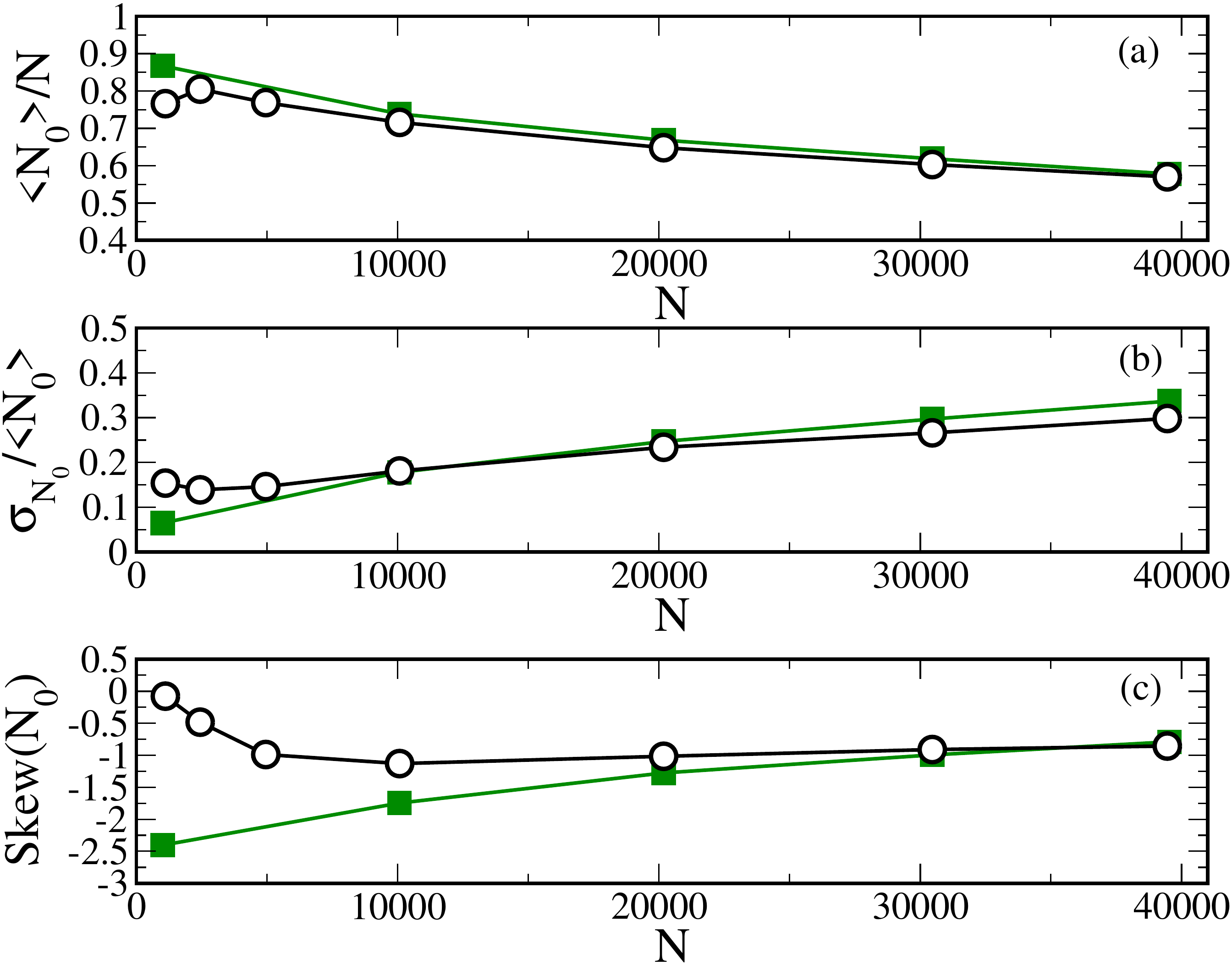}
  \caption[]{(Color online) 
  First three moments of $P(N_c)$ vs. total particle number 
  $\langle N \rangle$
            of sGPe (open circles) vs. theory of Ref.\cite{Svidzinsky2006} (filled green squares):
	    (a) mean value $\langle N_c \rangle$ scaled to $\langle N \rangle$ 
	    (condensate fraction),
	    (b) relative standard deviation $\sigma( N_c ) / \langle N_c \rangle$,
	    (c) skewness or third centered moment,
	    ${\rm skew}( N_c ) 
	    = \langle (N_c - \langle N_c \rangle)^3 \rangle / \sigma^{3}( N_c )$.
  The total particle number $N$ is calculated over the 
  region $|z|<2R$.}
\label{fig:Scully-sGPe_moments}
\end{figure}

Reasonable agreement between the stochastic simulation methods 
and S\&{}Co (dot-dashed green)
is apparent at all three temperatures shown in Fig.\ref{fig:counting-statistics}.
At the lowest temperature [Fig.\ref{fig:counting-statistics}(a)],
the sGPe approach (solid black) gives a slightly a 
broader distribution. This broadening
becomes more pronounced if the atom number is lowered [panel (d)],
bringing the system closer to an ideal gas.
At the highest temperature considered, the ncB distribution (dashed red)
gives too much
weight on small condensate numbers.
This failure is particularly striking, given that the method of 
S\&{}Co uses a Bogoliubov description of the 
non-condensate particles that is fairly close to the ncB expansion.
There is one additional ingredient, 
however, namely the growth 
and depletion rates in the rate equation \emph{Ansatz} for the
counting statistics (see App. B for details):
these rates are calculated as a function of
$N_c$, while in the ncB expansion, the Bogoliubov spectrum
is calculated only for the extreme case $N_c = N$. We thus expect
that some effects of a strongly depleted condensate are not captured.
This illustrates again the importance of self-consistently
adjusting $N_{c}$ within the theory as temperature is varied, see
also Ref.\cite{Idziaszek2009}.

\subsubsection{Discussion}

It may come as a surprise that a grand-canonical approach like the
sGPe where the total atom number is not fixed (i.e., values
$N_c  > \langle N \rangle$ are not excluded), 
is able to reproduce
the counting statistics of number-conserving theories (like the ncB
and S\&{}Co).
We attribute this to the interatomic interactions in the system
that translate fluctuations in the condensate number into energetic 
changes. This makes the system ``stiffer'' and
suppresses number fluctuations relative to
the ideal Bose gas \cite{Wilkens1997,Kocharovsky2006}. 
A complementary explanation is based on the observation that the
condensate mode $a_c \phi_c$ is a low-energy subsystem of the 
total field (represented by $\sGPphi$), where the
non-condensate fraction can play the role of a particle reservoir.
This suggests that the condensate subsystem can be described 
within a grand-canonical scheme even if the total atom number is
fixed: for that it would be sufficient to consider a high enough 
temperature so that a large number of non-condensate atoms is present. 
Indeed, the width $\sigma( N_c )$ of the canonical
counting statistics translates
two physically different mechanisms: on the one hand, 
the statistical uncertainty of
the non-condensate (Bogoliubov) occupation numbers in the ncB 
expansion (exponentially distributed with mean $\BE( E_k )$),
and on the other hand, the dynamical particle exchange with the 
non-condensate modes due to interactions, similar to what is done 
between system and bath within the sGPe.

These considerations also suggest an explanation for the 
broader statistics that the sGPe method returns at low
temperatures and small numbers
{}[Fig.\ref{fig:cond_stats_summary}(d)].
It is symptomatic of the grand canonical ensemble 
which underlies the formulation of the sGPe, and leads to anomalously
large number fluctuations for this nearly ideal gas.
We have checked that the classical approximation 
underlying the sGPe is not in error here: indeed, the counting
statistics of the ncB method (canonical ensemble) is essentially the same
when Bose-Einstein occupation numbers are replaced by their classical 
(Rayleigh-Jeans) limit [Fig.\ref{fig:cond_stats_summary}(d), dotted blue curve]. 

Figure~\ref{fig:Scully-sGPe_moments} shows 
the moments of the sGPe counting statistics as a function
of the total particle number. This is compared to the
theory of S\&Co (canonical ensemble).
We vary $\langle N \rangle$ 
over more than one order of magnitude,
as a way to change the importance of particle interactions. 
The temperature is kept at a fixed ratio $T/\Tc = 0.23$,
where $\Tc = \Tc( N )$ is the critical temperature for an ideal gas 
(see Eq.(\ref{eq:def-T_phi-and-T_c}) and Fig.~\ref{fig:TN-plane}).
The mean values [Fig.\ref{fig:Scully-sGPe_moments}(a)]
agree well between the ensembles,
as expected \cite{Kocharovsky2006}, except perhaps at the smallest 
particle numbers. The standard deviation
{}[Fig.\ref{fig:Scully-sGPe_moments}(b)] 
is larger at small $\langle N \rangle$ where one is closer to the ideal gas,
but converges to S\&Co theory for larger systems.
In the third moment [Fig.\ref{fig:Scully-sGPe_moments}(c)],
which measures the deviation from Gaussian statistics, 
we see that, at rather small numbers, the sGPe predicts a more symmetric distribution
compared to the negatively skewed distribution obtained in the 
canonical ensemble. 
This suggests that in weakly interacting systems, the small non-condensate 
fraction cannot protect the condensate, like a ``buffer'',
against the Gaussian noise in the stochastic dynamics.
The skewness builds up at higher particle numbers where also more modes
are highly occupied, which is of course the regime where the sGPe should
perform well.

\subsubsection{Fragmentation}
\label{subsec:fragmentation}


For $T \ll T_{\phi}$, one mode dominates the system, as many atoms
are condensed and phase coherent (nearly pure condensate), whereas 
at higher temperatures, 
many modes become appreciably occupied (quasi-condensate). 
This behaviour can be 
made quantitative by considering the set of eigenvalues of the one-body
density matrix
$\langle \psi^{*}(z) \psi(z') \rangle$, 
i.e., the average occupations $N_k$
of the corresponding modes ($k = 1, \ldots, {\cal M}$).
We recall that the eigenmodes of the one-body density matrix 
are distinct from those of the Hamiltonian, since the latter is not 
quadratic in the field.

\begin{figure}[b!]
  \centering
  \includegraphics[scale=0.205,clip]{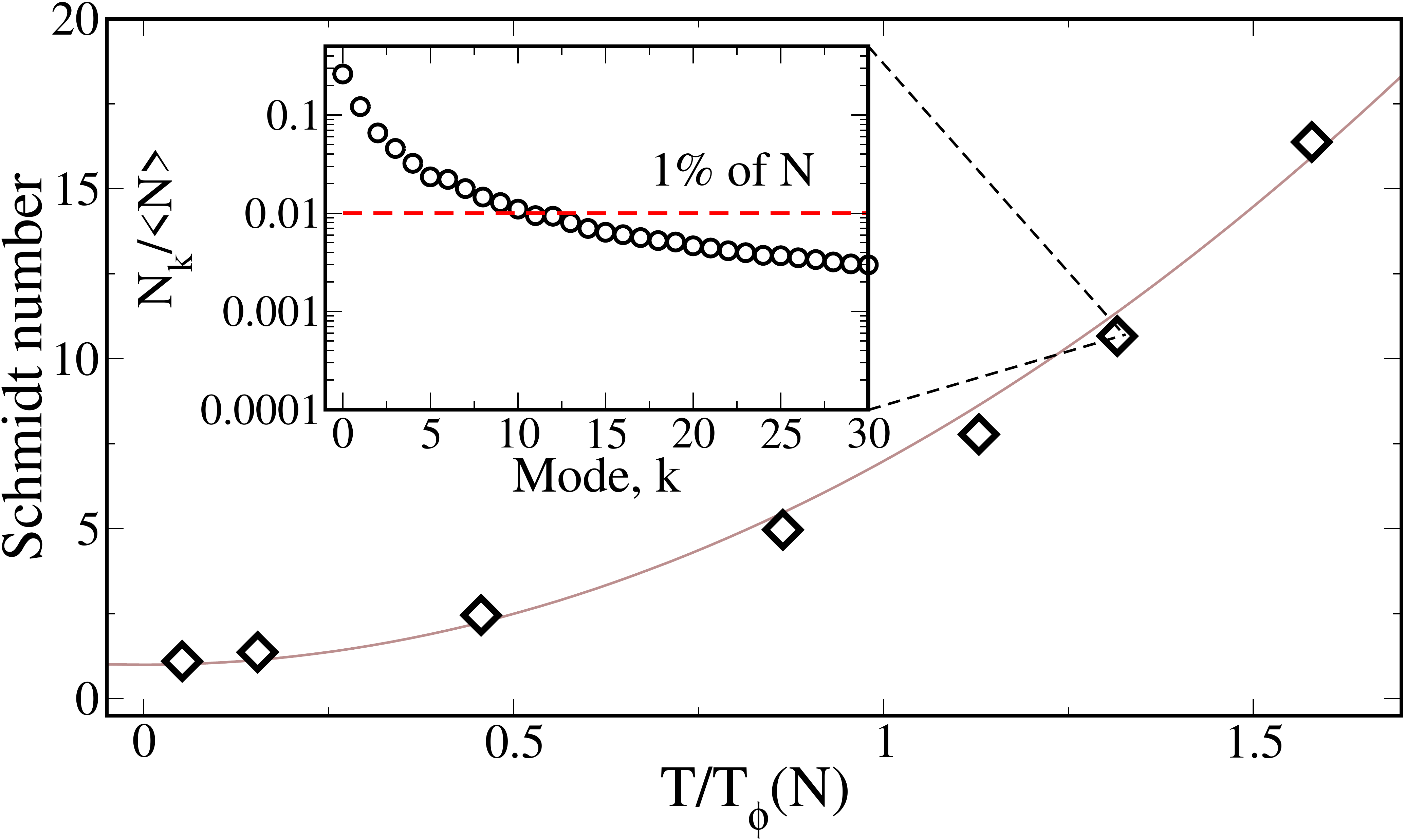}
  \caption{(Color online) Schmidt number versus scaled temperature
           for the sGPe simulations. Inset: at the
	   temperature $T \approx 1.3\,T_\phi$,
	   the fractional occupation $N_{k} / \langle N \rangle$ for the $30$ 
           lowest modes with the largest occupation; 
   condensate fraction (in $k=0$ mode above) 
   $\langle N_c \rangle / \langle N \rangle \approx 0.2$,
   and $\langle N \rangle \approx 23\,800$.
           }
  \label{fig:Schmidt}
\end{figure}

The modes for $k \leq 30$ are shown for the system at $T = 1.3\,T_{\phi}$ 
in the inset of Fig.\ref{fig:Schmidt}.
The low-lying modes ($k < 10$) share a significant fraction of the
total occupation; this is a consequence
of the short-range phase coherence or quasi-condensation in the system,
and similar to a ``fragmented'' condensate where many modes share a
macroscopic occupation \cite{Leggett2001,Cederbaum2004}. A quantitative measure of
how many modes contribute with a significant occupation can be given
in terms of the Schmidt number $S$ defined as
\cite{Grobe1994}
\begin{equation}
S^{-1} = \sum_{k=0}^{{\cal M}} f_{k}^{2}
\end{equation}
where $f_{k}$ is the fractional mode occupation given by
\begin{equation}
f_{k} = N_{k} / \sum_{k=0}^{{\cal M}} N_{k}.
\end{equation}
The Schmidt number versus temperature extracted from sGPe simulations 
is shown in the main plot of
Figure \ref{fig:Schmidt}. In the limit of zero temperature, 
it tends towards $1$: this is the signature of
a pure Bose-Einstein condensate. 
For temperatures approaching $T_\phi$, it increases quickly
(a rather good fit is 
$S \approx 1 + 6(T /  T_{\phi})^2$, the black dotted line). 
At temperatures where the phase coherence length is shorter
than the condensate size, one may expect a scaling $S \propto
R(T) / L_\phi(T)$ which would be slower than linear. Modes
outside the condensate region therefore contribute as well.
 


\section{Condensation vs. quasi-condensation}
\label{sec:quasi-cond}

Due to the 1D nature of the system we consider, 
fluctuations in the density and phase are suppressed
at different characteristic temperatures \cite{PopovBook}.
In this section, therefore, we highlight the distinction between the
phase and density coherent portions of the gas, for which we will use the
terminology ``condensate'' and ``quasi-condensate'', respectively.
This is an important problem in its own right, since stochastic theories 
(such as the sGPe \cite{Stoof1999,Gardiner2003}, or the pGPe \cite{Davis2001})
automatically generate {\it total} densities 
of the field. The 
condensate mode is extracted from these either using bimodal fits 
(not commonly done, but experimentally well-known), or via the 
Penrose-Onsager (PO) prescription. This leads to a `gap' between
stochastic approaches and theories where a
symmetry-breaking argument (or a variant of it) assigns a special
``condensate mode'' from the outset \cite{ZNGbook}. 

To investigate this issue further, we wish 
to establish here a more direct link between the PO
condensate and the quasi-condensate often calculated in 
low-dimensional systems; we do this by directly comparing 
the PO mode of the sGPe
data to the {\it ab initio} prediction of
the modified Popov theory of 
Refs.\cite{Andersen2002, AlKhawaja2002, AlKhawaja2002erratum}
outlined in Sec.\ref{subsec:mod_Popov}. As we shall show, this
link has the advantage of providing an approximate PO condensate
density without performing additional manipulations of the
stochastic data like the diagonalization of the one-body
density matrix.

\subsection{Identifying the quasi-condensate}
\label{subsec:quasi-vs-PO-condensate}

\begin{figure}[bth] 
  \centering
  \includegraphics[scale=0.310,clip]{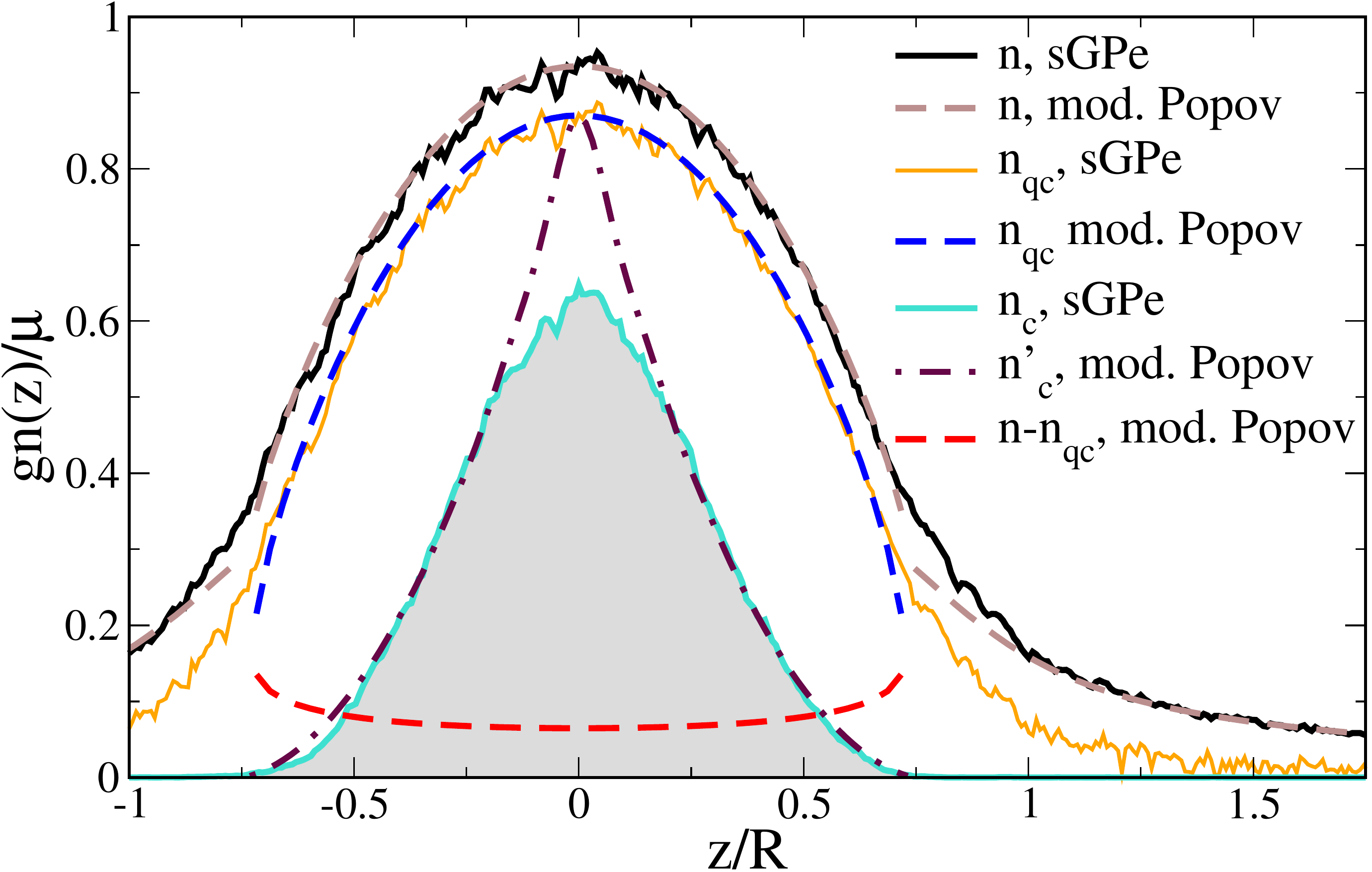}
  \caption{(Color online) Normalized density profiles showing: 
           Total density (sGPe: Solid black, noisy; 
           modified Popov: dashed, brown),   
           quasi-condensate density (sGPe: solid orange, noisy; modified 
           Popov: dashed, blue) and 
           (phase coherent) condensate densities (sGPe PO: 
           noisy, turquoise; modified Popov: 
           dot-dashed/dotted, maroon). The modified Popov condensate density is 
           shown dotted at small distances from the trap centre, where the 
           relation $n_c'( z ) = n_c( z )$
           [see Eq.(\ref{eq:qc_to_PO_trap})] breaks down.
           The grey shaded region shows the Penrose-Onsager condensate density. 
           The dashed red line 
           shows $n(z)-n_{\rm qc}(z)$ of the modified Popov theory. Here
           $T = 1.3 \,T_{\phi}=0.63 \,\Tc$ and $\langle N \rangle
	   \approx 23\,800$. 
	   There are no ncB data because the ncB expansion
           no longer works at this temperature.
           }
  \label{fig:qc_vs_PO}
\end{figure}

In Fig.\ref{fig:qc_vs_PO}, we compare the quasi-condensate
calculated from the modified Popov theory
(within the local density approximation, see 
Sec.\ref{subsec:mod_Popov}) to the definition~(\ref{eq:nqc_svist})
below, 
based on the density correlation function (Sec.\ref{s:g2}).
In order to reveal the different physics contained in each of
the approaches, in
this figure, we have chosen a relatively
high temperature ($T\approx 1.3 \, T_{\phi} \approx 0.63\, \Tc$)
than for the previous comparison (chemical potential and
interaction constant are kept the same). 
The breakdown of the ncB initial state in that regime constrains 
us to use only the sGPe stochastic data for this comparison.
Due to the mapping between moments of the Bose field operator 
to the stochastic field (Sec.~\ref{sec:methods}), one may
extract a quasi-condensate density within these approaches
in the following way
\begin{equation}
n_{\rm qc}^2(z) = 2 \langle
	\hat\psi^\dag( z )\hat\psi( z )
  \rangle^2
-
\langle
	\hat \psi^\dag( z )\hat\psi( z )
	\hat\psi^\dag( z )\hat\psi( z )
\rangle
.
\label{eq:nqc_svist}
\end{equation}
This definition has been put forward 
in Ref.\cite{Svistunov2001}, and implemented in Refs.%
\cite{Proukakis2006b, Bisset2009a}. Its equivalent form
\begin{equation}
	n_{\rm qc}( z ) = n( z ) \sqrt{ 2 - \gTwo( z ) }
	,
	\label{eq:def-quasi-condensate-density-a-la-Nick}
\end{equation}
has also been used at lower temperatures with
the aim of extracting the (conventional) condensate mode in a 3D 
system~\cite{Dodd1997}, where phase fluctuations were not expected
to contribute significantly. 
\begin{figure*}[htb!]
  \centerline{
  \includegraphics[scale=0.39,clip]{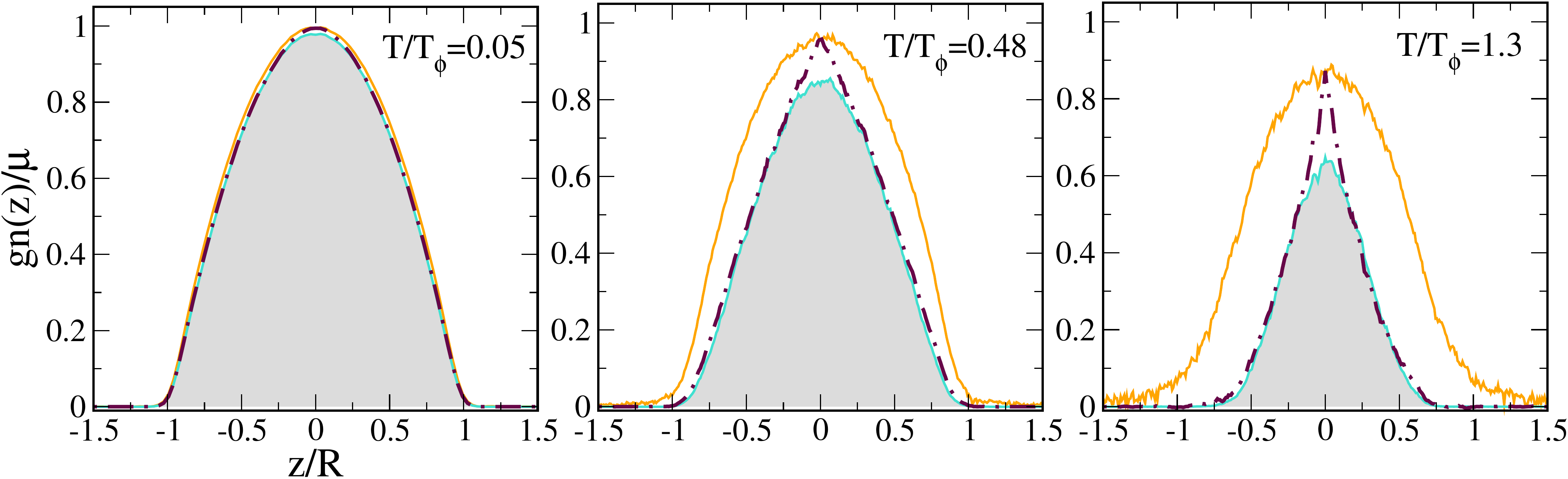}
  } 
  \caption{(Color online) Comparison between Eq. \eqref{eq:qc_to_PO_trap} 
           (dot-dashed, maroon) and the PO mode due to diagonalizing
           the density matrix (solid turquoise and shaded) at three  
           temperatures. Also shown is the quasi-condensate density
           from 
           Eq.\eqref{eq:def-quasi-condensate-density-a-la-Nick}
           (noisy orange curve), used to generate the dot-dashed maroon densities. 
           The data here is extracted 
           from the sGPe simulations.
           }
  \label{fig:sGPe_approx_vs_dm}
\end{figure*}
The comparison of both the quasi-condensate (solid orange and dashed 
blue in Fig.\ref{fig:qc_vs_PO}) 
and the total densities (solid black and dashed brown) 
between sGPe and modified Popov (respectively)
gives very good agreement \cite{Proukakis2006a}.
The quasi-condensate density profile is at this temperature
clearly distinguishable from both the total density and the 
PO condensate (greyed area). 
The physical meaning of the quasi-condensate in modified Popov theory is 
thus that part of the 
system where density fluctuations are reduced such that
$\gTwo( z ) \approx 1$, as is typical for a single-mode coherent
state. The plateau with $\gTwo( z ) = 2$ outside the quasi-condensate
in Fig.\ref{fig:g2} is, on the other hand, characteristic for 
a ``chaotic'' (or multi-mode) field \cite{MeystreSargent,Kheruntsyan2005}.
Note that this definition of the quasi-condensate is immune to 
phase fluctuations by construction \cite{PopovBook}.
Figure~\ref{fig:qc_vs_PO}
shows that the density correlations obtained within the sGPe
[Eq.\eqref{eq:def-quasi-condensate-density-a-la-Nick}]
indeed capture a quasi-condensate density consistent with that 
of Popov theory.

\subsection{Identifying the Penrose-Onsager condensate density}


The modified mean field theory of Andersen {\it et al.}
 \cite{Andersen2002,AlKhawaja2002}
splits the system into a quasi-condensate and other modes
and thus avoids the ``problem'' of assuming the existence of
long-range phase coherence. This makes the theory 
valid in arbitrary dimensions and at all temperatures. The approach can also
capture the (conventional, PO) condensate mode by calculating
the long-range limit of the one-particle density matrix
\cite{AlKhawaja2002erratum} (see also Ref.\cite{StoofBook2009}).
In a homogeneous system, this leads to the definition
\begin{equation}
n_{c} = \displaystyle \lim_{x \to \infty} 
	n_{\rm qc} \,
	e^{- \frac12 
		\langle \left[\hat{\theta}(x) - \hat{\theta}(0)\right]^{2} \rangle }.
\label{qc_to_PO}
\end{equation}
This procedure recovers exactly the Popov results for quantum
depletion in two  and three dimensions, as pointed out
in Ref.\cite{AlKhawaja2002erratum}. 

We wish to adapt Eq.\eqref{qc_to_PO} to the trapped case
and construct the quantity
\begin{eqnarray}
n'_c( z )  & :=
& n_{\rm qc}(z) \,
e^{ - \frac12 
	\langle \left[\hat{\theta}({z})-\hat{\theta}({0})\right]^{2} \rangle 
  }
\nonumber
\\
	&= & n(z) \sqrt{ 2 - \gTwo( z ) } \, g^{(1)}(0,z)
.
\label{eq:qc_to_PO_trap}
\end{eqnarray}
We expect that this agrees well with the PO condensate density
$n_c( z )$ for large $|z|$.
As appears plausible on physical grounds, Eq.\eqref{eq:qc_to_PO_trap} 
involves a combination of density and phase correlation functions.
This leads us to the second interesting feature of Fig.\ref{fig:qc_vs_PO}:
the density $n'_c( z )$ defined by Eq.(\ref{eq:qc_to_PO_trap})
(dot-dashed and dotted, maroon) 
coincides, at large $|z|$, with the condensate density $n_c( z )$
obtained from the PO analysis of the one-body density matrix (solid light blue and shaded). 
This is important, as it illustrates that the PO procedure produces
a condensate that coincides with that fraction of atoms for which both 
phase and density
fluctuations are reduced, i.e. the `true' condensate.
%
Equation~(\ref{eq:qc_to_PO_trap}) is also appealing from a practical
point of view, as the extraction of the PO mode   
from stochastic data usually requires 
diagonalization of the one-body density matrix, 
which for 
very large systems can become a 
significant computational task.
In contrast, the correlation functions $g^{(1)}(0,z)$ and $g^{(2)}(z)$ 
that enter Eq.(\ref{eq:qc_to_PO_trap}) 
are very straightforward to calculate, making the analysis much quicker.

The condensate density predicted by Eq.(\ref{eq:qc_to_PO_trap}) 
agrees well with the one obtained by Penrose-Onsager analysis also
for a range of temperatures, as shown in Fig.\ref{fig:sGPe_approx_vs_dm}.
This figure also illustrates how the distinction between quasi-condensate
and PO condensate becomes more noticeable as $T \sim T_\phi$.
For all temperatures, Eq.\eqref{eq:qc_to_PO_trap} gives a
good approximation to the PO density, with an increasing difference
evident only in the central region of the trap. 
The size of this region at $T=1.3T_{\phi}$ 
corresponds roughly to the 
extent over which the non-condensate part of $g^{(1)}(z)$
is positively correlated (see Eq.(\ref{eq:break-g1-in-c-and-nc})
and the dot-dashed red curve in Fig. \ref{fig:g1}(d)). The origin
of the `spike', then, lies in the fact that the thermal component 
is coherent over this small region too,
so it is not just the condensate which contributes to $g^{(1)}(z)$
at short scales.
Again from  Fig. \ref{fig:g1}, it is interesting to note that the
non-condensate contribution to $g^{(1)}(z)$ is actually significant 
at larger $|z|$ where it reduces the condensate contribution 
(dashed blue curve).
%

\subsection{Application: (quasi) condensate fraction}
\begin{figure}[hb!]
  \centerline{
  \includegraphics[scale=0.31,clip]{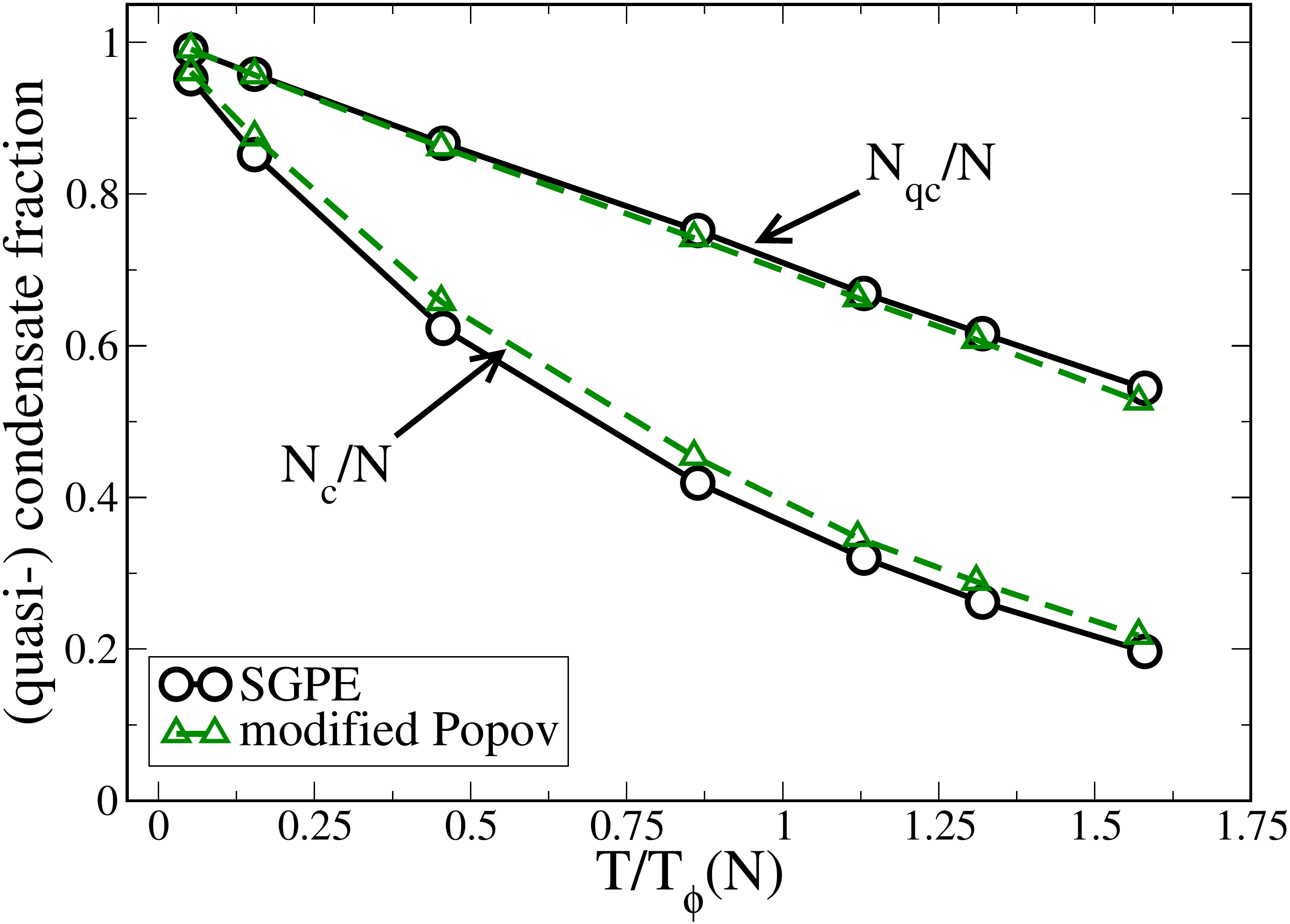}
  }
  \caption{(Color online) (1d) Quasi-condensate and PO condensate numbers from 
           the sGPe (black circles) 
           and modified Popov theory 
           (hollow green
           triangles).
The sGPe results are based on the condensate mode given by PO analysis
of the one-body density matrix (Sec.\ref{subsec:PO-densities}) and on
Eq.(\ref{eq:nqc_svist}) for the quasi-condensate. For the integration over
$n_{\rm qc}( z )$, only the region $| z | \le R(T)$ is taken into account. 
The modified Popov data are based on the quasi-condensate density
described in Sec.\ref{subsec:mod_Popov} and on the condensate 
density $n'_c( z )$ given in Eq.(\ref{eq:qc_to_PO_trap}), where 
$\gOne$(0,z) is calculated from Eq.\eqref{eq:log_g1}.
           }
  \label{fig:PO_qc_Popov_SGPE}
\end{figure}
%
\begin{figure*}[htb!]
  \centering
  \includegraphics[scale=0.45,clip]{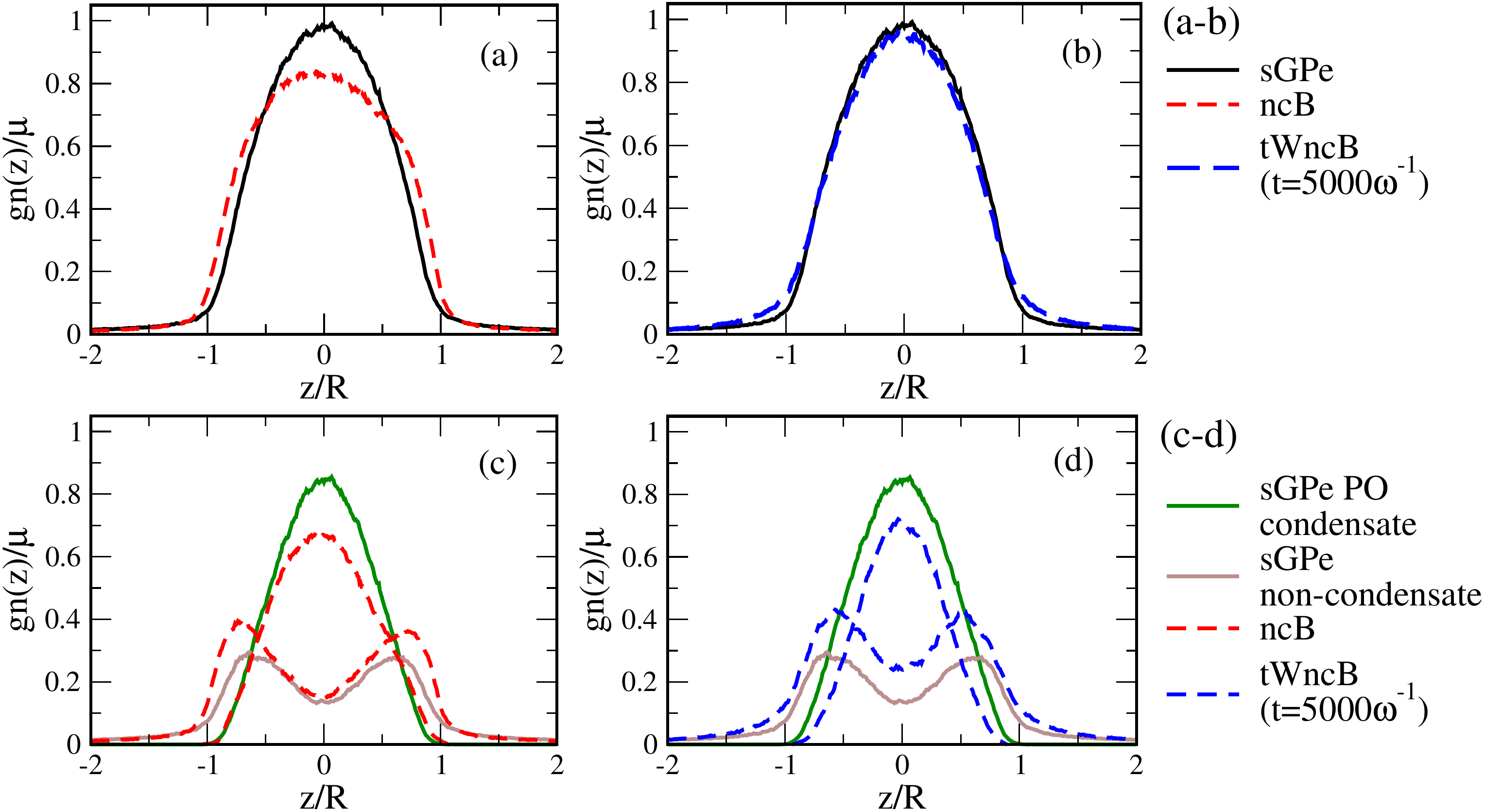}
\caption{(Color online) Top row (a, b): 
total atomic density for the sGPe (solid, black) at equilibrium
and tWncB (a) before (dashed red)
and (b) after (dashed blue) evolution 
via the GPe. 
The tWncB data contain
the correction to normal order; the evolution period is 
$5000\,\omegaZ^{-1}$.
Bottom row (c, d): condensate/thermal density 
(sGPe:solid green/brown; tWncB: dashed), as obtained by
Penrose-Onsager analysis [see Fig.\ref{fig:PO_densities}], 
(c) before (dashed red) and (d) after (dashed blue) 
evolution of the ncB state over $5000\omegaZ^{-1}$. 
Upper curves: condensate density, lower curves with maxima
at $|z| \lesssim R$: thermal density.
	    }
\label{fig:erg_densities}
\end{figure*}

To test the reliability of the condensate density suggested in
Eq.(\ref{eq:qc_to_PO_trap}), we have calculated the fraction of
atoms in the (quasi-) condensate over a range of 
temperatures.
The data are shown in Fig.\ref{fig:PO_qc_Popov_SGPE}. The two
upper curves give the quasi-condensate fraction which is systematically
larger. At low temperatures, $T \ll T_\phi$, 
condensate and quasi-condensate numbers are very similar, 
as already noted in \cite{Prokofev2002, Blakie2005, Proukakis2006b}.
In this range, we also get a good agreement between stochastic data
and the modified Popov theory.
As $T$ increases towards $T_{\phi}$, however, the deviation between
condensate and quasi-condensate increases. This is clearly due to the shrinking 
of the PO condensate at its borders,
as seen in Fig.\ref{fig:sGPe_approx_vs_dm}. 
The difference among the two (PO) condensate numbers is due 
to the central ``spike'' that follows from Eq.(\ref{eq:qc_to_PO_trap}):
at $z = 0$, one has necessarily $g^{(1)}(0) = 1$ and $n'_c( 0 ) =
n_{\rm qc}( 0 )$,  but diagonalizing the one-body density
matrix of the sGPe data yields
$n_{c}(0) < n_{\rm qc}(0)$ [Fig.\ref{fig:sGPe_approx_vs_dm}]. Despite
this difference, the two condensate numbers agree relatively well and
show the same qualitative temperature dependence.

The approximate condensate density $n'_c( z )$ of  Eq.\eqref{eq:qc_to_PO_trap}
can be evaluated in a simpler way by using 
Eq.\eqref{eq:log_g1} for the calculation of the phase correlation
function $g^{(1)}(0,z)$. This means in practice that one only needs to
extract the parameter $R(T)$ from the modified Popov density profiles.
We have already shown in Fig.\ref{fig:PO_dens_g1cond}(d) that this
simplified expression works well to approximate $g^{(1)}(0,z)$.

\section{Slow thermalization of the initial state}
\label{sec:tWncB-thermalization}

\begin{figure*}[htb!]
  \centering
$\begin{array}{l r}
  \includegraphics[scale=0.4,clip]{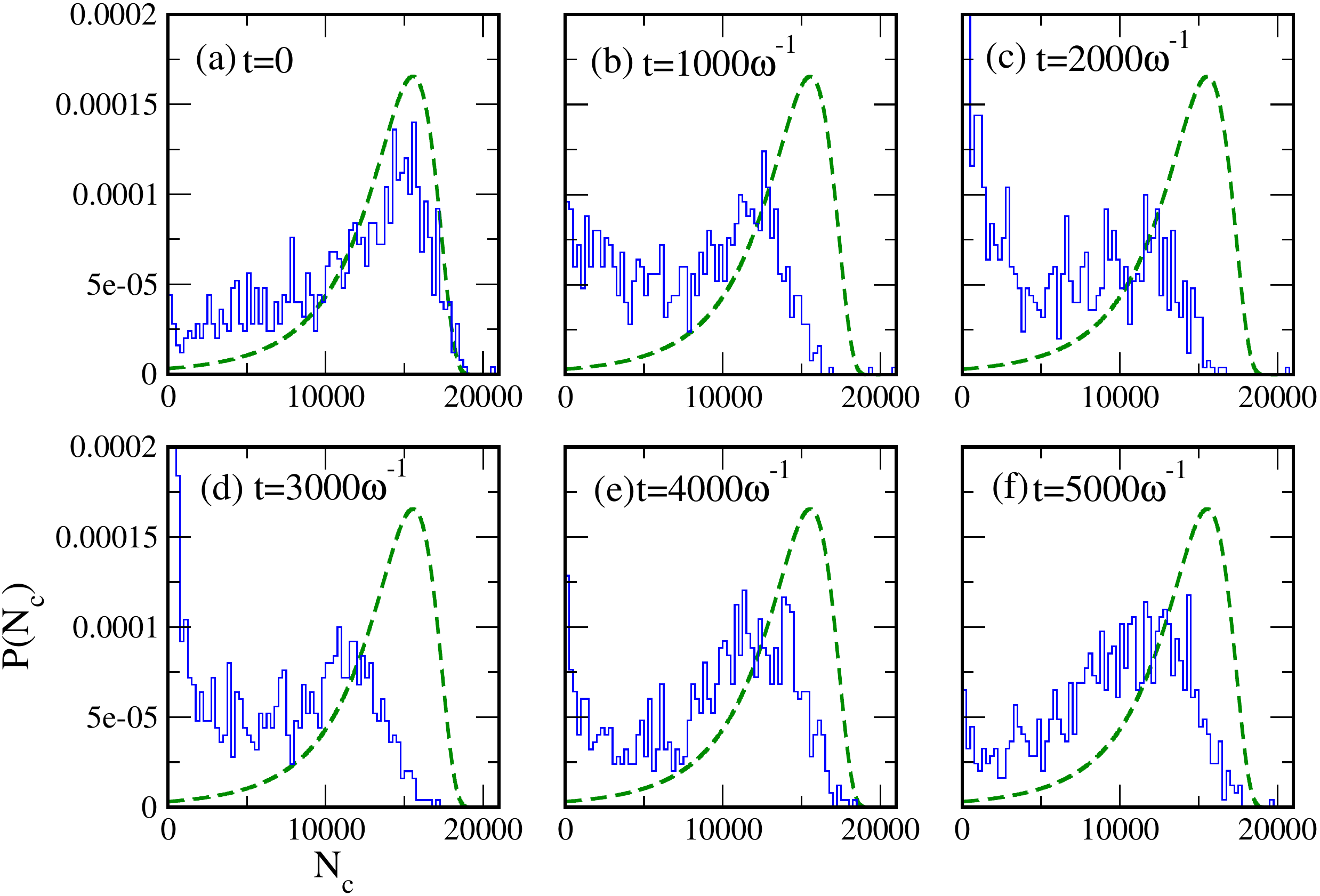} &
  \includegraphics[scale=0.30,clip]{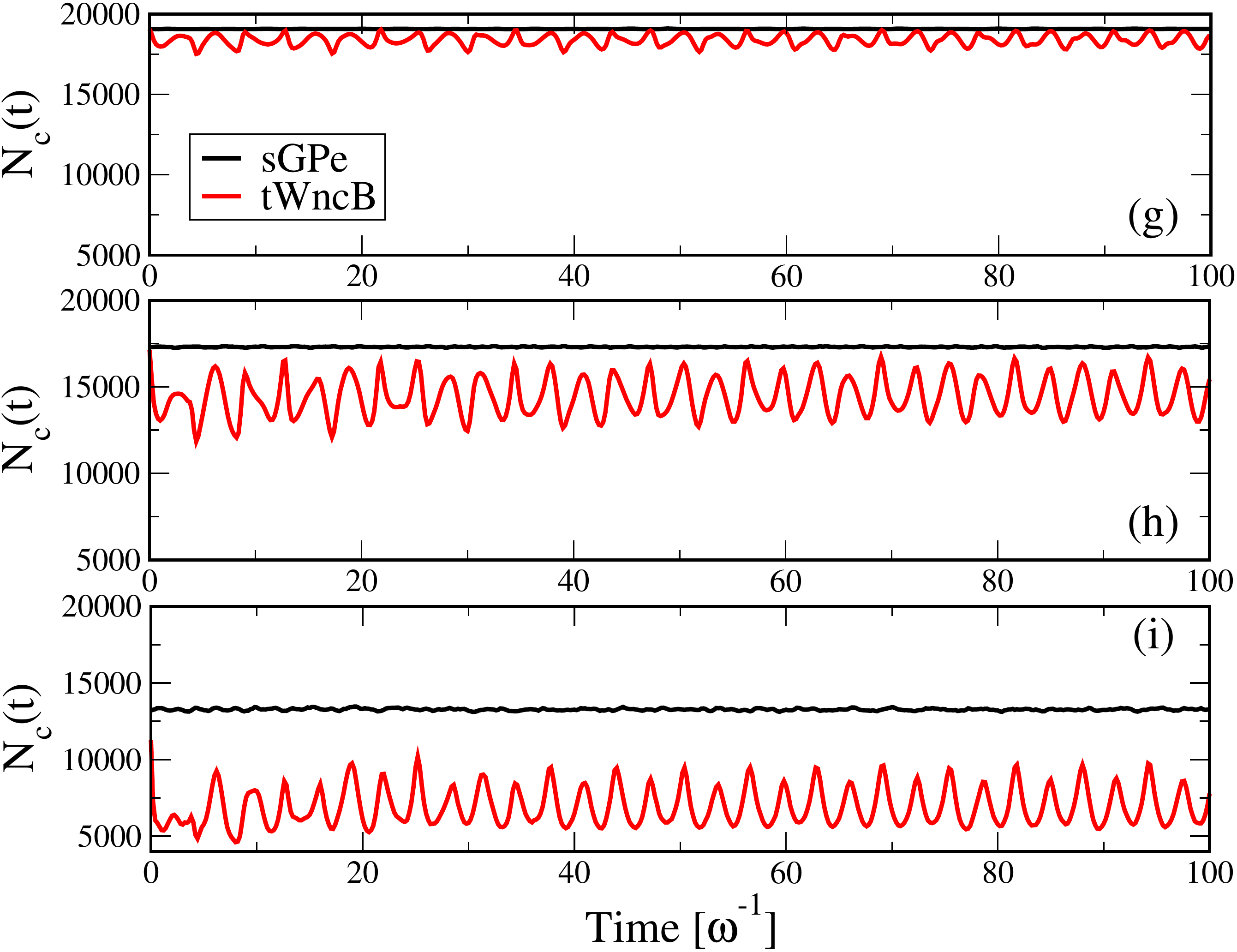} \\
\end{array}$
\caption{(Color online) Left: condensate statistics at different evolution
times under the GPe, at initial nominal temperature 
$T \approx 0.48 \,T_\phi$. Solid blue: tWncB data, dashed green:
equilibrium S\&co theory \cite{Svidzinsky2006}. The sGPe data at equilibrium 
are not shown,
as the condensate statistics essentially stays the same as in 
Fig.\ref{fig:counting-statistics}.
Right: mean condensate number $\langle N_{c}( t ) \rangle$for sGPe (upper, black line),  
             tWncB (lower, red curve) at the temperatures of 
             Table~\ref{tab:temp_context}, with temperature increasing from top (g) 
             to bottom (i).
	    }
\label{fig:erg-condstats}
\end{figure*}

We have seen the importance of a consistent treatment of phase and
density fluctuations in calculating accurately a finite-temperature
initial state for our (quasi)-one-dimensional system.
From the physical observables probed, 
the picture emerging so far is that equilibrium properties, like total 
densities, agree quite well over a range of temperatures well beneath 
$T_{\phi}$ and in the spatial range where highly occupied modes are
dominant.
The stochastic ensembles prepared by the two methods are
considered in this section as an initial condition for 
the Gross-Pitaevskii equation, which
will be used to describe the subsequent dynamics.
We take a temperature
$\approx 0.48\, T_\phi$ where the ncB data is clearly not in equilibrium, 
and address the question whether these data evolve dynamically into a thermal equilibrium state.
We shall find that even after a fairly long evolution time
($ > 5000 \,\omegaZ^{-1}$), the system is not yet stationary.
This may be related to the absence of thermalization in integrable 
homogeneous 
1D systems \cite{Kinoshita2006,Rigol2009,
Mazets2010,Cazalilla2010}.


The thermalization study presented here 
provides a link to other classical field methods.
These approaches, 
for example the pGPe of Davis, Blakie and co-workers \cite{Davis2001}, 
that employed by Berloff and Svistunov \cite{Berloff2002}
and the approach of the Polish group \cite{Brewczyk2007}, 
typically use 
a single field realization with suitably randomized initial conditions 
as an input to evolution under the GPe (with the possible
addition of a projector \cite{Blakie2008}),
rather than an ensemble of initial states.
The system then corresponds to a microcanonical ensemble, as both
particle number and total energy are fixed in this scenario. The
dynamics acquires an irreversible character by spatial or temporal
coarse-graining \cite{Rzazewski2002}. This can be 
mapped to a Boltzmann equation that yields
an irreversible evolution where 
the system thermalizes to an equilibrium state with Rayleigh-Jeans 
statistics \cite{Davis2001,Connaughton2005}.


Figure~\ref{fig:erg_densities} shows the density profile and its
resolution into condensate and thermal density, before (left panels) and 
after (right panels) GPe evolution. While the total 
densities (top row) agree well between the two methods, the ncB data evolve
towards a smaller condensate density (bottom row), suggesting the system to be
thermalized at a higher temperature.
 

The temporal evolution of the condensate statistics is illustrated
in Fig.~\ref{fig:erg-condstats} (left). This data was extracted as follows:
we obtain the one-body density matrix at the indicated times
and get the condensate mode by Penrose-Onsager diagonalization.
Projecting the ensemble of wave functions onto this mode, one
gets a few snapshots of the evolving condensate statistics
$P( N_c, t )$. While this distribution shows essentially no variation
for the sGPe initial data,
the ncB case shows a significant evolution. The peak at 
$\langle N_c \rangle$ disintegrates quite rapidly, and the condensate
is re-formed gradually, with a broader peak re-appearing from smaller to
higher numbers.
At the final time, the distribution does yet not appear to have reached
a stable distribution.

The right panel in Fig.~\ref{fig:erg-condstats} illustrates that over short
time scales, the average condensate number 
oscillates for the ncB data, in contrast to the sGPe
(note the shorter time scale compared to the left panels, 
in order to resolve the oscillations in the ncB data). 
This oscillation at roughly $2\,\omegaZ$, whose amplitude
increases with temperature, may be due to a nonlinear locking
between the condensate mode and its low-lying, highly excited excitations.
%



%

\begin{figure*}[htb!]
  \centering
  \includegraphics[scale=0.45,clip]{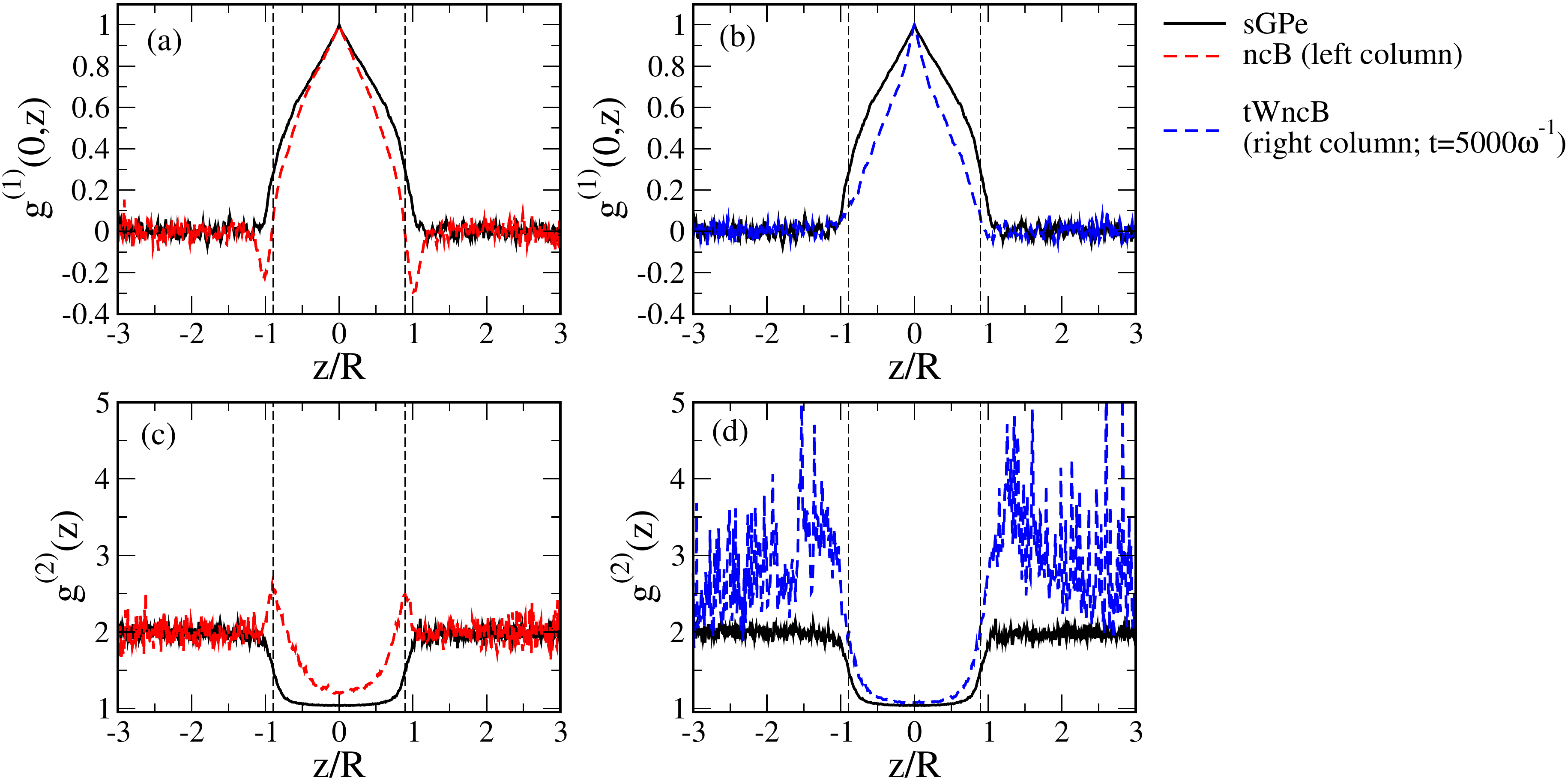}
  \caption{(Color online) 
           Top row: $g^{(1)}(z)$ for the sGPe (solid black), 
           (a) tWncB initially at $(t=0)$ (dashed red) and (b)
           tWncB after GPe evolution up to $t=5000\omegaZ^{-1}$ (dashed blue); 
           bottom row: as per top row for $g^{(2)}(z)$.
           In each plot, the vertical dashed line indicates
           $R(T)$ at $T=430\hbar\omega$.
	    }
\label{fig:erg_g1g2}
\end{figure*}

The evolution of the coherence functions is illustrated in 
Fig.\ref{fig:erg_g1g2} for 
the $\gOne$ and $\gTwo$ functions introduced in 
Secs.\ref{s:g1}, \ref{s:g2}. The kinks in Fig.~\ref{fig:erg_g1g2}(a)
could be due to a mode coupling between the condensate and
low-energy excitations whose mode functions have nodes and are slightly
broader. But it is unclear whether this picture may explain 
the strong density fluctuations (panel (d)). The average density is
quite low for $|z| > R$, which amplifies sampling errors, however.

\begin{figure}[b!]
  \centering
  \includegraphics[scale=0.335,clip]{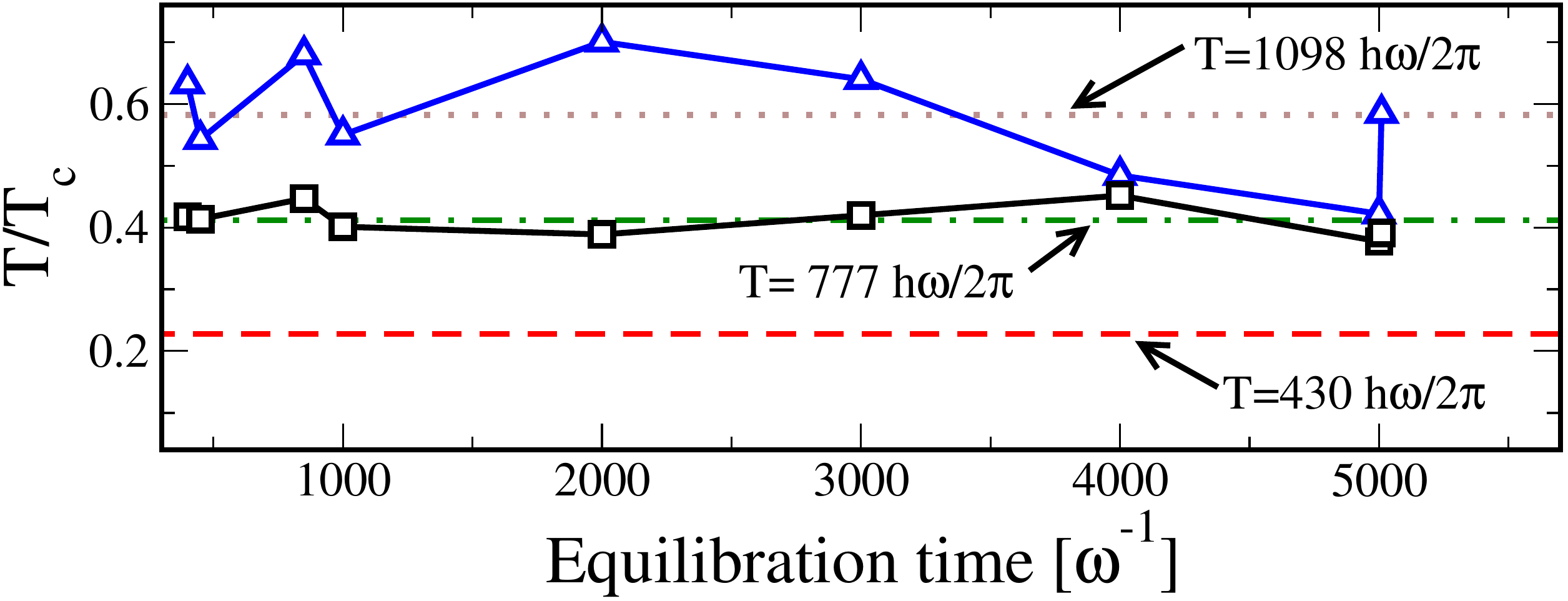}
  \caption{(Color online) Temperature measurement using correlation functions as a 
             function of GPe evolution time: (upper data) 
             $g^{(1)}(z)$, for $0<z<R/2$ and 
             Eq. \eqref{eq:approx-model-for-g1-correlation} (blue triangles);
             (lower data)
             $g^{(2)}(0)$ and Eq. \eqref{eq:g2-Kheruntsyan-our-units} (black squares).
             Initial temperature (dashed red);
             Best horizontal fit through $g^{(1)}(z)$ data (dotted brown) 
             and through $g^{(2)}(0)$ data (dot-dashed green); temperatures 
             corresponding to the horizontal lines are indicated.
	    }
\label{fig:erg_g1_gradient}
\end{figure}

%
%
Recalling the discussion in Secs.\ref{s:g1}, \ref{s:g2} 
[Eqs.\eqref{eq:approx-model-for-g1-correlation}
and~(\ref{eq:g2-Kheruntsyan-our-units})], we can use the correlation
functions to measure the temperature of the ensemble. This may
not yield a consistent picture, since the system is not (yet)
thermalized. At least, we can place bounds on the temperature range
that the system might thermalize to, albeit after some longer time. 
The results of this are summarized in 
Fig.~\ref{fig:erg_g1_gradient}. 
We probe the system properties at non-constant time intervals, 
in order to account for the possibility of periodic behaviour
in the correlation functions.
For example, the last two data points on the upper curve 
are separated by only $10\,\omegaZ^{-1}$ units of time.
The lower dashed red line indicates the input temperature ($0.48\,T_\phi$),
but both $\gOne$ and $\gTwo$ yield higher numbers.
Notice that $\gOne$ gives a higher temperature, as might be
expected in our regime due to pronounced phase fluctuations.
The temperature extracted from $\gTwo$ is 
more stable in time, which may suggest a faster damping rate
for modes with density fluctuations.
This would be consistent with Landau-Beliaev 
damping, see, e.g., Refs.\cite{Sinatra2002, Brewczyk2007}.

%


To summarize this discussion, we emphasize that in this example
thermalization proceeds quite slowly. The 1D character of the system
that is nearly integrable probably plays a role here, but also relevant
are the large fluctuations that are present in the initial state 
produced by the ncB method. One may think of the $1/2$ ``quantum atoms'' 
per mode that are included in the
truncated Wigner sampling: the observed temperature increase 
is roughly comparable to the ``classical thermalization'' of these atoms, 
as simple estimates show \cite{Sinatra2002}. 
The different temperatures extracted from phase and 
density correlations, 
however, are probably related to the wrong account of 
phase fluctuations: they are mis-interpreted in terms of non-condensate density, 
as illustrated in Fig.\ref{fig:point-cloud-2}. This may be cured by formulating
the stochastic
scheme in terms of phase and density variables, instead of the
non-condensate field $\psi_\perp( z )$, similar to the analysis
of Ref.\cite{Mora2003}.

%


\section{Conclusions}
\label{sec:conc}

We have analyzed the equilibrium properties
of a weakly interacting, trapped quasi-one-dimensional
Bose gas at finite temperatures,
which we modelled as an
effective one-dimensional system.
The predictions of a number of independent 
finite-temperature theories have been compared.
We focussed in particular 
on two methods incorporating phase and density 
fluctuations in a stochastic manner: a number conserving Bogoliubov 
approach with stochastic sampling and the stochastic Gross-Pitaevskii equation.

At low temperatures, we found average quantities,
such as total density profiles, condensate fractions
and first and second order spatial correlation functions
to give good agreement between the two theories. These properties were additionally 
found to coincide with other theoretical predictions from the 
literature, which we used in order to `benchmark' our findings.
As higher temperatures were probed, though still within the
regime $T<T_{\phi}$, the ncB initial state was 
found to give predictions for equilibrium properties
in disagreement with both the
sGPe and the results of other, 
`benchmark' theories, including the modified Popov theory of 
\cite{Andersen2002},
while
the latter two showed good agreement.
We attribute this failure to the use of the $T=0$ Bogoliubov
spectrum in the ncB expansion, i.e. the condensate number
is assumed equal to the total particle number at all temperatures,
and to the overestimation of the non-condensate density due to
spurious contributions of phase fluctuations at higher orders.
The `point cloud' in Fig. \ref{fig:point-cloud-2} 
illustrates the enhanced phase fluctuations 
at higher temperatures, which mean that density
fluctuations in modes above the condensate were no longer suppressed
in the ncB method,
as would be expected at temperatures much below the `degeneracy' temperature  ($T_{d}$).
A procedure taking condensate depletion into account in a temperature
dependent way, would likely improve the high-temperature behaviour of this
method. 
%
%

We have also probed quantities involving higher-order
statistical moments than just the density or its correlation,
in particular the full distribution function of the condensate number.
At low temperatures, and for not too small atom numbers (where the assumption of a `classical' occupation of modes fails), both ncB and sGPe were found to produce the correct statistics, in perfect agreement to
the theory of Svidzinsky and Scully 
\cite{Svidzinsky2006}.
However, for small total particle numbers, and at low temperature, the sGPe results were 
found to be broader than the ncB statistics, whereas the latter were
found to agree well with 
those of Ref. \cite{Svidzinsky2006}.
The incorrect sGPe prediction
at low particle numbers was attributed
to the onset of anomalously large number fluctuations, 
familiar from the grand-canonical analysis of the ideal gas.
As the importance of
interactions within the system was increased, 
i.e.\ by increasing particle number with all other parameters fixed, 
the sGPe and Svidzinsky and Scully results were then found to match well.

We further propagated the ncB data via the (ordinary) GPe, as done in 
the truncated Wigner approach. 
We found that this
leads to the correct profiles for the total system density,
but fails to predict all other features accurately, due to its
attempt to thermalize to a higher temperature classical field.
This thermalisation was found to take extremely long here,
due to our 1d system configuration.


%
Finally, we have illustrated the conceptual difference between the
(phase-coherent) condensate and the (density-coherent)
quasi-condensate. The former is usually obtained by the Penrose-Onsager
analysis of the one-body density matrix 
while the latter appears e.g. in the context of the
modified Popov theory of Refs.\cite{Andersen2002,AlKhawaja2002}. 
Building on the identification of
the (PO) condensate for a homogeneous 
system \cite{AlKhawaja2002erratum}, we have provided an alternative,
numerically very efficient means of extracting information about the
condensate density that involves only first- and second-order correlation
functions, as obtained from the sGPe simulations. 
Although some issues remain to be improved near the trap
centre, 
condensate density and fraction are perfectly matched to the conventional
PO approach over a broad range of parameters. 
We believe that this identification, along with the systematic benchmarking
of observables to alternative theories for finite-$T$ Bose gases
will provide a better understanding
of the links between stochastic theories
and thermodynamics based on mean field theories.

\bigskip\

\paragraph*{Acknowledgments.}

We acknowledge financial support from the EPSRC (SPC, NPP), 
the Lundbeck Foundation (AN), the Deutsche Forschungsgemeinschaft within the grants SFB/TRR21 (AN)
and He-2849/3 (CH),
 the Marie Curie Intra-European Fellowship within the 7th European Community Framework Programme (AN), and the 
Forschungsbonus der Universit\"at Ulm und der Ulmer 
Universit\"atsgesellschaft (AN).
AN acknowledges fruitful discussions with
A. Sinatra, Y. Castin, U. V. Poulsen, and I. Carusotto
on the truncated Wigner method. NPP acknowledges stimulating
discussions with Matt Davis, Allan Griffin, Tod Wright 
and Eugene Zaremba
on the links between `classical field' and kinetic equations
based on symmetry-breaking. CH thanks M. Abel for inspiring comments.


\appendix

\section{Sampling the number-conserving Bogoliubov state}
\label{a:tWncB}

The condensate mode function $\phi_c(z)$
contains two terms 
\begin{equation}
	\phi_c(z) = \frac{ \NCphiZero(z) + \NCphiSecond( z ) / N }{
	(1 + \Vert \NCphiSecond( z ) / N\Vert^2 )^{1/2} } 
	\,, 
\label{eq:psi-c-in-ncB}
\end{equation}
that arise in zero'th and second order of the expansion.
The lowest-order contribution $\NCphiZero( z )$ 
solves the stationary GPe
\begin{eqnarray}
H_{\rm GP}[N|\NCphiZero|^2]\NCphiZero = \mu \NCphiZero
.
	\label{Eqn:Hgp}
\end{eqnarray}
Note that the `chemical potential' $\mu$ emerges here as the lowest 
eigenvalue of a nonlinear eigenproblem:
it depends on the product $g N$ and the trapping potential $V( z )$.
Equation~(\ref{Eqn:Hgp}) is conveniently solved by propagating the wave function
in imaginary time. 

For the correction $\NCphiSecond( z )$, one needs the non-condensate
field $\psi_{\perp}( z )$, and we return to it in Eq.(\ref{eq:phi2-correction}). 
The non-condensate atoms populate
Bogoliubov mode functions $u_k( z )$ and $v_k( z )$. 
These are eigenfunctions of the (projected) Bogoliubov--de Gennes 
operator
\begin{eqnarray}
	\label{eq:def-LQ}
\mathcal{L}_Q &=
\left( 
	\begin{array}{cc} \mathcal{Q} & 0 \\ 0 & \mathcal{Q}^* \end{array} 
\right)
\mathcal{L} 
\left( 
	\begin{array}{cc} \mathcal{Q} & 0 \\ 0 & \mathcal{Q}^* \end{array} 
\right)
	\\
	\label{Eqn:Lbg}
\mathcal{L} &= \left(
\begin{array}{cc}
H_{\rm GP}[2 N|\NCphiZero|^2 ] - \mu & \hspace*{-0.5em}
g N \NCphiZero^2 \\
-g N \NCphiZero^{*2} & 
\hspace*{-0.5em}
- H_{\rm GP}[2 N|\NCphiZero|^2] + \mu
\end{array}
\right),
\end{eqnarray}
where $\mathcal{Q}$ ($\mathcal{Q}^*$)
is the projector orthogonal to $\NCphiZero( z )$ (to $\NCphiZero^*( z )$),
respectively. 
On the spatial grid, its matrix elements are
\begin{equation}
	\mathcal{Q}_{zz'} = \delta_{zz'} 
	- \Delta z\,\NCphiZero(z)\,\NCphiZero^*(z')
	.
	\label{eq:matrix-elements-Q}
\end{equation}
The numerical diagonalization of $\mathcal{L}_Q$ may be 
approached simply with a Fourier grid method \cite{Marston1989}, 
for example.
We need in Eq.(\ref{Eqn:psiper}) only those modes with
$\epsilon_{k} > 0$ that can be normalized to
\begin{equation}
	\Delta z \sum_z\left[ u_k^*( z ) u_l( z ) 
	- v_k^*( z ) v_l( z ) \right] = \delta_{kl}
	\label{eq:uv-orthogonal}
\end{equation}
where the sum represents the Bogoliubov scalar product on the spatial 
grid (spacing $\Delta z$).
For a grid
of length $L$, the number of Bogoliubov modes is 
${\cal M} = L / \Delta z - 1$.

The quantity $\mathcal{A}(\{ b_k \})$ in Eq.(\ref{Eqn:N0}) 
is calculated as
\begin{eqnarray}
&&\mathcal{A}(\{ b_k \}) =  \sum_{k} 
\frac{ |b_k|^2 - \sigma_k^2 }{ 4 \sigma_k^4 }
+
\sum_{k,q}
\frac{ \Delta z }
{2\,\sigma_k^{2} \sigma_q^{2} }\times
\nonumber
\\
&& \quad
\sum_z \big\{
{\rm Re}(b_k\,b_q^* - \delta_{kq} \sigma_k^2 ) 
v_k^*( z ) v_q( z )
\nonumber
\\
&& \quad - 
{\rm Re}(b_k\,b_q)
u_k^*( z ) v_q( z )
\big\}
,
\label{Eqn:A}
\end{eqnarray}
where $\sigma_k^2$ is the expectation value of $|b_k|^2$, so
that $\mathcal{A}(\{ b_k \})$ averages to zero. This term 
encodes the Wigner correction for getting the (normally ordered)
non-condensate particle number out of the semiclassical Wigner
functions, at the level of the corresponding number variances. 

The second-order correction $\NCphiSecond(z)$ to the 
condensate mode $\NCphiZero(z)$ is due to thermal depletion. 
It contributes in the ncB expansion
at the same order as the non-condensate modes
to typical observables like the average density and the one-body density 
matrix. It
must be orthogonal to $\NCphiZero(z)$ (see \cite{Castin1998}) and can 
hence be expanded over the Bogoliubov modes
\begin{equation}
	\NCphiSecond(z) = \sum_{k} 
        \left[c_k  u_k(z) + c_k^* v_k^*(z)\right].
	\label{eq:phi2-correction}
\end{equation}
The coefficients $c_k$ are found by solving an inhomogeneous
linear equation for the Bogoliubov-de Gennes operator. For the convenience
of the reader, we reproduce here
Eqs.(71)-(75) provided in Ref.\cite{Sinatra2000}.
The Wigner field $\psi_{\perp}$ represents 
a non-condensate field operator denoted
$\hat\Lambda$ in Ref.\cite{Sinatra2000}.
The second-order correction $\NCphiSecond$ solves
the stationary equation
\begin{equation}
\mathcal{Q} (H_{\rm GP}[2 N|\NCphiZero|^2 ] - \mu) \NCphiSecond
+ g N \mathcal{Q} \NCphiZero^2 \NCphiSecond^* = - \mathcal{Q} R
	\label{eq:phi-2-equation}
\end{equation}
with a ``source term''
\begin{eqnarray}
	R(x) &=& - g N |\NCphiZero(x)|^2 \NCphiZero(x) (1 
	+ \langle \hat N_{\rm th} \rangle ) 
	\label{eq:source-term-phi-2}
\\
	&&
	+ 2 g N \langle \hat\Lambda^\dag( x ) \hat\Lambda( x ) \rangle \NCphiZero(x)
	+ g N \NCphiZero^*(x) \langle \hat\Lambda( x ) \hat\Lambda( x ) \rangle
	\nonumber\\
	&& - g N \int\!{\rm d}y\, |\NCphiZero(y)|^2 \langle
	( \hat\Lambda^\dag( y )\NCphiZero(y) + \NCphiZero^*(y) \hat\Lambda( y ) )
	\hat\Lambda( x ) \rangle
	\nonumber
	\end{eqnarray}
where $\hat N_{\rm th} = \int\!{\rm d}x \, 
\hat\Lambda^\dag( x ) \hat\Lambda( x )$ is the operator for the 
non-condensate atom number. 
The expectation values of $\hat\Lambda$
are translated into Wigner averages of $\psi_{\perp}$
according to the symmetrization rule
\begin{equation}
	\langle \hat\Lambda^\dag( x ) \hat\Lambda( x' ) \rangle = 
	\langle \psi_{\perp}^*( x ) \psi_{\perp}( x' ) \rangle_W - 
	\frac{ 1 }{ 2 } \mathcal{Q}( x, x' )
	\label{eq:operator-ordering}
\end{equation}
where the projector
$\mathcal{Q}(x,x')$
appears because the fields live in the subspace orthogonal 
to the zero'th order condensate mode $\NCphiZero( z )$.
Equation~(\ref{eq:phi-2-equation}) is solved by constructing an imaginary
time evolution that leads to $\NCphiSecond( z )$ as a stationary solution.
%
%
This is then plugged into Eq.(\ref{eq:psi-c-in-ncB}) to complete the
construction of the normalize condensate mode $\phi_c( z )$.

With this normalisation, the stochastic matter wave field $\psi( z )$ 
gives access to the total particle density $n( z )$ as 
[cf. Eq.(\ref{eq:mapping-correlation-function})]
\begin{equation}
	n( z ) = \langle |\psi(z)|^2 \rangle_W - n_q
	\label{eq:get-particle-density-in-Wigner}
\end{equation}
where $n_q$ is the quantum density~(\ref{eq:def-quantum-density}).
It is hence normalized such that
$\langle \Vert \psi \Vert^2 \rangle_W = N + {\cal M}/2$. 
The statistics of the condensate atom number $N_{c}$ is determined
by the non-condensate field $\psi_\perp$, via Eq.(\ref{Eqn:N0}).

The propagation in time of the matter field $\psi(z, t )$ is found by solving
the GPe~(\ref{eqn:GPE_TW}). This turns out to be more accurate,
in particular at long times, than to propagate separately the condensate
mode and $\psi_{\perp}$,
using the time-dependent versions of the
equations they solve in the respective orders of the expansion
in $(\delta N / N)^{1/2}$ (see, e.g., \cite{Sinatra2001}).
All numerical simulations in this paper (for both methods)
are based on the Crank-Nicholson method for time stepping with
the results averaged over at least $1000$ 
realisations of the initial conditions.

\section{Canonical counting statistics of a Bose gas}
\label{a:Scully-theory}

The group of Scully and co-workers (S\&{}Co)
has developed theoretical models to calculate
the counting statistics $P( N_c )$ of a Bose condensate, building
on the canonical ensemble where the operators $\hat N_c$ and
$\hat N_{\rm th}$ must sum up to the (fixed) total number $N$.
Therefore, the counting statistics is the ``mirror image'' of the
probability distribution $P( N_{\rm th} )$. The latter can be calculated
when the non-condensate number $\hat N_{\rm th}$ splits into a sum 
of statistically
independent terms. In an ideal Bose gas, this would be the occupation
numbers $\hat n_k$ of single-particle modes (quantum number $k$).
For a weakly interacting, homogeneous Bose gas, the non-condensate atoms
are clearly those in non-zero momentum modes, and hence
\begin{eqnarray}
	\hat N_{\rm th} 
	&=&
	\sum_{p > 0}
	\{
	(u_p^2 + v_p^2) (	\hat b_p^\dag \hat b_p^{\phantom\dag}
	+
		\hat b_{-p}^\dag \hat b_{-p}^{\phantom\dag})
\nonumber\\
	&& \quad {}
	+ 2 u_p v_p 
	(	\hat b_p^\dag \hat b_{-p}^{\dag}
	+
		\hat b_{p}^{\phantom\dag} \hat b_{-p}^{\phantom\dag})
	+ 2 v_p^2
	\}
	\label{eq:split-non-condensate-number}
\end{eqnarray}
where $u_p$ and $v_p$ are real-valued Bogoliubov coefficients
[Eq.(258) of Ref.\cite{Kocharovsky2006}], normalized to 
$u_p^2 - v_p^2 = 1$, and where the $b_p$ are the bosonic
operators for Bogoliubov quasi-particles.
This Bogoliubov spectrum depends on the condensate occupation
number $N_c$. S\&{}Co make the approximation that
the dependence is weak enough so that $N_c$ can be replaced by
its average value $\langle N_c \rangle$ (i.e., by the first moment
of $P( N_c )$). This is consistent if $P( N_c )$ is narrow enough.
The operator identity~(\ref{eq:split-non-condensate-number}) is
also based on the assumption of a negligibly small 
probability $P( N_c = 0 )$ of finding 
the condensate mode empty [Eq.(210) 
of Ref.\cite{Kocharovsky2006}].

The quasi-particle operators are constructed such that 
the sum~(\ref{eq:split-non-condensate-number}) contains
mutually commuting operators (only momenta $p$ and $-p$ are
correlated), and the probability distribution 
$P( N_{\rm th} )$ can be found with standard techniques. The
following moments are found, for example
\begin{eqnarray}
	\langle N_{\rm th} \rangle &=&
	\sum_{p \ne 0} \{ 
	\BE_p u_p^2 + 
	(\BE_p + 1 ) v_p^2
	\}
\label{eq:average-Nth}
\\
	\sigma^2(N_{\rm th}) &=&
	\sum_{p \ne 0} \{ 
	\BE_p (\BE_p + 1 ) [
	1 + 8 u_p^2 v_p^2 
	]
	+ 2 u_p^2 v_p^2
	\} \qquad
\label{eq:variance-Nth}
\\
	\mu_3 &=&
	- \sum_{p \ne 0}
	(u_p^2 + v_p^2 ) \{ 
	\BE_p ( 2 \BE_p^2  + 3 \BE_p + 1 ) 
	\nonumber\\
	&& {} \times [
	1 + 16 u_p^2 v_p^2 
	]
	+ (2 \BE_p + 1)
	4 u_p^2 v_p^2
	\}
\label{eq:skewness-Nth}
\end{eqnarray}
where $\BE_p = \BE( \epsilon_p )$ is the Bose-Einstein statistics
and where 
$\mu_3 = \langle (\hat N_{\rm th} - 
\langle \hat N_{\rm th} \rangle )^3 \rangle$
is the third central moment. Its nonzero value is a clear indication of
non-Gaussian statistics. These results are obtained within the
Bogoliubov approximation with a weak thermal fraction, 
$\langle N_{\rm th} \rangle \ll N$, hence at low temperatures. 

Svidzinsky and Scully
\cite{Svidzinsky2006} have generalized this approach to
any temperature, 
using a rate equation \emph{Ansatz}
similar to quantum laser theory~\cite{SargentScully}.
The growth and loss rates for the condensate depend on $N_c$
and are described by a rational (Pad\'e) approximation.
This leads to a closed formula for the stationary $P( N_c )$. 
The Pad\'e parameters are matched to the low-temperature limit
and can be expressed in terms of the 
moments~(\ref{eq:average-Nth}--\ref{eq:skewness-Nth}). The
equation $\langle N_c \rangle = N - \langle N_{\rm th} \rangle$
must be solved iteratively since the Bogoliubov modes in
Eq.(\ref{eq:average-Nth}) depend themselves on $\langle N_c \rangle$.
It has been pointed out in Ref.\cite{Idziaszek2009} that the dependence
of the Bogoliubov spectrum on the condensate number,
$\epsilon_p( N_c ) \ne \epsilon_p( \langle N_c \rangle )$,
actually leads to an observable difference in the counting statistics,
although it makes the calculations much more involved.

We calculate the counting statistics in 
Figs.\ref{fig:counting-statistics},
\ref{fig:Scully-sGPe_moments}
using the theory of 
Ref.\cite{Svidzinsky2006} and making the identification
$u_p^2 \mapsto \Vert u_k \Vert^2$ to the Bogoliubov modes
in a trap. The moments of $P( N_c )$ in 
Fig.\ref{fig:Scully-sGPe_moments} are calculated from
Eqs.(\ref{eq:average-Nth}--\ref{eq:skewness-Nth}).

\clearpage
\bibliographystyle{apsrev}
\bibliography{twa-sgpe}

\end{document}